\documentclass[12pt]{article}
\pdfoutput=1

\renewcommand{\arraystretch}{1.2}

\usepackage[english]{babel}

\usepackage[frozencache,cachedir=.]{minted}
\usepackage[sort&compress,numbers,merge]{natbib}
\usepackage[dvipsnames]{xcolor}
\usepackage{needspace} %So that we add needspace to "\contribution" ensuring it doesn't go at the bottom of a page

\definecolor{red}{rgb}{0.6,.0706,.1373}
\definecolor{blue}{rgb}{0,0.396,0.741}
\definecolor{teal}{rgb}{0.11,0.6,0.6}
\colorlet{blueref}{blue!80!black}
\colorlet{bluelink}{blue!90!black}
\usepackage{hyperref}
\hypersetup{
	colorlinks, 
	bookmarksopen, 
	bookmarksnumbered,
	citecolor=blueref, 		%color of links to bibliography
	linkcolor=bluelink,	%color of internal links
	urlcolor=teal,			%color of external links 
	pdftitle={Computing Tools for Effective Field Theories},    %   - title (PDF meta)
}

\usepackage{latexsym,mathrsfs,dsfont}

\usepackage{microtype,xspace}
\usepackage[utf8]{inputenc}	% Allows for writing special charachters in the tex-file 
\usepackage[T1]{fontenc}
\usepackage{fontawesome}
\usepackage{subfig}
\usepackage[normalem]{ulem}
 
\usepackage{amsfonts,amsmath,amssymb,bm,mathrsfs,mathtools,dsfont} 	% Standard mathematics 
\usepackage{slashed}
\numberwithin{equation}{section}
\allowdisplaybreaks %Allows pagebreak during multiline math enviroments

\RequirePackage{multirow}
\RequirePackage{xcolor,colortbl}

\usepackage{tabularx,tabulary,booktabs}
\newcolumntype{L}[1]{>{\raggedright\let\newline\\\arraybackslash\hspace{0pt}}m{#1}}
\newcolumntype{C}[1]{>{\centering\let\newline\\\arraybackslash\hspace{0pt}}m{#1}}
\newcolumntype{R}[1]{>{\raggedleft\let\newline\\\arraybackslash\hspace{0pt}}m{#1}}
\makeatletter
\def\hlinewd#1{%
\noalign{\ifnum0=`}\fi\hrule \@height #1%
\futurelet\reserved@a\@xhline}
\makeatother

\makeatletter
\g@addto@macro\bfseries{\boldmath}
\makeatother

\usepackage{mmacells}

\usepackage{listings}
\definecolor{shaded}{RGB}{245,245,245}
\lstdefinestyle{mathematica}{
  basicstyle=\ttfamily\mdseries,
  backgroundcolor=\color{shaded},
  language=Mathematica,
  frame=false	
}

\definecolor{codegreen}{rgb}{0,0.6,0}
\definecolor{codegray}{rgb}{0.5,0.5,0.5}
\definecolor{codegris}{rgb}{0.92,0.92,0.92}
\definecolor{codepurple}{rgb}{0.58,0,0.82}
\definecolor{backcolour}{rgb}{0.95,0.95,0.92}

\lstdefinestyle{mystyle}{
    backgroundcolor=\color{backcolour},   
    commentstyle=\color{codegreen},
    keywordstyle=\color{magenta},
    numberstyle=\tiny\color{codegray},
    stringstyle=\color{codepurple},
    basicstyle=\ttfamily\footnotesize,
    breakatwhitespace=false,         
    breaklines=true,                 
    captionpos=b,                    
    keepspaces=true,                 
    numbers=left,                    
    numbersep=5pt,                  
    showspaces=false,                
    showstringspaces=false,
    showtabs=false,                  
    tabsize=2
}

\newcommand{\mme}{\texttt{matchmakereft}\xspace}
\newcommand{\Mme}{\texttt{Matchmakereft}\xspace}
\newcommand{\feynrules}{\texttt{feynrules}\xspace}

\DeclareMathOperator{\Tr}{Tr}

\newcommand{\contribution}[3][]{
    \needspace{5\baselineskip}
    \subsection{#2}
    \vspace{-.5em}
    \noindent
    \begin{minipage}[t]{.4\textwidth}
    \vspace{0pt} % Required for the vertical alignment... For some reason!
    #1 \hfill
    \end{minipage}% This must go next to `\end{minipage}`
    \begin{minipage}[t]{.6\textwidth}
    \vspace{0pt} % Required for the vertical alignment... For some reason!
    \begin{flushright} \textit{#3} \end{flushright}
    \end{minipage}
    \vspace{1em}
}

\newcommand{\chpt}{$\chi$PT}
\newcommand{\nn}{\nonumber\\ }
\renewcommand{\O}{\mathcal{O}}
\renewcommand{\L}{\mathcal{L}}
\NewDocumentCommand{\lwc}{ m m O{} o }{
	L^{\ifblank{#3}{}{#3,}#2 }_{\IfNoValueTF{#4}{#1}{\substack{#1\\#4}}}
}

\def\ChiQED/{$\chi$QED}
\usepackage{ifthen}
\newcommand{\dInt}[2][]{%
    \ifthenelse{\equal{#1}{}}
    {\ensuremath{\operatorname{d}{#2}\;}}
    {\ensuremath{\operatorname{d}^{#1}{#2}\;}}
}

\renewcommand{\imath}{i} % {\operatorname{i}}
\newcommand{\hc}{\text{h.c.}}

\newcommand{\Proj}[1]{{\mathbb{P}_\text{#1}}}

\usepackage[capitalise]{cleveref}
\newcommand{\ii}{\ensuremath{\mathrm{i}}}

\newcommand{\matchete}{\texttt{Matchete}\xspace}
\mmaSet{
  morefv={gobble=2},
  linklocaluri=mma/symbol/definition:#1,
  morecellgraphics={yoffset=1.9ex},
  moredefined={DefineGaugeGroup, DefineField, DefineCoupling,
    Fermion, Scalar, Charges, Mass, SelfConjugate, Heavy, 
    H, U1, SU, fund, eps, CG, l, EFTOrder, Indices, LoopOrder, ModelParameters,
    FreeLag, NiceForm, PlusHc, Bar, PR, PL, Match, GreensSimplify, EOMSimplify, LoadModel,
    Match, CovariantLoop, HcSimplify, Contract, Light, m, SelectOperatorClass,
    FundAlphabet, AdjAlphabet, IndexAlphabet, DefineFlavorIndex, Chiral, LeftHanded, RightHanded, RelabelIndices, CheckLagrangian
  }
}

\definecolor{prule}{gray}{0.80}
\definecolor{pbg}{gray}{0.98}
\definecolor{pcomment}{gray}{0.50}
\definecolor{pstring}{HTML}{AA1111}
\definecolor{pkw}{HTML}{00BB00}
\definecolor{poutput}{HTML}{000099}

\lstdefinelanguage{ppython}{
  language=Python,
  basicstyle=\footnotesize\ttfamily,
  commentstyle=\color{pcomment}\ttfamily,
  stringstyle=\color{pstring}\ttfamily,
  keywordstyle=\color{pkw}\ttfamily,
  rulecolor=\color{prule},
  frame=single,
  backgroundcolor=\color{pbg},
}
\lstdefinelanguage{poutput}{
  morekeywords={Output},
  basicstyle=\footnotesize\ttfamily,
  keywordstyle=\it\vspace{3pt},
  rulecolor=\color{prule},
  frame=single,
  backgroundcolor=\color{white},
  aboveskip=-6pt
}

\usepackage{lipsum}
\usepackage{bbold}

\newcommand{\cL}{\mathcal{L}}
\newcommand{\cC}{\mathcal{C}}
\newcommand{\cO}{\mathcal{O}}
\newcommand{\cA}{\mathcal{A}}
\newcommand{\cF}{\mathcal{F}}

\newcommand{\cN}{\mathcal{N}}

\newcommand{\cpp}{\texttt{C++}\xspace}

\newcommand{\EOS}{\texttt{EOS}\xspace}
\newcommand{\FlavBit}{\texttt{FlavBit}\xspace}
\newcommand{\HEPfit}{\texttt{HEPfit}\xspace}
\newcommand{\flavio}{\texttt{flavio}\xspace}
\newcommand{\Jupyter}{\texttt{Jupyter}\xspace}

\newcommand{\Python}{\texttt{Python}\xspace}

\newcommand{\SuperIso}{\texttt{SuperIso}\xspace}

\def\sfr{{\tt SmeftFR}~}
\def\frules{{\tt FeynRules}~}
\def\webpage{\url{www.fuw.edu.pl/smeft}}

\mmaSet{
  morefv={gobble=2},
  linklocaluri=mma/symbol/definition:#1,
  morecellgraphics={yoffset=1.9ex},
  moredefined={
  EventYield, CrossSection, DifferentialCrossSection, 
  ChiSquareLHC, 
  MatchToSMEFT, SubstituteFF, SelectTerms, LHCSearch, ParallelHighPT, GetParameters, DefineParameters, PythonExport, WCxf, InitializeModel, HighPTLogo, DefineBasisAlignment,
  e, \(\nu\), u, d, WC, Coupling, FF, Mass, CKM, Vckm, Width, Param, Coefficients, EFTorder, OperatorDimension, OutputFormat, PTcuts, MLLcuts, Luminosity, SM, CombineBins, RescaleError, EFTscale, Mediators,
  Scalar, Tensor, Vector, DipoleL, DipoleQ
  }
}

\def\hc{\text{h.c.}}
\newcommand{\pkg}[1]{{\tt #1}\xspace}
\newcommand{\dsix}{\pkg{DsixTools}}
\newcommand{\dsixbf}{\pkg{\bf DsixTools}}

\newcommand{\mmebf}{\pkg{\bf Matchmakereft}}

\begin{document}

\thispagestyle{empty}
\begin{minipage}{15.5cm}
\vspace{-0.7cm}
\begin{flushright}
{\footnotesize \itshape
TUM-HEP-1465/23
}
\end{flushright}

%\vspace{-2.1cm}

\end{minipage}

\begin{flushleft}
{
\bf\LARGE 
Computing Tools for Effective Field Theories
}
\\[3mm]
{{\bf SMEFT-Tools 2022 Workshop Report}, 14-16th September 2022, Zürich}
\\[5mm]

{\it Editors:}\\[2mm]

Jason Aebischer${}^a$, Matteo Fael${}^b$, Javier Fuentes-Martín${}^c$,\\
Anders Eller Thomsen${}^d$, Javier Virto${}^{e,f}$ \\[3mm]

{\it Contributors:}\\[2mm]

{
    Lukas Allwicher${}^a$,
    Supratim Das Bakshi${}^c$,
    Herm\`{e}s B\'{e}lusca-Ma\"{i}to${}^g$,
    Jorge de Blas${}^c$,
    Mikael Chala${}^c$, 
    Juan Carlos Criado${}^c$,
    Athanasios Dedes${}^h$, 
    Renato M. Fonseca${}^c$,
    Angelica Goncalves${}^i$,
    Amon Ilakovac${}^g$,
    Matthias König${}^j$,
    Sunando Kumar Patra${}^k$,
    Paul K\"{u}hler${}^l$,
    Marija Ma\dj{}or-Bo\v{z}inovi\'{c}${}^g$,
    Miko{\l}aj Misiak${}^m$,
    Víctor Miralles${}^n$,
    Ignacy Na{\l\c{e}}cz${}^m$,
    M\'eril Reboud${}^o$,
    Laura Reina${}^i$,
    Janusz Rosiek${}^m$, 
    Michal Ryczkowski${}^m$, 
    Jos\'e Santiago${}^c$,
    Luca Silvestrini${}^n$,
    Peter Stangl${}^b$, 
    Dominik St\"{o}ckinger${}^l$,
    Peter Stoffer${}^{a,p}$,
    Avelino Vicente${}^{q,r}$,
    Matthias Wei\ss{}wange${}^l$
}\\[5mm]

{\scriptsize
${}^a$
Physik-Institut, Universit\"at Z\"urich, CH-8057 Z\"urich, Switzerland\\
${}^b$ CERN, Theoretical Physics Department, CH-1211 Geneva 23, Switzerland\\
${}^c$
Departamento de F\'isica Te\'orica y del Cosmos, Universidad de Granada, Campus de Fuentenueva, E–18071 Granada, Spain\\
${}^d$ Department of Physics, University of Basel, Klingelbergstrasse 82, CH-4056 Basel, Switzerland\\
${}^e$ Departament de F\'isica Qu\`{a}ntica i Astrof\'isica, Universitat de Barcelona,
Mart\'i i Franqu\'es 1, 08028 Barcelona, Catalunya\\
${}^f$ Institut de Ci\`{e}ncies del Cosmos (ICCUB), Universitat de Barcelona,
Mart\'i i Franqu\'es 1, 08028 Barcelona, Catalunya\\
${}^g$
Department of Physics, Faculty of Sciences, University of Zagreb,
Bijeni\v{c}ka cesta 32, HR-10000 Zagreb, Croatia\\
${}^h$ Department of Physics, University of Ioannina, GR 45110, Ioannina, Greece\\
${}^i$ Physics Department, Florida State University,
Tallahassee, FL 32306-4350, USA \\
${}^j$ Physik Department T31, Technische Universit\"at M\"unchen, James-Franck-Str. 1, D-85748 Garching, Germany\\
${}^k$ Bangabasi Evening College, Kolkata, India\\
${}^l$ Institut f\"{u}r Kern- und Teilchenphysik, TU Dresden,
Zellescher Weg 19, DE-01069 Dresden, Germany\\
${}^m$ Faculty of Physics, University of Warsaw, Pasteura 5, 02-093 Warsaw, Poland\\
${}^n$ INFN, Sezione di Roma, Piazzale A. Moro 2, I-00185 Roma, Italy \\
${}^o$ Institute for Particle Physics Phenomenology and Department of Physics, 
Durham University, Durham DH1 3LE, UK \\
${}^p$ Paul Scherrer Institut, 5232 Villigen PSI, Switzerland\\
${}^q$ Instituto de F\'isica Corpuscular, CSIC-Universitat de Val\`{e}ncia, 46980 Paterna, Spain \\[-6pt]
${}^r$ Departament de F\'isica Te\`{o}rica, Universitat de Val\`{e}ncia, 46100 Burjassot, Spain
}

\end{flushleft}

\vspace{3mm}

\noindent
{\bf Abstract}\\
In recent years, theoretical and phenomenological studies with effective field theories have become a trending and prolific line of research in the field of high-energy physics. In order to discuss present and future prospects concerning automated tools in this field, the SMEFT-Tools 2022 workshop was held at the University of Zurich from 14th-16th September 2022. The current document collects and summarizes the content of this workshop.

\setcounter{page}{0}
\thispagestyle{empty}
\newpage

{
    \hypersetup{linkcolor = black}
    \setcounter{tocdepth}{2}
    \tableofcontents
}

\newpage 
\section*{Preface by the Editors}
\addcontentsline{toc}{section}{\protect\numberline{}Preface by the Editors}

The current developments in beyond-the-Standard-Model (BSM) phenomenology point to an ever greater use of Effective Field Theories (EFTs). With no concrete hints of a forthcoming discovery of on-shell new physics (NP), we see no reason that this trend should change any time soon. In fact, even a discovery of new high-energy resonances would call for the use of EFTs to study many of their observable effects. The role of computer tools is central to the successful use of EFT in BSM physics: simply put, the amount of repetitive computations is all but impossible to perform on a case by case basis without them. For these reasons, we gathered creators and developers of EFT tools for another workshop three years after the first SMEFT-Tools workshop~\cite{Proceedings:2019rnh}.

The SMEFT-Tools 2022 workshop received contributions from a large part of the EFT theory community, especially, as it pertains to the computer tools of the field. As a result, this review is a comprehensive, if not quite complete, report on the current status of the tools available in the field. Additionally, several speakers at the workshop were presenting new results dealing with the more formal theory aspects. These results frame the current and future developments of EFT tools and are crucial to the ever growing capabilities of the tools. However, the theory developments included in this report are merely a sample of what is being undertaken in the field as a whole; the field is simply too active to include all, or even most, of the developments here. 

The introductory section provides some context and motivation for the use of EFTs and discusses some of the trends in EFT used for BSM. We have grouped the other contributions in two main sections: On the one hand, Section~\ref{section:matchingcodes} details computer tools for the study of ultraviolet (UV) models using EFTs. This section describes tools for matching ultraviolet (UV) models to EFTs, automating the renormalization-group (RG) evolution of EFT coefficients, generating EFT operator bases, and a proposal for a unified format for the storage of EFT matching results. On the other hand,  Section~\ref{section:pheno} describes computer tools necessary for the phenomenological study of EFTs. These tools are equally invaluable for bottom--up or top--down analyses. This section contains four different tools for the key task of performing global fits to experimental data, along with a code for automatically deriving the Feynman rules of the Standard Model Effective Theory (SMEFT). In both sections, contributions covering recent theory developments relevant for future implementations or practical applications of EFT tools are also included.

\newpage
\section{Introduction and Motivation}
\begin{flushright} \hfill \textit{José Santiago and Peter Stoffer}\end{flushright}
\vspace{1em}

EFTs have been a basic tool in particle physics for many years. In most cases EFTs were used in the context of well-defined, usually renormalizable, models, either because they were the only way to compute certain observables (for example, due to the strong coupling of QCD at low energies) or because their use greatly simplified the calculation of interest (gluon-fusion Higgs production at a high perturbative order in the infinite mass limit is a clear example). Their application to the study of physics beyond the Standard Model (SM), while already present in the past, has experienced an exponential increase in the last decade. The reasons for this are two-fold: first, the LHC and other experiments are producing increasingly better limits on the mass of new particles, searching in a multitude of different channels, which seems to clearly indicate the presence of a mass gap between the scale of new physics and the energies at which most experimental observables are measured; second, and this is especially relevant for this workshop, the last few years have seen the appearance of a plethora of new computer tools, that simplify, and in many cases fully automate, the tedious calculations needed to apply EFTs to new physics searches. 

\subsection*{Connecting Theory and Experiment via EFTs}
\addcontentsline{toc}{subsection}{\protect\numberline{}Connecting Theory and Experiment via EFTs}

The problem of obtaining the implications of experimental data on models of new physics is highly non-trivial. The vast number of observables measured experimentally has to be computed, via complicated, sometimes multi-loop calculations, for each particular model of new physics. These difficult calculations have to be repeated for every experimental observable and every model with the added complication that, despite the very large number of new physics models developed by theorists, we are not guaranteed that the true description of Nature falls into one of these models.

EFTs simplify the problem of obtaining the phenomenological implications of experimental data on new models by splitting the calculation in two (mostly independent) steps. In the first one, the bottom-up approach, the experimental observables are computed, to the required order in perturbation theory, in terms of the Wilson coefficients (WCs) of the corresponding effective Lagrangian. This process can be performed with no mention of any new physics model and therefore represents a mostly model-independent parametrization of experimental data in the form of global fits (or rather a global likelihood), see Section~\ref{section:pheno}. In the second step, the top-down approach, the WCs of the effective Lagrangian are computed in terms of the couplings and masses of specific UV models, that complete the EFT at high energies. This calculation, called matching, has to be done for every model of new physics (but not any more for every observable) but, thanks to recently developed tools, it can be fully automated (see Section \ref{section:matchingcodes}). When the bottom-up and top-down approaches are combined one can obtain the phenomenological implications of any experimental observable in any UV model and, thanks to existing computer tools, in a mostly automated way.

%%%%%%%%%%%%%%%%%%%%%%%%%%%%%%%%%%%%%%%%%%%%%%%%%%%%%%%%%%
\subsection*{The SMEFT and the LEFT}
\addcontentsline{toc}{subsection}{\protect\numberline{}The SMEFT and the LEFT}

The absence of evidence for physics beyond the SM in direct LHC searches suggests that new particles are either very weakly coupled~\cite{Bauer:2020jbp} or much heavier than the electroweak scale. In the latter scenario, their effects at energies below the scale of new physics can be described by an EFT. Depending on the assumption about the nature of the Higgs particle, this is either the Standard Model Effective Field Theory (SMEFT)~\cite{Buchmuller:1985jz,Grzadkowski:2010es} or Higgs Effective Field Theory (HEFT)~\cite{Feruglio:1992wf,Grinstein:2007iv}.  In particular, the SMEFT is the most general EFT invariant under the SM gauge symmetry, $SU(3)_c\times SU(2)_L\times U(1)_Y$, involving only SM particles with the Higgs field taken as an $SU(2)_L$ doublet. 

The SMEFT Lagrangian up to dimension-six operators is given by
\begin{align}
{\mathcal L}_{\rm SMEFT}^{d\leq6} &= {\mathcal L}_{\rm SM} + \sum_{k} C_k^{(5)} Q_k^{(5)}+ \sum_{k} C_k^{(6)}Q_k^{(6)}\,,
\end{align}
with ${\mathcal L}_{\rm SM}$ being the SM Lagrangian. There is only one term at dimension five corresponding to the Weinberg operator~\cite{Weinberg:1979sa}. This operator violates baryon number in two units and yields Majorana masses for the neutrinos after electroweak symmetry breaking. At dimension six, there are $59$ terms that preserve baryon number and another $5$ that violate baryon and lepton numbers in one unit. These are commonly presented in the so-called \textit{Warsaw basis}~\cite{Grzadkowski:2010es}. The complete set of RG equations for the dimension-six SMEFT in the Warsaw basis has been calculated in~\cite{Jenkins:2013zja,Jenkins:2013wua,Alonso:2013hga,Alonso:2014zka}. As we describe in the sections below, these advances, together with simultaneous theoretical and computational developments towards the automation of one-loop matching calculations, pave the way to the systematic use of EFT methods in the analysis of NP models.

For processes below the electroweak scale, another EFT should be used, wherein the heavy SM particles, i.e., the top quark, the Higgs scalar, and the heavy gauge bosons, are integrated out. This low-energy effective field theory (LEFT) is a gauge theory invariant only under the unbroken SM groups $SU(3)_c \times U(1)_{\rm em}$, i.e., QCD and QED augmented by a complete set of effective operators. If matched to the SM at the electroweak scale, it corresponds to the Fermi theory of weak interaction~\cite{Fermi:1934sk}, but when all operators invariant under the unbroken gauge groups are included, it also describes the low-energy effects of arbitrary heavy physics beyond the SM.

The LEFT is defined by the Lagrangian
\begin{align}
	\label{eq:LEFTLagrangian}
	\L_{\rm LEFT} = \L_{\rm QCD+QED} + \L^{(3)}_{\slashed L} + \sum_{d\ge5} \sum_i L_i^{(d)} \O_i^{(d)} \, ,
\end{align}
where the QCD and QED Lagrangian is given by
\begin{align}
	\label{eq:qcdqed}
	\L_{\rm QCD + QED} &= - \frac14 G_{\mu \nu}^A G^{A \mu \nu} -\frac14 F_{\mu \nu} F^{\mu\nu} + \theta_{\rm QCD} \frac{g^2}{32 \pi^2} G_{\mu \nu}^A \widetilde G^{A \mu \nu} +  \theta_{\rm QED} \frac{e^2}{32 \pi^2} F_{\mu \nu} \widetilde F^{\mu \nu} \nn
		&\quad + \sum_{\psi=u,d,e,\nu_L}\overline \psi i \slashed{D} \psi   - \left[ \sum_{\psi=u,d,e}  \overline \psi_{Rr} [M_\psi]_{rs} \psi_{Ls} + \text{h.c.} \right] \, .
\end{align}
The additional operators are the Majorana-neutrino mass terms $\L_{\slashed L}^{(3)}$ at dimension three, as well as operators at dimension five and above. At dimension five, there are photonic dipole operators for all the fermions (including a lepton--number-violating neutrino dipole operator) as well as gluonic dipole operators for the up- and down-type quarks. At dimension six, there are the CP-even and CP-odd three-gluon operators and a large number of four-fermion operators. The entire list of operators up to dimension six can be found in~\cite{Jenkins:2017jig}, including operators that violate baryon and lepton number.

This theory has been extensively studied in the context of $B$ physics. The operator basis relevant for $B$-meson decay and mixing has been constructed in~\cite{Aebischer:2017gaw}. The complete LEFT operator basis up to dimension six in the power counting has been derived in~\cite{Jenkins:2017jig}, where also the tree-level matching to the dimension-six SMEFT above the weak scale was provided. By now, the LEFT operator basis is known up to dimension 9~\cite{Liao:2020zyx,Murphy:2020cly,Li:2020tsi}. Recently, the tree-level matching to the SMEFT has been extended to dimension eight in the SMEFT power counting~\cite{Hamoudou:2022tdn}. Partial results for lepton--flavor-violating operators were given already in~\cite{Ardu:2021koz}.

The complete one-loop LEFT RG equations were derived in~\cite{Jenkins:2017dyc}. Partial results for the RG equations were known before and have been studied to higher loop orders~\cite{Altarelli:1980fi,Buras:1989xd,Buras:1992tc,Ciuchini:1993vr,Buchalla:1995vs,Ciuchini:1997bw,Buras:2000if,Misiak:2004ew,Czakon:2006ss,Cirigliano:2012ab,Dekens:2013zca,Heeck:2013rpa,Pruna:2014asa,Bhattacharya:2015rsa,Aebischer:2015fzz,Davidson:2016edt,Feruglio:2016gvd,Crivellin:2017rmk,Bordone:2017anc,Misiak:2017woa,Cirigliano:2017azj,Fuentes-Martin:2020zaz,Aebischer:2017gaw,Gonzalez-Alonso:2017iyc,Falkowski:2017pss,Panico:2018hal}.
Within the SMEFT/LEFT framework, the one-loop RG equations at the high scale~\cite{Jenkins:2013zja,Jenkins:2013wua,Alonso:2013hga}, the tree-level matching~\cite{Jenkins:2017jig}, and the RG equations below the weak scale~\cite{Jenkins:2017dyc} allow one to resum the leading logarithms and to describe the indirect low-energy effects of heavy physics beyond the SM within one unified framework. The RG and matching equations have been implemented in several software tools, many of which were presented at the SMEFT-Tools workshops. Consistent EFT analyses at leading-log accuracy that combine constraints from experiments at very different energy scales are becoming standard.

For certain high-precision observables at low energies it is desirable to extend the analysis beyond leading logarithms. Steps in this direction have been taken, e.g., in~\cite{Pruna:2014asa,Crivellin:2017rmk,Panico:2018hal}. Partial results for the matching at the weak scale at one loop were derived in the context of $B$ physics in~\cite{Aebischer:2015fzz,Hurth:2019ula}. The complete one-loop matching between the SMEFT and the LEFT at dimension six was calculated in~\cite{Dekens:2019ept}. It can be used for fixed-order calculations at one-loop accuracy in cases where the logs are not large, and it is an ingredient in next-to-leading-log analyses within a resummed framework. Several tools are being developed that automate the one-loop matching between the EFT framework and UV models for new physics~\cite{Carmona:2021xtq,Fuentes-Martin:2022jrf}. 

At energies as low as the hadronic scale, additional complications appear due to the non-perturbative nature of QCD. In these low-energy processes, one should not work with perturbative quark and gluon degrees of freedom but rather perform either direct non-perturbative calculations of hadronic matrix elements of effective operators or switch to another effective theory in terms of hadronic degrees of freedom, i.e., chiral perturbation theory (\chpt{})~\cite{Weinberg:1968de,Gasser:1983yg,Gasser:1984gg}. In~\cite{Dekens:2018pbu}, the matching of semileptonic LEFT operators to \chpt{} has been discussed, which can be obtained within standard \chpt{} augmented by tensor sources~\cite{Cata:2007ns}. The chiral realization of four-quark operators was studied in~\cite{Pich:2021yll}, while~\cite{Akdag:2022sbn} analyzed $C$- and CP-odd LEFT operators up to dimension 8. If lattice QCD is employed to deal with the non-perturbative effects at low energies, one faces the problem that the EFT framework requires matrix elements of dimensionally renormalized operators. This necessitates another matching calculation to a scheme amenable to lattice computations. This matching has to be performed at a scale of a few GeV, which is already accessible to lattice computations but at the same time sufficiently high that perturbation theory can be assumed to work reasonably well. Traditionally, these matching calculations are based on regularization-independent momentum-subtraction (RI-MOM)  schemes~\cite{Martinelli:1994ty,Aoki:2007xm,Sturm:2009kb,Bhattacharya:2015rsa,Cirigliano:2020msr}, whereas in recent years the gradient flow~\cite{Luscher:2010iy,Luscher:2013cpa} has received attention~\cite{Rizik:2020naq,Mereghetti:2021nkt,Harlander:2022tgk,Buhler:2023gsg,Crosas:2023anw}.

\subsection*{Going Beyond}
\addcontentsline{toc}{subsection}{\protect\numberline{}Going Beyond}

%No evidence of direct production of new fundamental particles has been found so far. 
The great sensitivity achieved in the search for CP violation, rare meson decays, magnetic and electric dipole moments and lepton-flavor-violating processes requires improvements in the theoretical precision with EFTs. 
The need to include higher-order corrections is twofold. On the one hand, the inclusion of higher-order corrections, especially in QCD, allows to better assess
the uncertainties in the theoretical calculation. On the other hand, some new-physics effects are only generated once higher-loop effects have been accounted for in certain UV completions, thus including them naturally yields to better constraints on the underlying theory. 

In fact, it is often the case that the leading effects of new physics are due to loop-level processes. The last decade has seen results for one-loop running in the SMEFT~\cite{Jenkins:2013zja,Jenkins:2013wua,Alonso:2013hga,Alonso:2014zka,Machado:2022ozb} and the LEFT~\cite{Jenkins:2017dyc} and the one-loop SMEFT to LEFT matching~\cite{Dekens:2019ept}. Likewise, as we highlight in this manuscript, there has been recent substantial progress in the connection of NP models to their EFTs. Going beyond the leading logarithm effects requires systematic treatment of RG effects in the EFTs because of the scheme dependence of the anomalous dimension matrix and the matching coefficients appearing, for instance, in the chosen prescription for $\gamma_5$ in $d$ dimensions~\cite{tHooft:1972tcz,Breitenlohner:1977hr,Korner:1989is,Ferrari:1994ct,Trueman:1995ca,Jegerlehner:2000dz,Boughezal:2019xpp,Belusca-Maito:2020ala,Belusca-Maito:2021lnk,Cornella:2022hkc} and the definition of evanescent operators~\cite{Buras:1989xd,Dugan:1990df,Herrlich:1994kh,Dekens:2019ept,Aebischer:2022aze,Fuentes-Martin:2022vvu}.
To this end, consistent calculations across different EFTs and bases have given cause for a new look at the role of evanescent contributions~\cite{Fuentes-Martin:2022jrf,Aebischer:2022aze}. In the perspective of systematic multi-loop computations within EFTs, there has been a recent interest in the proper treatment of $ \gamma_5 $~\cite{Cornella:2022hkc,Belusca-Maito:2023wah,Naterop:2023dek}, a notorious stumbling block in dimensional regularization.

Higher-order anomalous dimensions have been calculated for subsets of dimension-six operators~\cite{Altarelli:1980fi,Buras:1989xd,Buras:1992tc,Ciuchini:1993vr,Buchalla:1995vs,Ciuchini:1997bw,Buras:2000if,Misiak:2004ew,Czakon:2006ss,Cirigliano:2012ab,Dekens:2013zca,Heeck:2013rpa,Pruna:2014asa,Bhattacharya:2015rsa,Aebischer:2015fzz,Davidson:2016edt,Feruglio:2016gvd,Crivellin:2017rmk,Bordone:2017anc,Misiak:2017woa,Cirigliano:2017azj,Fuentes-Martin:2020zaz,Aebischer:2017gaw,Gonzalez-Alonso:2017iyc,Falkowski:2017pss,Panico:2018hal}, but due to the hard technical nature of the calculations the complete matrix is not known. Given the large number of operators of the SMEFT and LEFT, it may be more convenient to consider first a generic EFT with an arbitrary number of real scalars and left-handed fermions and compute the anomalous dimensions and the RG in such theory invariant under 
a generic gauge group. Results for NLO running of the SMEFT or LEFT WCs can be then extracted in a second step by specifying the field content and the gauge group.

There has also been recent progress in expanding the EFT formulation beyond dimension-six operators, sparking the formulation of geometric EFTs~\cite{Alonso:2015fsp,Helset:2020yio} and the determination of higher-dimension bases~\cite{Liao:2020zyx,Murphy:2020rsh,Chala:2021cgt}, as well as the  counting of the EFT operators through the use of Hilbert series~\cite{Henning:2015alf}. Recent work has also started constraining the effects from higher-dimensional operators through the use of unitarity bounds, e.g.,~\cite{Distler:2006if,Vecchi:2007na,Low:2009di,Bellazzini:2014waa,Zhang:2018shp,Englert:2019zmt,Englert:2019zmt,Remmen:2019cyz,Zhang:2020jyn,Bonnefoy:2020yee,Gu:2020ldn,Alberte:2020bdz,Zhang:2021eeo,Henriksson:2021ymi,Davighi:2021osh,Chala:2021wpj,Fernandez:2022kzi,Haring:2022sdp}. 

\newpage
%%%%%%%%%%%%%%%%%%%%%%%%%%%%%%%%%%%%%%%%%%%%%%%%%%%%%%
%%%%%%%%%%%%%%%%%%%%%%%%%%%%%%%%%%%%%%%%%%%%%%%%%%%%%%
\section{Effective Field Theory Matching and Running
\label{section:matchingcodes}}
%%%%%%%%%%%%%%%%%%%%%%%%%%%%%%%%%%%%%%%%%%%%%%%%%%%%%%
%%%%%%%%%%%%%%%%%%%%%%%%%%%%%%%%%%%%%%%%%%%%%%%%%%%%%%

Despite the usefulness of the EFT approach, the interpretation of data in terms of NP models requires a direct connection between those models and their EFT description. This typically involves the calculation of sequential matching steps at the relevant mass thresholds, and RG equations between these thresholds and the scale of the observables. In recent years, many tools that (at least partially) automate these calculations have been developed. 

In the absence of light particles beyond those in the SM, the necessary calculations for RG running and matching below the NP mass threshold are known up to dimension-six operators~\cite{Jenkins:2013zja,Jenkins:2013wua,Alonso:2013hga, Alonso:2014zka,Aebischer:2015fzz,Jenkins:2017jig,Dekens:2019ept,Jenkins:2017dyc}. These results have been implemented into several computer tools including \texttt{DsixTools}~\cite{Celis:2017hod, Fuentes-Martin:2020zaz} (which we describe in Section~\ref{sec:DsixTools}), \texttt{wilson}~\cite{Aebischer:2018bkb}, and \texttt{RGESolver}~\cite{DiNoi:2022ejg}. The SMEFT RG evolution has also been incorporated~\cite{Aoude:2022aro} into the \texttt{MadGraph} Monte Carlo generator~\cite{Alwall:2014hca}. As far as tree-level matching is concerned, the Python package \texttt{MatchingTools}~\cite{Criado:2017khh} allows to perform a fully automated matching computation for arbitrary heavy particles and gauge groups. Furthermore, the matching code \texttt{CoDEx} (see Section~\ref{sec:Codex}) implements formulae based on path-integral methods~\cite{Henning:2014wua,Drozd:2015rsp,Fuentes-Martin:2016uol,Ellis:2017jns} to automate the matching of some NP models into the dimension-six SMEFT. Further matching tools that rely on functional methods are \texttt{SuperTracer}~\cite{Fuentes-Martin:2020udw} and \texttt{STrEAM}~\cite{Cohen:2020fcu}.

Although it might be tempting to think of the target EFT as the SMEFT, many realistic BSM constructions contain several energy scales, calling for intermediate EFTs, or feature additional light states, such as axion-like or dark-matter particles, thus, demanding extensions of the SMEFT (see, e.g.,~\cite{Criado:2021trs,Aebischer:2022wnl,Chala:2020wvs,Bauer:2020jbp,Galda:2021hbr}). Furthermore, some phenomenological studies require extending EFT calculations beyond dimension-six operators (see, e.g.,~\cite{Chala:2021wpj,Banerjee:2022thk,Dawson:2022cmu,Chala:2023jyx,Banerjee:2023iiv} for recent literature examples). Reflecting on this, a new generation of tools is now aiming at solving the more general problem of completely automating one-loop matching and RG evolution of arbitrary weakly-coupled models. The most notable examples in this direction are \texttt{matchmakereft} and \texttt{matchete}, described in Sections~\ref{sec:MME} and~\ref{sec:matchete}, respectively. 

Additional developments to assist matching calculations are also described in this section. In particular, the computer tool \texttt{Sym2Int} (see Section~\ref{sec:Sym2Int}),\footnote{Other tools that allow for finding higher-dimensional operators in the SMEFT are \texttt{BasisGen}~\cite{Criado:2019ugp}, \texttt{DEFT}~\cite{Gripaios:2018zrz} and \texttt{ABC4EFT}~\cite{Li:2022tec}.} which automates the construction of EFT basis, and the MatchingDB format (see Section~\ref{sec:MatchingDB}), aimed at standardizing the storage of matching results.

%%%%%%%%%%%%%%%%%%%%%%%%%%%%%%%%%%%%%%%%%%%%%%%%%%%%%%%%%%%%%%%%%%%%%%%%%%%%%%%%%%%%%%%%%%%%%%%%
\contribution[{\includegraphics[width=3cm]{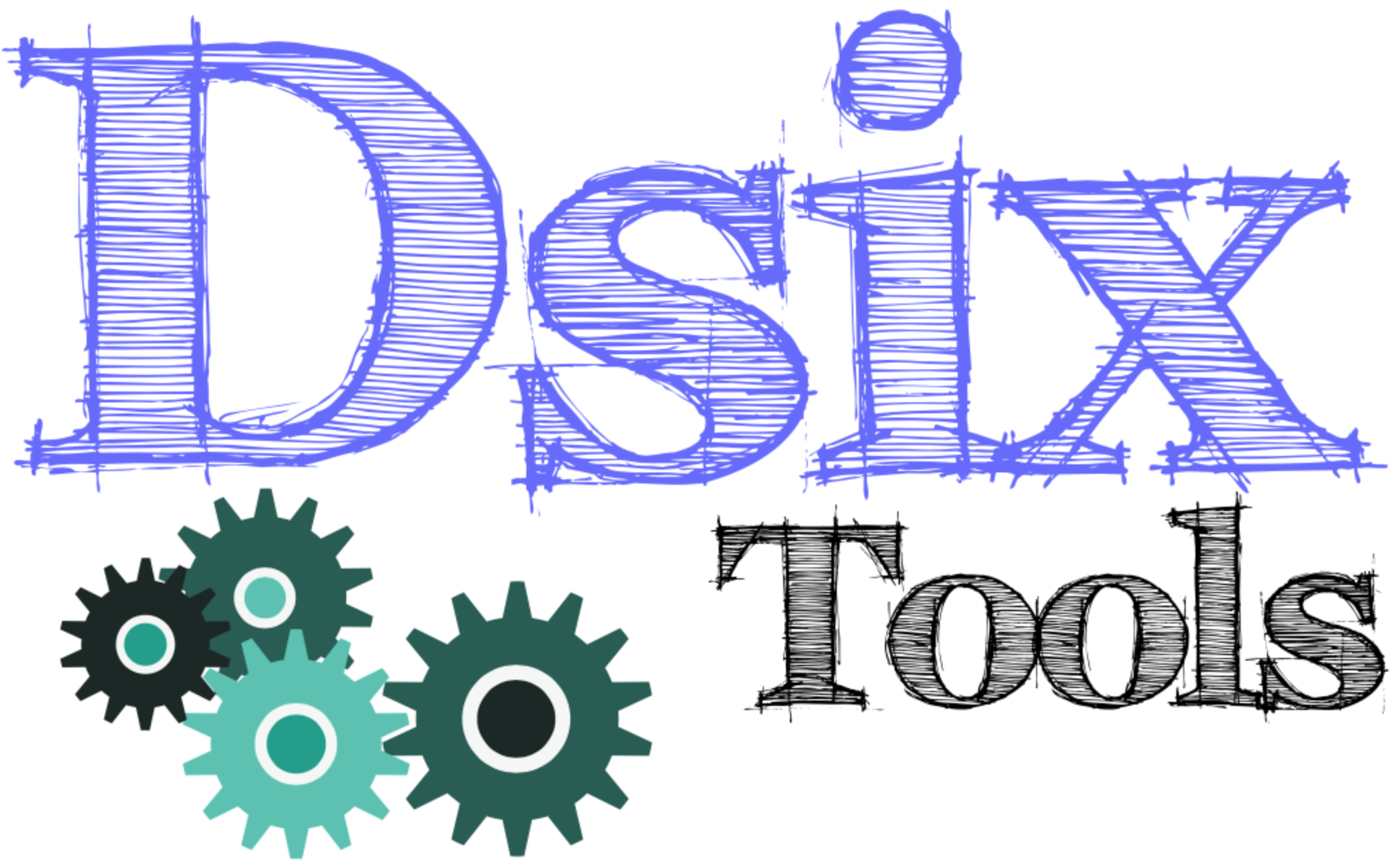}}]{\texttt{DsixTools}: The Effective Field Theory Toolkit}{Avelino Vicente}{}\label{sec:DsixTools}
%%%%%%%%%%%%%%%%%%%%%%%%%%%%%%%%%%%%%%%%%%%%%%%%%%%%%%%%%%%%%%%%%%%%%%%%%%%%%%%%%%%%%%%%%%%%%%%%

\dsix is a {\tt Mathematica}~\cite{Celis:2017hod, Fuentes-Martin:2020zaz} package for the matching and renormalization group evolution from the NP scale to the scale of low energy observables. The current version of \dsix fully integrates the SMEFT and the LEFT, treating both theories on an equal footing. It allows the user to perform the full one-loop renormalization group evolution of the WCs in the SMEFT and in the LEFT (with SM $\beta$ functions up to 5-loop order in QCD), and the full one-loop SMEFT-LEFT matching at the electroweak scale. Therefore, the user can start with some numerical values for the SMEFT WCs at the high-energy scale $\Lambda_{\rm UV}$, in principle obtained after matching to a specific NP model, and translate them into numerical values for the LEFT WCs at the low-energy scale $\Lambda_{\rm IR}$, where some observables of interest can be computed. This is achieved by adopting some conventions and implementing some results in the recent literature:
\begin{itemize}
\item The Warsaw basis~\cite{Grzadkowski:2010es} for the SMEFT, for which full one-loop RG equations~\cite{Jenkins:2013zja,Jenkins:2013wua,Alonso:2013hga,Alonso:2014zka,Antusch:2001ck} are known.
\item Full one-loop SMEFT-LEFT matching~\cite{Jenkins:2017jig,Dekens:2019ept}.
\item The San Diego basis~\cite{Jenkins:2017jig} for the LEFT, for which full one-loop RG equations~\cite{Jenkins:2017dyc} are known.
\end{itemize}
All these results can be used in a visually accessible and operationally convenient way thanks to \dsix. In addition to running and matching numerical routines, it also includes several functions for analytical applications, as well as user-friendly SMEFT/LEFT dictionary tools. Since version 2.1, \dsix also admits input obtained with \mme, thus extending its capabilities.

The simplest way to download and install \dsix is to run the following command in a {\tt Mathematica} session:\footnote{It requires {\tt Mathematica} version 9.0 (or newer).}
\begin{lstlisting}[style=mathematica]
Import["https://raw.githubusercontent.com/DsixTools/DsixTools/
master/install.m"];
\end{lstlisting}
This will download and install \dsix, activate the documentation
and load the package. Alternatively, \dsix can also be installed
manually. Finally, \dsix can also be loaded (once installed) by
running the usual
\begin{lstlisting}[style=mathematica]
Needs["DsixTools`"]
\end{lstlisting}

\bigskip

\subsubsection{What \dsixbf can do for you}
\label{sec:sec2}

For a full and updated list with all the tools provided by \dsix we
refer to the manual on the package website~\cite{website}. We will now
concentrate on some useful features that illustrate what \dsix can do
for you in practice. A \textit{demo notebook} with these and other
examples of use is also provided at~\cite{notebook}. 

\bigskip
{\bf User-friendly SMEFT \& LEFT information.} \dsix contains several
routines and functions that allow one to use the tool as a SMEFT/LEFT
dictionary. For instance, one can load \dsix and execute the command
\begin{lstlisting}[style=mathematica]
ObjectInfo[CHl1]
\end{lstlisting}
to print many details about the $C_{\varphi \ell}^{(1)}$ SMEFT WC. One
may learn the definition of the associated $Q_{\varphi \ell}^{(1)}$
operator, its dimensionality and type (2-fermion in this case). In
case of WCs carrying flavor indices, such as this one, this command
also prints information about the possible symmetries under exchange
of indices, the number of independent coefficients or the relations
(due to Hermiticity, for example) among them. This information is
displayed in a user-friendly way. Similarly, with
\begin{lstlisting}[style=mathematica]
ObjectInfo[LG]
\end{lstlisting}
one would get the same information about the LEFT WC $L_G$. Finally,
with the functions
\begin{lstlisting}[style=mathematica]
  SMEFTOperatorsGrid
  SMEFTOperatorsMenu
\end{lstlisting}
and the analogous ones in the LEFT, \dsix shows a visual grid or a
dropdown menu with all the WCs of the theory. The user can now click
on any of them to run the {\tt ObjectInfo} function on the selected WC
and obtain all its properties.

\bigskip
{\bf Introducing and changing input values.} There are two methods to
introduce input values in \dsix: with the {\tt NewInput} routine or
from a file. In the latter case one can choose between a native \dsix
format or the {\tt WCxf} format~\cite{Aebischer:2017ugx}. Let us focus
on the former case. With the {\tt NewInput} routine the user loads the
input values directly in the {\tt Mathematica} notebook. Only the
non-zero WCs must be given. The rest will be assumed to vanish. For
instance, the command
\begin{lstlisting}[style=mathematica]
NewInput[{Clq1[1, 1, 1, 2] -> 1, Clq1[1, 1, 2, 1] -> 1,
CHBtilde -> -0.5}];
\end{lstlisting}
sets $[C_{\ell q}^{(1)}]_{1112} = [C_{\ell q}^{(1)}]_{1121} = 1$
GeV$^{-2}$ and $C_{\varphi \widetilde B} = - 0.5$ GeV$^{-2}$. We note
that dimensionful quantities in \dsix are always given in GeV to the
proper power. In \dsix, the input values for the parameters of the
effective theory at work (SMEFT or LEFT) are stored as replacement
rules in a dispatch variable called {\tt Input Values}. Then, after
defining an input, the user can easily read it as
\begin{lstlisting}[style=mathematica]
Clq1[1, 1, 1, 2] /. InputValues
\end{lstlisting}
Once the input values have been set, the user can change them
individually at any moment in the notebook. This is done with the {\tt
  ChangeInput} routine. For example, the line
\begin{lstlisting}[style=mathematica]
ChangeInput[{CHBtilde -> 0.6}]
\end{lstlisting}
changes the value of $C_{\varphi \widetilde B}$ to $0.6$
GeV$^{-2}$. Finally, \dsix produces a warning message when the WCs
provided by the user lead to an invalid set of input values. There are
two possible reasons for this:
\begin{enumerate}
\item {\bf Non-Hermiticity errors}: Some WCs are related due to the Hermiticity of the Lagrangian. For instance, $[C_{\ell q}^{(1)}]_{1112} = [C_{\ell q}^{(1)}]^\ast_{1121}$ must necessarily hold.
\item {\bf Antisymmetry errors}: Some LEFT WCs are antisymmetric under the exchange of two flavor indices. For instance, $[L_{\nu \gamma}]_{11} = 0$ must necessarily hold.
\end{enumerate}
When the user's input is not consistent with any of these
restrictions, a warning is issued and \dsix corrects the input by
replacing it by a new one that ensures a complete consistency of the
Lagrangian. The list of invalid input values can be seen by clicking
on the button {\tt Input errors}. We note, however, that in some cases
other WCs, related to these by the two reasons given above, may be
modified too.

\bigskip
{\bf A simple \dsixbf program.} Let us illustrate how easily one can
use \dsix with a simple but complete program, given by the following
three lines after opening {\tt Mathematica} and loading \dsix:
%
% \begin{widetext}
\begin{lstlisting}[style=mathematica]
NewInput[{Clq1[2, 2, 3, 3]->1.0/HIGHSCALE^2}, HIGHSCALE->10^4];
RunDsixTools;
D6run[Clq1[2, 2, 3, 3]]/.\[Mu]->EWSCALE
\end{lstlisting}
% \end{widetext}
%
Here we consider an example SMEFT input with $[C_{\ell
    q}^{(1)}]_{2233} = 1/\Lambda_{\rm UV}^2$, given at $\Lambda_{\rm
  UV} = 10$ TeV. The rest of the SMEFT WCs are assumed to vanish at
$\Lambda_{\rm UV}$. Notice that input for the energy scales must be
given too. However, $\Lambda_{\rm EW}$ and $\Lambda_{\rm IR}$ are
taken to be equal to $m_W$ and $5$ GeV by default, and then only
$\Lambda_{\rm UV}$ must be provided. In the first line of this
program, the {\tt NewInput} routine is used to introduce the SMEFT WCs
as well as the NP energy scale $\Lambda_{\rm UV}$. In the second line
we make use of {\tt RunDsixTools}, one of the most important routines
in \dsix. It runs the SMEFT RG equations, it matches the resulting SMEFT
Lagrangian at the electroweak scale onto the LEFT one and runs down to
$\Lambda_{\rm IR}$ with the LEFT RG equations. The results of this process can
be obtained by means of the {\tt D6run} function, which returns
interpolating functions that can be evaluated for any value of the
energy scale $\mu$. For instance, in this program we choose to print
$[C_{\ell q}^{(1)}]_{2233}$ at the electroweak scale. Last but not
least, we emphasize that \dsix not only provides numerical
routines. In fact, all the analytical information in the code can be
printed and used in {\tt Mathematica} sessions. There are plenty of
examples of this. For instance, with the command
\begin{lstlisting}[style=mathematica]
LeuVLL[2, 2, 1, 1] /. MatchAnalytical
\end{lstlisting}
the user can display the analytical expression for the WC
$[L_{eu}^{V,LL}]_{2211}$ of the LEFT after matching at one-loop to the
SMEFT WCs. Similarly, the SMEFT and LEFT $\beta$ functions can be
readily accessed as
\begin{lstlisting}[style=mathematica]
  \[Beta][gs]
  \[Beta][LeuVLL[2, 2, 1, 1]]
\end{lstlisting}
We refer to the demo notebook~\cite{notebook} for examples of use of
other \dsix routines and functions.

\bigskip
{\bf Using \mmebf results.} The first step in the study of specific NP
models with the tools described here is the matching of the model to
an EFT. If the NP degrees of freedom lie at high energies, this EFT is
generally the SMEFT. Even though this theory is very well known
nowadays, the calculation might be hard, especially if done at
one-loop. Since version 2.1, \dsix admits input obtained with
\mme~\cite{Carmona:2021xtq}, a fully automated {\tt Python} code to
compute the tree-level and one-loop matching of arbitrary models onto
arbitrary EFTs. Its use is very simple. Let us illustrate it with an
example NP model that extends the SM field inventory with a
right-handed neutrino $N \sim (\mathbf{1},\mathbf{1})_0$ and a scalar
leptoquark $S \sim (\mathbf{3},\mathbf{2})_{\frac{1}{6}}$, where we
denote their representations under $(\rm SU(3)_c, SU(2)_L)_{\rm
  U(1)_Y}$. The NP Lagrangian contains the pieces
\begin{align}
  \mathcal{L}_N &= i \overline N \, \gamma_\mu D^\mu N - \frac{1}{2} M_N \, \overline{N^c} \, N \, , \\
  \mathcal{L}_S &= D_\mu S^\dagger \, D^\mu S - M_S^2 \, S^\dagger \, S \, , \\
  \mathcal{L}_{SH} &= - \lambda_2 \, H^\dagger H \, S^\dagger S - \lambda_3 \, H^\dagger S \, S^\dagger H \, , \\ 
  \mathcal{L}_{\rm Y} &= - Y_N^\alpha \, \overline N \, \ell_L^\alpha H - Y_S^\alpha \, \overline q_L^\alpha \, N S + \hc \, ,
\end{align}
where $\alpha = 1,2,3$ is a flavor index. This model can be easily
implemented and matched onto the SMEFT at the one-loop level with
\mme. The results are saved in text file called {\tt
  MatchingResult.dat}, which can be loaded into \dsix with the command
\begin{lstlisting}[style=mathematica]
  NewInput[{MMEfile -> "SN_MM/MatchingResult.dat", MN -> 1000, MS -> 1200, lam2 -> 0.1, lam3 -> 0.1, YN[a_] :> YNnum[[a]], YS[a_] :> YSnum[[a]]}, HIGHSCALE -> 10^4]
\end{lstlisting}
With this line, the user not only loads the analytical information in
{\tt MatchingResult.dat}, but also gives numerical values for the NP
parameters. After executing this command, the user can check some
input values for the SMEFT WCs at $\Lambda_{\rm UV} = 10$ TeV. Their
analytical expressions in terms of the parameters of the UV model can
also be printed thanks to the dispatch {\tt MatchAnalyticalUV}. For
instance, the analytical expression and numerical value of $C_\varphi$
can be printed with
\begin{lstlisting}[style=mathematica]
  CH /. MatchAnalyticalUV
  CH /. InputValues
\end{lstlisting}
With the \dsix SMEFT input fully generated, one can now proceed and
use the {\tt RunDsixTools} routine. Therefore, thanks to this novel
functionality, the user can easily combine \dsix and \mme to study NP
models using the full power of EFTs.

\subsubsection{Summary}
\label{sec:sum}

Some of the most common tasks in the SMEFT and in the LEFT require the
handling of a large number of WCs and/or the resolution of a huge set
of coupled RG equations. These can be automatized with the help of \dsix, a
{\tt Mathematica} package designed to provide a simple and
user-friendly experience. \dsix contains many routines and functions
to deal with the SMEFT or the LEFT, both at the algebraic and
numerical levels. Some examples of use that illustrate the capabilities
of \dsix are given here. We refer to the manual on the package
website~\cite{website}, as well as to the comprehensive reference and
documentation environment provided with \dsix, for further information
on the tool.

%%%%%%%%%%%%%%%%%%%%%%%%%%%%%%%%%%%%%%%%%%%%%%%%%%%%%%%%%%%%%%%%%%%%%%%%%%%%%%%%%%%%%%%%%%%%%%%%
\contribution[{\includegraphics[width=0.4\textwidth]{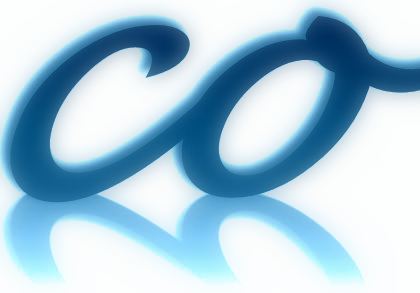}}]{\texttt{CoDEx:} Matching BSMs to SMEFT}{Supratim Das Bakshi and Sunando Kumar Patra}{}\label{sec:Codex}
%%%%%%%%%%%%%%%%%%%%%%%%%%%%%%%%%%%%%%%%%%%%%%%%%%%%%%%%%%%%%%%%%%%%%%%%%%%%%%%%%%%%%%%%%%%%%%%%

\texttt{CoDEx}~\cite{DasBakshi:2018vni} is a Mathematica package that computes WCs for SMEFT effective operators up to one-loop level and mass dimension six in terms of UV model parameters. The computation of WCs is based on the evaluation of effective action formulae derived using functional methods~\cite{Henning:2014wua,Drozd:2015rsp,Fuentes-Martin:2016uol,Ellis:2017jns}. The package is applicable to  BSM scenarios containing single or multiple mass-degenerate heavy fields of spin 0, $\frac{1}{2}$, and 1.\footnote{The effective action formula used in \texttt{CoDEx} for 1-loop generation of Wilson coefficients assumes the degenerate masses for heavy fields. This utility will be extended to non-degenerate masses in subsequent versions.} It computes the effective operators in both Strong Interacting Light Higgs (SILH)~\cite{Giudice:2007fh,Elias-Miro:2013mua} and Warsaw~\cite{Buchmuller:1985jz,Grzadkowski:2010es} bases. The code also provides an option to perform the RG evolution of these operators in the Warsaw basis, using the anomalous dimension matrix computed in \cite{Jenkins:2013zja,Jenkins:2013wua,Alonso:2013hga}. Thus, one can get all effective operators at the EW scale, generated from any such BSM theory. To run the program, it requires very minimal input within a user-friendly format. The user needs to provide only the relevant part of the BSM Lagrangian that involves the heavy field(s) to be integrated-out. \texttt{CoDEx}, with its installation instructions, web documentation, and model examples, is available on GitHub \href{https://effexteam.github.io/CoDEx/}{\faGithub}.\footnote{Send bug reports and questions at \href{mailto:effex.package@gmail.com}{\texttt{effex.package@gmail.com}}.}

\begin{figure}[ht]
	\begin{center}
		\vspace{-2em}%
		\includegraphics[width=0.5\textwidth]{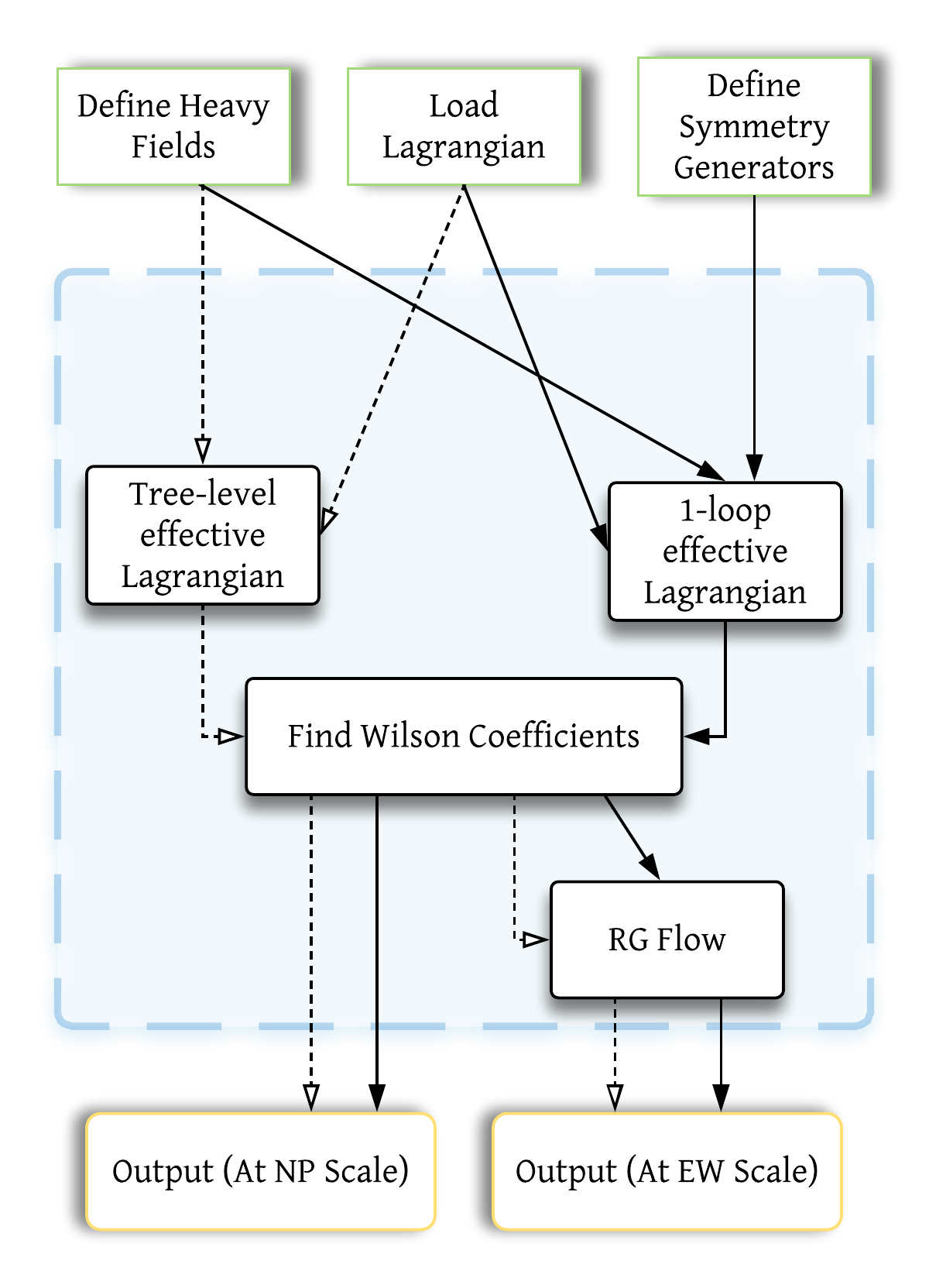}%
		\vspace{-2em}%
	\end{center}
	\caption{\texttt{CoDEx} Flowchart}
	\label{fig:flowchart}       % Give a unique label
\end{figure}

\subsubsection{ User Inputs \& CoDEx Outputs}

\indent The input information for any BSM to implement in \texttt{CoDEx} is minimal. Here, we depict a step-by-step procedure to compute the effective operators and the internal computation that is carried out at each step in \texttt{CoDEx} (see also the flowchart in Fig.~\ref{fig:flowchart}):

\begin{itemize}\setlength\itemsep{1em}
	\item The users need to provide the following information (quantum numbers) about the heavy field(s): Color, Isospin, Hypercharge, Mass, and Spin, based on which the representation(s) of the heavy field(s) are evaluated by the package internally. As mentioned, the SM gauge group quantum numbers of the BSM field are needed as input.  On top of that, the relevant part of the BSM Lagrangian that contains the heavy field(s) must be supplied by the user. The code automatically builds the heavy field kinetic (derivative and mass) terms, which are not required from the user. The SM Lagrangian is also appended by default.
	
	Let us consider an example here: we have only one heavy field -- a real singlet scalar (\mmaInlineCell{Code}{color} $\rightarrow$ 1, \mmaInlineCell{Code}{isospin} $\rightarrow$ 1, \mmaInlineCell{Code}{hypercharge} $\rightarrow$ 0, \mmaInlineCell{Code}{spin} $\rightarrow$ 0). Let us denote \mmaInlineCell{Code}{fieldName} $\rightarrow$ `hf' and \mmaInlineCell{Code}{mass} $\rightarrow$ `m'.
	This represents the field content of our model in the correct way:
	
	\,
	
\noindent\begin{mmaCell}[index=1]{Code}
  fields={{(*heavy field name *) hf,
  (*color*) 1,
  (*isospin*) 1,
  (*hypercharge*)  0,
  (*spin*) 0,
  (*mass*) m}
  };

\end{mmaCell}
\,

From this input, we construct the field representation that is needed to write the BSM Lagrangian and the \texttt{CoDEx} internal functions recognise this representation for further analysis. First, load the package:

\,

\begin{mmaCell}{Input}
  Needs["CoDEx`"]
 
\end{mmaCell}

\begin{mmaCell}{Code}
  hfvecsewrss=\mmaDef{defineHeavyFields}[\mmaDef{fields}]
  
\end{mmaCell}
\begin{mmaCell}{Output}
  \{\{\{hf[1,1]\}\}\}
\end{mmaCell}
\,

To write the Lagrangian in a compact form one can define the heavy field {\it ${\mathcal{S}}$} as:

\,

\begin{mmaCell}{Input}
  \mmaUnd{\(\pmb{\mathcal{S}}\)} = \mmaDef{hfvecsewrss}[[1,1,1]]
  
\end{mmaCell}
\begin{mmaCell}{Output}
  hf[1,1]
  
\end{mmaCell}

\,

Then we need to build the relevant part of the  Lagrangian (involving the heavy field only). Note that we do not need to construct the heavy field kinetic term (the covariant derivative and the mass terms) in the \texttt{CoDEx} Lagrangian. Thus, the only part of the Lagrangian we need here is:\footnote{If there is a tadpole term in the input BSM Lagrangian, the user has
to perform field redefinition before implementing the Lagrangian in CoDEx functions}

\,

\begin{mmaCell}{Input}
  Lpotenewrss=Expand[-\mmaSub{c}{a}*\mmaSup{\mmaDef{abs}[\mmaDef{H}]}{2}*\mmaDef{\(\pmb{\mathcal{S}}\)}\
  -\mmaFrac{\mmaUnd{\(\pmb{\kappa}\)}}{2}*\mmaSup{\mmaDef{abs}[\mmaDef{H}]}{2}*\mmaSup{\mmaDef{\(\pmb{\mathcal{S}}\)}}{2}-\mmaFrac{1}{3!} \mmaUnd{\(\pmb{\mu}\)}*\mmaSup{\mmaDef{\(\pmb{\mathcal{S}}\)}}{3}-\mmaFrac{1}{4!} \mmaUnd{\(\pmb{\lambda}\)}*\mmaSup{\mmaDef{\(\pmb{\mathcal{S}}\)}}{4}];
  
\end{mmaCell}

\,

\item Next, we need to load the symmetry generators for computing loop-level WCs:

\,

\begin{mmaCell}{Input}
  \mmaDef{initializeLoop}["ewrss",\mmaDef{fieldewrss}]
  
\end{mmaCell}

	\item Based on these inputs, one can generate the tree- and one-loop-level WCs. The \texttt{CoDEx}-functions for generating WCs are listed in Table\,\ref{tab:mainFuncs}.
	
\begin{table}[ht]
\begin{center}
	\scriptsize
	\renewcommand*{\arraystretch}{2}
	\begin{tabular}{|cl|}
		\rowcolor[gray]{.92}
		\hline
		Function &  Details \\ 
		\hline
		\rowcolor[gray]{.92}
		\mmaInlineCell{Code}{treeOutput} & \text{Calculates tree-level WCs.} \\
		\rowcolor[gray]{.92}
		\mmaInlineCell{Code}{loopOutput} & \text{Calculates one-loop-level WCs.} \\
		\rowcolor[gray]{.92}
		\mmaInlineCell{Code}{codexOutput} & \text{Generic function for WCs calculation up to one loop.} \\
		\hline
	\end{tabular}
	\caption{\texttt{CoDEx} functions for computing WCs.}\label{tab:mainFuncs}
\end{center}\vspace{-1em}%
\end{table}

\begin{mmaCell}{Input}
  \mmaDef{initializeLoop}["ewrss",\mmaDef{fieldewrss}]
  
\end{mmaCell}
\begin{mmaCell}[label={\mmaShd{$ \gg $}}]{Output}
  Isospin Symmetry Generators for
  the field `hf' are isoewrss[1,a] = 0
  
\end{mmaCell}
\begin{mmaCell}[label={\mmaShd{$ \gg $}}]{Output}
  Color Symmetry Generators for
  the field `hf' are colewrss[1,a] = 0
  
\end{mmaCell}
(See the documentation of \mmaInlineCell{Input}{initializeLoop} for details.)

\,

\item The last step is the computation of effective operators and associated WCs:

\,

\begin{mmaCell}{Input}
  res1=\mmaDef{codexOutput}[\mmaDef{Lpotenewrss},
    \mmaDef{fieldewrss},model\(\pmb{\to}\)"ewrss",
    outRange\(\pmb{\to}\)"Tree"];
    \mmaDef{formPick}["Warsaw","Detailed2",\mmaDef{res1},
    FontSize\(\pmb{\to}\)\mmaDef{Medium},
    FontFamily\(\pmb{\to}\)"Times New Roman",
    Frame\(\pmb{\to}\)All]
  
\end{mmaCell}

\,

\begin{table*}[ht]
	\caption{\footnotesize Effective operators and WCs for Real Singlet Scalar model. These results are calculated in $\overline{\text{MS}}$ renormalization scheme. The one-loop result depends on the choice of renormalization scheme, e.g. in this particular case, we have noted  differences with results given in Ref.~\cite{Henning:2014wua} where different renormalization scheme has been considered. Here,  we have highlighted the extra terms obtained in our calculation in $\overline{\text{MS}}$ scheme in red color. We have further cross-checked these results in the other scheme, adopted in Ref.~\cite{Henning:2014wua}. Warsaw basis WCs are consistent with that reported in Refs.~\cite{deBlas:2017xtg,Jiang:2018pbd,Haisch:2020ahr}.}
	\label{tab:realSing}
	\centering
	\small
	\renewcommand{\arraystretch}{1.6}
	\subfloat[SILH  (Tree level)]{
		\begin{tabular}{|*{2}{c|}}
			\hline
			$O_H$  &  $\frac{c_a^2}{m^4}$  \\
			\hline
			$O_6$  &  $\frac{\mu  c_a^3}{6 m^6}-\frac{\kappa  c_a^2}{2 m^4}$\\
			\hline
		\end{tabular}
		\label{tab:realSingST}		
	}
	\subfloat[SILH  (one-loop level)]{
		\begin{tabular}{|*{2}{c|}}
			\hline
			$O_H$  & $ \frac{\kappa ^2}{192 \pi ^2 m^2}  \textcolor{red}{-\frac{11\mu ^2 c_a^2}{192 \pi ^2 m^6}+\frac{5\kappa  \mu  c_a}{96 \pi ^2 m^4}+\frac{\lambda  c_a^2}{16 \pi ^2 m^4}}$  \\
			\hline
			$O_6$  & $-\frac{\kappa ^3}{192 \pi ^2 m^2}  \textcolor{red}{-\frac{c_a^2 \kappa \lambda}{32 \pi^2 m^2}-\frac{c_a \kappa^2 \mu}{64 \pi^2 m^4} +\frac{c_a^3 \lambda \mu}{48 \pi^2 m^6}+\frac{c_a^2 \kappa \mu^2}{32 \pi^2 m^6}-\frac{c_a^3 \mu^3}{96 \pi^2 m^8}
			}$  \\
			\hline
		\end{tabular}
		\label{tab:realSingSL}	
	}\\
	\subfloat[Warsaw (Tree level)]{
		\begin{tabular}{|*{2}{c|}}
			\hline
			$Q_H$  &  $\frac{\mu  c_a^3}{6 m^6}-\frac{\kappa  c_a^2}{2 m^4}$  \\
			\hline
			$Q_{\text{H}\square}$  &  $-\frac{c_a^2}{2m^4}$  \\
			\hline
		\end{tabular}
		\label{tab:realSingWT}
	}
	\subfloat[Warsaw  (one-loop level)]{
		\begin{tabular}{|*{2}{c|}}
			\hline
			$Q_H$  &  $-\frac{\kappa ^3}{192 \pi ^2 m^2} -\frac{c_a^2 \kappa \lambda}{32 \pi^2 m^2}-\frac{c_a \kappa^2 \mu}{64 \pi^2 m^4} +\frac{c_a^3 \lambda \mu}{48 \pi^2 m^6}+\frac{c_a^2 \kappa \mu^2}{32 \pi^2 m^6}-\frac{c_a^3 \mu^3}{96 \pi^2 m^8}$  \\
			\hline
			$Q_{\text{H}\square}$  &  $-\frac{\kappa^2}{384 \pi^2 m^2} -\frac{c_a^2 \lambda}{32 \pi^2 m^4}-\frac{5 c_a \kappa \mu}{192 \pi^2 m^4}+\frac{11 c_a^2 \mu^2}{384 \pi^2 m^6}$  \\
			\hline
		\end{tabular}
		\label{tab:realSingWL}
	}
\end{table*}

\item The output is obtained in \mmaInlineCell{Code}{"Warsaw"} basis and is formatted as a detailed table in 
\newline
\mmaInlineCell{Code}{ TraditionalForm .}
 There is provision to export the result in LaTeX~ format. Table~\ref{tab:realSingWT} is actually obtained from the output of the code above.
We can compute the same in \mmaInlineCell{Code}{"SILH"} basis as well and for that we have to use:

\,

\begin{mmaCell}{Input}
  res2=\mmaDef{codexOutput}[\mmaDef{Lpotenewrss},
    \mmaDef{fieldewrss}model\(\pmb{\to}\)"ewrss",
    operBasis\(\pmb{\to}\)"SILH",
    outRange\(\pmb{\to}\)"Tree"];
  \mmaDef{formPick}["SILH","Detailed2",
    \mmaDef{res2},FontSize\(\pmb{\to}\)\mmaDef{Medium},
    FontFamily\(\pmb{\to}\)"Times New Roman",
    Frame\(\pmb{\to}\)All]
  
\end{mmaCell}

\,

Output of this can be found in Table~\ref{tab:realSingST}. Similarly, one-loop results can be obtained by changing the option value of `\mmaInlineCell{Input}{\mmaDef{outRange}}' to \mmaInlineCell{Code}{"Loop"}. The default value of \mmaInlineCell{Input}{\mmaDef{outRange}} is \mmaInlineCell{Code}{"All"}, which combines both tree and one-loop results. These resulting WCs can then be run down to the electro-weak scale, using \mmaInlineCell{Code}{RGFlow}. This function computes RG evolution for \texttt{Warsaw} basis effective operators (in leading log approximation) using anomalous dimension matrices available in Refs.~\cite{Jenkins:2013wua,Jenkins:2013zja,Alonso:2013hga}. 

\item  Detail model building guide is available on the package web-documentation available here \href{https://effexteam.github.io/CoDEx/}{\faGithub} . Moreover, SMEFT matching results for multiple scalar extension are available in these articles \cite{Anisha:2021hgc,Bakshi:2020eyg}. \vspace{2em}
\end{itemize}

\subsubsection{Developers' version: yet to be released}

\begin{itemize}\setlength\itemsep{1em}
		\item {\bf Heavy-light mixed WCs and dimension-8:} 
		
		A module for incorporating effects from the mixed processes at one-loop including heavy fields and light fields is included. We generate these contributions by expanding the UV action around the light field classical solution obtained using the onshell relations of light fields. We implement the universal effective action formulae for the mixed heavy-light contributions and agree with that of Ref.~\cite{Ellis:2017jns} (see Tables 1–5 in there). We evaluated this formula in \texttt{CoDEx} along with 16 BSM models to generate the mixed heavy-light WCs \cite{Bakshi:2020eyg,Anisha:2020ggj,Anisha:2021hgc}. Modules for evaluation of one-loop processes involving fields with non-identical spins and incorporating SMEFT operators up to dimension-8 will be released shortly \cite{Banerjee:2022thk}.
			
	\item {\bf WCxF \cite{Aebischer:2017ugx}:} 
	
	There are multiple packages available with different applications for the EFT matching and running of the WCs, and mapping these WCs to the observables \cite{Dawson:2022ewj}. It is desirable to have a data/result exchange format among these packages. WCxF is such a data exchange format widely used among EFT packages, see Ref.~\cite{Aebischer:2017ugx}. \texttt{CoDEx} has two functions for exporting and importing data in WCxF. These functions are \mmaInlineCell{Input}{wcxfIn} and \mmaInlineCell{Input}{wcxfOut}. We briefly discuss the utilities of these functions below.
	
\begin{mmaCell}[moredefined={wcxfOut}]{Input}
  ??wcxfOut
  
\end{mmaCell}

\,

\begin{mmaCell}{Output}
  This function prepares an output format
  of WCs and effective
  operators that can be exported to .json
  file using Mathematica Export[] function.
  
\end{mmaCell}

\,

\begin{mmaCell}{Print}
  Attributes[wcxfOut]=\{Locked,
  Protected,ReadProtected\}
  Options[wcxfOut]=\{eft\(\to\)SMEFT,
  basis\(\to\)Warsaw\}
  
\end{mmaCell}

\,

\begin{mmaCell}[moredefined={result,wcxfOut}]{Input}
  (*A sample input --*)
  wcxfOut[246 (*scale*),
  result (* Numerical WCs*)]

\end{mmaCell}

\,

The output (which is suppressed here) is in WCxF template, and it can be interfaced with other available programs. In Ref.~\cite{Anisha:2021hgc}, we have validated this by interfacing a .json file generated by \mmaInlineCell{Input}{\mmaDef{wcxfOut}} to DSixTools\cite{Celis:2017hod} package successfully.

\,

The function \mmaInlineCell{Input}{\mmaDef{wcxfIn}} takes WCxF files as input and provides output in \texttt{CoDEx} data format.

\,

\begin{mmaCell}[moredefined={wcxfIn}]{Input}
  ??wcxfIn
  
\end{mmaCell}

\begin{mmaCell}{Output}
  This function takes the WCs and effective operators
  imported from .json file
  (using Mathematica Import function),
  which is taken as input
  and for that the output is
  WCs and effective
  operators in \mmaDef{CoDEx}
  result conventions.

\end{mmaCell}

\begin{mmaCell}{Print}
  Attributes[wcxfIn]=\{Locked,
    Protected,ReadProtected\}

\end{mmaCell}

\begin{mmaCell}{Input}
  Import["sample_result_1.json"];
  
\end{mmaCell}

\begin{mmaCell}[moredefined={wcxfIn}]{Input}
  wcxfIn[246 (*scale*),%(*wcxf data*)]
  
\end{mmaCell}

\begin{mmaCell}{Output}
  \{\{q1Hl[1,1],-3.53103*10^-12\},...\}
  
\end{mmaCell}

\,

This output is in \texttt{CoDEx} readable format and the ellipses above represent other WCs in the list.

	\item {\bf Identities:} The effective action evaluation for a BSM generates gauge-invariant structures, which do not directly map to the desired effective operator basis. We implement operator identities and equations of motion on the derived effective Lagrangian to cast the gauge-invariant terms to desired structures. These identities depends on the choice of the effective operator basis. The transformations like Fierz identities, SM field equations of motion, and SMEFT dimension-six operator identities are introduced in the developer version of \texttt{CoDEx}. In future developments, new modules will be available to capture the evanescent operator effects \cite{Aebischer:2022rxf,Aebischer:2022aze,Aebischer:2022tvz,Aebischer:2023djt}, as well as these identities will be extended to incorporate effects from SMEFT dimension-8 operators \cite{Banerjee:2022thk}.
\end{itemize}
%

%%%%%%%%%%%%%%%%%%%%%%%%%%%%%%%%%%%%%%%%%%%%%%%%%%%%%%%%%%%%%%%%%%%%%%%%%%%%%%%%%%%%%%%%%%%%%%%%
\contribution[{\includegraphics[width=\textwidth]{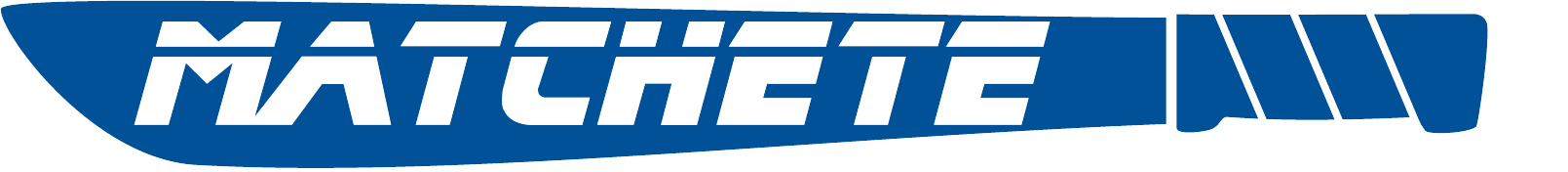}}]{\texttt{Matchete:} Matching Effective Theories Efficiently}{Matthias König}{}\label{sec:matchete}
%%%%%%%%%%%%%%%%%%%%%%%%%%%%%%%%%%%%%%%%%%%%%%%%%%%%%%%%%%%%%%%%%%%%%%%%%%%%%%%%%%%%%%%%%%%%%%%%

\matchete (MATCHing Effective Theories Efficiently) is a \texttt{Mathematica} package fully automating matching computations up to one-loop order, utilizing functional methods. The user supplies the UV theory by first defining the symmetry groups (both local and global), the fields and the coupling parameters with simple commands. With these definitions in place, the Lagrangian can be written in \texttt{Mathematica} language in a simple way and then passed to the matching function to integrate out fields that the user has defined as ``heavy". As a consequence of the functional matching procedure, no prior knowledge of the operators basis of the resulting effective theory is required. \matchete automatically generates the full set of effective operators and reduces it to a basis without any user input.

The \matchete package is free software under the terms of the GNU General Public License~v3.0 and is publicly available in the following GitLab repository:
\begin{center}
\href{https://gitlab.com/matchete/matchete}{\texttt{https://gitlab.com/matchete/matchete}}
\end{center}
This note only serves as a brief overview of the package; we refer the reader to Ref.~\cite{Fuentes-Martin:2022jrf} for details.

\subsubsection{Matching Strategy}
\paragraph{Functional Methods and Expansion by Regions}
\matchete achieves the matching by directly computing the Wilsonian effective action, i.e. the contribution to the generating functional encoding only short-distance physics, corresponding to energy scales $E>\Lambda$, where $\Lambda$ is the matching scale (or in BSM contexts often called ``new-physics scale'')~\cite{Fuentes-Martin:2016uol,Zhang:2016pja,Cohen:2020fcu,Fuentes-Martin:2020udw,Cohen:2020qvb}. To this end, one first splits the field content of the theory into Fourier modes with frequencies above (\emph{hard}) and below (\emph{soft}) the scale $\Lambda$:
\begin{align}
    \phi = \phi_H+\phi_S\,.
\end{align}
Low-energy matrix elements are computed from the generating functional:
\begin{align}
    Z[J_S] = \int \mathcal D \phi_S\,\mathcal D\phi_H\,
    \exp\left\{ iS(\phi_S,\phi_H)+i\int d^4z \, J_s(z)\phi_S(z) \right\}\,,
\end{align}
from which the Wilsonian effective action $S_\Lambda$ is defined:
\begin{align}
    \int\mathcal D \phi_H \,\exp\left\{iS(\phi_S,\phi_H) \right\}\equiv \exp\left\{ iS_\Lambda(\phi_S)\right\} \,.
\end{align}
This object can be calculated directly by means of a background field expansion, meaning each field is further split into classical fields~$\hat{\phi}_i$ and quantum fluctuations~$\eta_i$,
\begin{align}
    \phi_i = \hat\phi_i + \eta_i\,,
\end{align}
 and an expansion of $S(\hat\phi_S+\eta_S,\hat\phi_H+\eta_H)$ in the quantum fields is performed. Collecting hard and soft modes into a single multiplet, i.e. $\hat\phi$ and $\eta$ for the classical field and quantum fluctuation, this expansion reads:
 \begin{align}
     S(\hat\phi+\eta) = S(\hat\phi) + \eta_i \left[ \frac{\delta S}{\delta\eta_i} \right]\!(\hat\phi) + \frac{1}{2}\,\bar\eta_i \left[\frac{\delta^2 S}{\delta\eta_j\delta\bar\eta_i} \right]\!(\hat\phi)\;\eta_j + \mathcal{O}(\eta^3)\,.
 \end{align}
The first term corresponds to the tree-level contributions, the second vanishes by virtue of the equations of motion and the third term encodes all the one-loop contributions. At tree level, the effective action is obtained by solving the equations of motions for the heavy fields, inserting them back into the full Lagrangian and expanding in the heavy mass. At one-loop level, the effective action is found by
 \begin{align}
     S_\Lambda^{(1)} = \pm\,\frac{i}{2}\int_h \frac{d^dk}{(2\pi)^d} \big\langle k \vert \mathrm{tr}\log Q \vert k \big\rangle \,,
 \end{align}
where $Q_{ij} = \frac{\delta^2 S}{\delta\bar\eta_i\delta\eta_j}$ is the fluctuation operator. The subscript $_h$ on the integral denotes the fact that the integral is taken in the hard region, meaning the integration momentum $k$ is assumed to be of the order of the hard scale $\Lambda$. In practice, this is implemented by assigning a power-counting to the soft scales (masses and momenta of soft modes) and expanding the \emph{integrand} systematically to the desired order in the EFT counting. This \emph{expansion by regions}~\cite{Beneke:1997zp,Jantzen:2011nz} simplifies the matching procedure as it computes directly the one-loop contributions to the matching coefficients without the need of having to evaluate matrix elements of the effective theory~\cite{Fuentes-Martin:2016uol}.

The results obtained with the method outlined above yields an effective Lagrangian that is not manifestly gauge-invariant, as it contains open covariant derivatives, meaning expression of the form 
\begin{align}
    \mathcal L_\mathrm{eff} \supset X^{\mu\nu} D_\mu D_\nu\,,
\end{align}
which cannot be dropped since the covariant derivatives commute non-trivially. This is mitigated by the so-called \emph{covariant derivative expansion}, for details of which we refer the user to the literature~\cite{Gaillard:1985uh,Chan:1986jq,Cheyette:1987qz}.

\paragraph{Basis Reduction}
After the matching procedure is performed, the obtained effective operators are not linearly independent. \matchete is able to handle the most common Lie algebras and performs simplifications of Dirac algebra using $d$-dimensional identities if they are available. To find a basis however, redundant operators still need to be eliminated. The first reduction technique is relating operators with covariant derivatives to each others by the means of integration-by-parts (IBP) identities, which can be derived by imposing total derivative operators to vanish, $D_\mu J^\mu = 0$. These IBP identities allow one to eliminate certain derivative operators in favor of others. As an example in a theory with charged Dirac fermions and a scalar, the following kind of reduction is achieved by this:
\begin{align}
    \phi\,\bar\psi \gamma^\mu\gamma^\nu \{D_\mu, D_\nu\}\psi 
    =-2(D_\mu\phi)(\bar\psi \gamma_\mu \slashed D\psi)-2 \phi (\bar\psi\overleftarrow{\slashed D}\slashed D\psi)+\phi F^{\mu\nu} (\bar\psi i \Gamma_{\mu\nu}\psi)\,.
\end{align}
The choice of which operators are preferred is not unique, but it is advantageous to favor operators proportional to the equations of motions of the fields. In the above expression, the derivatives have been either traded for a field-strength tensor or act on the field operators in such a way, that the Dirac equation can be used to further simplify them.

Operators with derivatives acting on fields in the way they appear in their equations of motion\footnote{One colloquially uses the phrase ``proportional to the equations of motion" even though this is not technically correct.} can be further reduced by the means of appropriate field redefinitions. For scalars $\phi$, fermions $\psi$ and vector fields $A^\mu$ these operators are of the form $J D_\mu D^\mu \phi$, $J \slashed D\psi$ and $J_\nu D_\mu F^{\mu\nu}$ respectively. Redefining the fields by shifts proportional to the coefficient operator $J$ then eliminates these operators while introducing operators with fewer derivatives as well as operators at higher mass dimension. Applying the procedure iteratively, order by order in power-counting, then fully eliminates all operators proportional to the field equations of motion.

\matchete applies all reductions described above fully automatically without the user having to derive and specify any operator reduction identities. As of the time of writing, the list of possible reductions is not completely implemented yet. In particular, the current version of \matchete does not yet implement Fierz reductions, as these require the proper treatment of evanescent operators~\cite{Fuentes-Martin:2022vvu}. This is left for a future release (see also Sec.~\ref{sec:matchete_outlook}).

\subsubsection{Usage Example}
In this section, we briefly outline a simple usage example. Once again, the reader is referred to Ref.~\cite{Fuentes-Martin:2022jrf} for a more detailed user manual of the package including an installation guide. To demonstrate the features of \matchete, we match a simple model in which we supplement the SM with a singlet real scalar field $\phi$, as has been discussed in Refs.~\cite{Jiang:2018pbd,Haisch:2020ahr}. 
The Lagrangian of this model reads:\footnote{\matchete also allows for the inclusion of tadpoles as long as they are suppressed in the EFT counting compared to heavy-mass scales. Otherwise, the user will have to redefine the fields and manually remove the tadpole before using \matchete.}
\begin{align}
 \mathcal L_\mathrm{UV} = 
 \mathcal{L}_\mathrm{SM} 
 + \frac{1}{2}(\partial_\mu\phi)^2
 - \frac{1}{2}M^2\phi^2 
 - \frac{\mu}{3!}\phi^3 
 - \frac{\lambda_\phi}{4!}\phi^4
 - A\phi |H|^2- \frac{\kappa}{2}\phi^2 |H|^2\,. \label{eq:matchete_Lagrangian}
\end{align}
\matchete provides the full definitions of the SM and its Lagrangian as a simple macro. 
After installing the package, it can be loaded via:
\begin{mmaCell}{Input}
  << \mmaDef{Matchete\(\,\grave\,\)}
\end{mmaCell}
Next, we load the SM Lagrangian from the predefined model file included with \matchete:
\begin{mmaCell}{Input}
  LSM = LoadModel["SM", ModelParameters -> \{"\(\mu\)" -> mH, "\(\lambda\)" -> \mmaUnd{\(\lambda\)h}\}];
\end{mmaCell}
where we rename the Higgs mass parameter to \mmaInlineCell[]{Input}{mH} and the quartic Higgs coupling to \mmaInlineCell[]{Input}{\mmaUnd{\(\lambda\)h}}. We then define the heavy scalar using the command:
\begin{mmaCell}{Input}
  DefineField[\mmaUnd{\(\phi\)}, Scalar, SelfConjugate -> True, Mass -> \{Heavy, \mmaUnd{M}\}]
\end{mmaCell}
The arguments supplied to the function indicate the spin of the field, the fact that it is real, the definition of the mass parameter and that it should be considered heavy. The remaining couplings in the Lagrangian~\eqref{eq:matchete_Lagrangian} are defined with:
\begin{mmaCell}{Input}
      DefineCoupling[\mmaUnd{A}, SelfConjugate -> True]\\DefineCoupling[\mmaUnd{\(\kappa\)}, SelfConjugate -> True]\\DefineCoupling[\mmaUnd{\(\mu\)}, SelfConjugate -> True]\\DefineCoupling[\mmaUnd{\(\lambda\phi\)}, SelfConjugate -> True]
\end{mmaCell}
\matchete defines simple shortcuts for the fields and couplings when these commands are evaluated. Even though the full objects are more complicated, the user can input them in a simple form:
\begin{mmaCell}{Input}
LUV = \mmaDef{LSM} + FreeLag[\(\phi\)] - \mmaFrac{1}{3!}\(\mu\)[]\,\mmaSup{\(\phi\)[]}{3} - \mmaFrac{1}{4!}\(\lambda\phi\)[]\,\mmaSup{\(\phi\)[]}{4} \\- \mmaDef{A}[]\,Bar[H[i]]H[i]\(\phi\)[] - \mmaFrac{1}{2}\(\kappa\)[]\,Bar[H[i]]H[i]\mmaSup{\(\phi\)[]}{2};
\end{mmaCell}
Note that the command \mmaInlineCell[]{Input}{FreeLag[\(\phi\)]} automatically generates the kinetic and mass terms for the new field and only interaction terms have to be written out. We are now ready to integrate out the scalar at one-loop order. This is achieved by running the \mmaInlineCell[]{Input}{Match} command with appropriate arguments:
\begin{mmaCell}{Input}
  LEFT = Match[\mmaDef{LUV}, LoopOrder -> 1, EFTOrder -> 6];
\end{mmaCell}
Here the first option defines the order at which the matching procedure is carried out while the second option denotes the fact that we want to obtain the effective Lagrangian up to dimension six. After this line is successfully evaluated, the object \mmaInlineCell[]{Input}{\mmaDef{LEFT}} contains redundant operators, as described earlier. Reductions using IBP identities and field redefinitions are then performed by running:
\begin{mmaCell}{Input}
  LEFTOnShell = \mmaDef{LEFT} //EOMSimplify;
\end{mmaCell}
The user can choose to only perform IBP identities without field redefinitions to obtain the Green's basis by using the \mmaInlineCell[]{Input}{GreensSimplify} command. The object \mmaInlineCell[]{Input}{\mmaDef{LEFTOnShell}} then contains the operators together with their matching coefficients. The full output is cumbersome, but individual contributions can be isolated using the \mmaInlineCell[]{Input}{SelectOperatorClass} command. As an example, to show only the leptonic four-fermion operator, one uses:
\begin{mmaCell}{Input}
  SelectOperatorClass[\mmaDef{LEFTOnShell},\{Bar[\mmaDef{l}],\mmaDef{e},Bar[\mmaDef{e}],l\},0] //NiceForm
\end{mmaCell}
\begin{mmaCell}{Output}
  \mmaFrac{1}{6} \!\!\(\hbar\)\,\mmaSup{\(\overline{Ye}\)}{rs}\mmaSub{Ye}{tp}\,\mmaSup{A}{2}\mmaFrac{1}{\mmaSup{M}{4}}\big(\mmaSup{\(\overline{e}\)}{s}\,\(\cdot\)\,\mmaSub{P}{L}\,\(\cdot\)\,\mmaSup{l}{ir}\big)\big(\mmaSubSup{\(\overline{l}\)}{i}{t}\,\(\cdot\)\,\mmaSub{P}{R}\,\(\cdot\)\,\mmaSup{e}{p}\big)
\end{mmaCell}
where the second argument specifies the field content of the operator(s) to be extracted, and the last argument gives the number of derivatives. Here $\hbar$ should be understood as a loop counting factor that is equal to $1/(16\pi^2)$. The matching example with more details is included with the \matchete package in the example notebook \texttt{Examples/Singlet\_Scalar\_Extension.nb}.

\subsubsection{Outlook}\label{sec:matchete_outlook}
We conclude this note with a roadmap for features intended for future releases:
\begin{itemize}
    \item The matching is currently done in strictly $d=4-2\epsilon$ dimensions. This prevents reductions of Dirac structures including Fierz rearrangements, because these hold only in four dimensions. When applied to $d$-dimensional operators, one has to account for evanescent contributions.
    \item After evanescent contributions can be handled automatically, \matchete will be able to produce output that, in the case of SMEFT computations, can be directly compared to the Warsaw basis. In the future, \matchete will be able to automatically perform this identification as well as output the result in the \texttt{WCxf}~\cite{Aebischer:2017ugx} format. An interface with other phenomenology codes and/or commonly used formats, such as \texttt{UFO}~\cite{Degrande:2011ua}, would be desirable as well.
    \item At this time, \matchete does not allow integrating out heavy vector fields at the loop level. The reason for this is that these cannot be generally written down in a renormalizable fashion. In weakly-coupled theories, heavy vectors must arise from spontaneous symmetry breaking. This results in a complicated interplay between vectors, ghosts, and Goldstone bosons, especially in the background field gauge. So as to avoid having to derive and input all interactions manually, we wish to provide (semi-)automated methods to determine the broken phase Lagrangian.    
    \item With small changes to the matching procedure, it is possible to determine the EFT counterterms and, thereby, the RG functions. Implementing this functionality in \matchete will allow for finding the RG functions for intermediate-scale EFTs and vastly simplify sequential matching scenarios. 
\end{itemize}
After the above list of features is included in the package, \matchete can be an integral part in a fully automated pipeline from a UV model down to phenomenology, and as such be a powerful tool for BSM phenomenology by taking away the laborious task of one-loop matching.

%%%%%%%%%%%%%%%%%%%%%%%%%%%%%%%%%%%%%%%%%%%%%%%%%%%%%%%%%%%%%%%%%%%%%%%%%%%%%%%%%%%%%%%%%%%%%%%%
\contribution[{\includegraphics[width=0.35\textwidth]{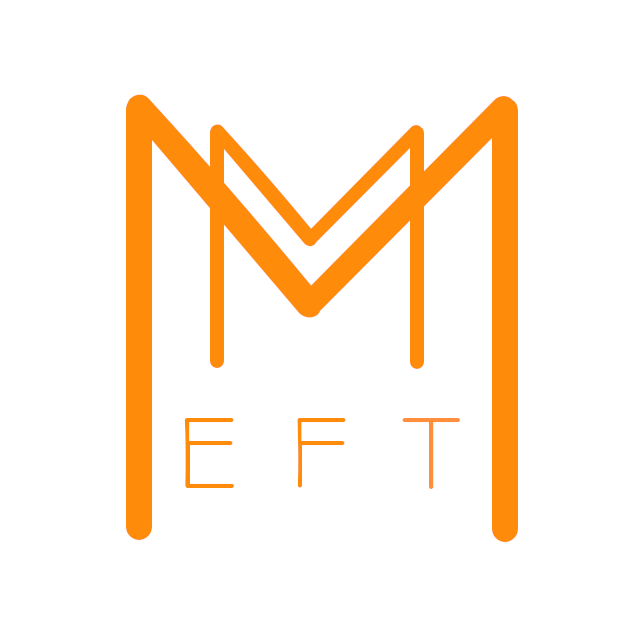}}]{\texttt{Matchmakereft}: a tool for tree-level and one-loop matching}{José Santiago}{}\label{sec:MME}
%%%%%%%%%%%%%%%%%%%%%%%%%%%%%%%%%%%%%%%%%%%%%%%%%%%%%%%%%%%%%%%%%%%%%%%%%%%%%%%%%%%%%%%%%%%%%%%%

\texttt{Matchmakereft} is a computer tool that automates the tree-level and one-loop matching of arbitrary weakly-coupled UV models onto their EFT. Due to lack of space, we refer the reader to the original publication~\cite{Carmona:2021xtq} and the manual that comes with the installation and summarize here its main features and newest developments, and provide a simple but illustrative example.

\texttt{Matchmakereft} is written in \texttt{python}, making it very easy to install in different platforms, and it uses well-tested tools that include \texttt{Feynrules}~\cite{Alloul:2013bka}, \texttt{QGRAF}~\cite{Nogueira:1991ex}, \texttt{Form}~\cite{Kuipers:2012rf} and \texttt{Mathematica} to perform an off-shell matching using diagrammatic methods in the background field gauge.

\texttt{Matchmakereft} takes advantage of the large degree of gauge and kinematic redundancy in off-shell matching to perform a significant number of non-trivial cross-checks, ensuring the validity of the resulting computation. It is also equipped to compute the RG equations of arbitrary EFTs and the off-shell (in)dependence of a set of local operators. \texttt{Matchmakereft} treats the kinematic and gauge dependence independently, leaving the latter arbitrary until the very end of the calculation. This increases its efficiency but it also makes \texttt{matchmakereft} an ideal tool to compute IR/UV dictionaries or to perform calculations in theories with arbitrary gauge structures.

Among the latest developments of \texttt{matchmakereft}, the calculation of amplitudes in chunks of a fixed number of diagrams and the ability to compute amplitudes in parallel, have significantly increased its efficiency (see the manual for details).

%%%%%%

Let us now demonstrate many of the features of {\tt matchmakereft} with a concrete example involving two scalar fields, a light, but not massless, field $\phi$ and a heavy field $\Phi$. Our model is described by the Lagrangian: 
\begin{equation}
    \mathcal{L} = \frac{1}{2}(\partial_\mu \phi)^2 -\frac{1}{2}m_L^2 \phi^2 + \frac{1}{2} (\partial_\mu \Phi)^2 - \frac{1}{2} M_H^2 \Phi^2 - \frac{\lambda_0}{4!}\phi^4 - \frac{\lambda_2}{4}\phi^2 \Phi^2 - \frac{\kappa}{2}\phi^2 \Phi,
    \label{lagrangian:uv}
\end{equation}
which we want to match to the EFT Lagrangian without the heavy scalar, 
\begin{equation}
    \mathcal{L}_{\rm EFT} = \frac{\alpha_{4k}}{2}(\partial_\mu \phi)^2 
    -\frac{\alpha_2}{2} \phi^2
    - \frac{\alpha_4}{4!}\phi^4 
    - \frac{\alpha_6}{6!}\phi^6 
    - \frac{\tilde{\alpha}_6}{4!}\phi^3 \partial^2\phi
    -\frac{\hat{\alpha}_6}{2}\left(\partial^2 \phi \right)^2.
\end{equation}
We will use this Lagrangian during off-shell matching. Subsequently, the kinetic term can be canonically normalized, and the redundant operators can be eliminated. Two of the three operators of dimension six are redundant. We choose $\phi^6$ as the independent operator. Using equations of motion we can readily find that:
\begin{align}
    \phi^3 \partial^2\phi &\to -\alpha_2\phi^4 - \frac{1}{3!}\alpha_4\phi^6, \\
    \left(\partial^2 \phi \right)^2 &\to  \alpha_2^2 \phi^2  + \frac{\alpha_2\alpha_4}{3}\phi^4 + \frac{\alpha_4^2}{36}\phi^6.
\end{align}
Eliminating these operators from the Lagrangian would induce the shifts
\begin{eqnarray}
\alpha_2 & \to & \alpha_2 + \alpha_2^2 \hat{\alpha}_6\\
\alpha_4 & \to & \alpha_4 - \tilde{\alpha}_6 \alpha_2 +4 \alpha_2\alpha_4\hat{\alpha}_6 \\
\alpha_6 & \to & \alpha_6 - 5\tilde{\alpha}_6\alpha_4 +10 \hat{\alpha}_6\alpha_4^2
\label{reductiontogreeneqs}
\end{eqnarray}
The coupling $\kappa$ of this model is a dimensionful coupling, and is expected to be parametrically of the order of the heavy mass scale $M_H$. Thus,  $\frac{\kappa}{M_H}$ is of $\mathcal{O}(1)$ and is kept throughout the matching procedure consistently. 

The \feynrules file for the UV model, saved at {\tt two\_scalars.fr}, is shown below.
\begin{lstlisting}[language=Mathematica]
(* --- Contents of Feynrules file two_scalars.fr --- *)
M$ModelName = "two_scalars";
(* **** Particle classes **** *)
M$ClassesDescription = {
S[1] == {ClassName -> phiH, SelfConjugate -> True, Mass -> MH,
         FullName -> "heavy"},
S[2] == {ClassName -> phi, SelfConjugate -> True, Mass -> mL,
         FullName -> "light"}
};
(* *****   Parameters   ***** *)
M$Parameters = {
MH == {ParameterType -> Internal, ComplexParameter -> False},
mL == {ParameterType -> Internal, ComplexParameter -> False},
V == {ParameterType -> Internal, ComplexParameter -> False},
lambda0 == {ParameterType -> Internal, ComplexParameter -> False},
kappa == {ParameterType -> Internal, ComplexParameter -> False},
lambda2 == {ParameterType -> Internal, ComplexParameter -> False}
};
(* *****   Lagrangian   ***** *)
Ltot := Block[{mu}, 
  + 1/2 * del[phi,mu] *  del[phi,mu]  + 1/2 *del[phiH,mu] * del[phiH,mu]
  - 1/2 * MH^2 * phiH^2 - 1/2 * mL^2 * phi^2 
  - lambda0 / 24 * phi^4 - kappa / 2  * phi^2 * phiH 
  - lambda2 / 4 * phi^2 * phiH^2
  ];
\end{lstlisting}

 Note that we use the keyword {\tt FullName} to characterize each field as {\tt "heavy"} or {\tt "light"}. This is mandatory: {\tt matchmaker} uses this keyword to distinguish between fields that are integrated out and those that are light and are also present in the EFT. Also note that all the parameters that are used in the Lagrangian, masses as well as couplings, must be declared. In this example all parameters are real. 

The \feynrules file for the EFT model, saved at {\tt one\_scalar.fr}, is: 
\begin{lstlisting}[language=Mathematica]
(* --- Contents of Feynrules file for the EFT model  one_scalar.fr --- *)
M$ModelName = "one_scalar";
(* **** Particle classes **** *)
M$ClassesDescription = {
S[2] == {ClassName -> phi, SelfConjugate -> True, Mass -> 0,
         FullName -> "light"}
};
(* *****   Parameters   ***** *)
M$Parameters = {
alpha4kin == {ParameterType -> Internal, ComplexParameter -> False},
alpha2mass == {ParameterType -> Internal, ComplexParameter -> False},
alpha4 == {ParameterType -> Internal, ComplexParameter -> False},
alpha6 == {ParameterType -> Internal, ComplexParameter -> False},
alpha6Rtilde == {ParameterType -> Internal, ComplexParameter -> False},
alpha6Rhat == {ParameterType -> Internal, ComplexParameter -> False}
};
(* *****   Lagrangian   ***** *)
Ltot := Block[{mu,mu2},
  1/2 * alpha4kin * del[phi,mu] * del[phi,mu] 
  - 1/2 * alpha2mass * phi^2
  - alpha4/24 * phi^4 
  - alpha6 * phi^6/720 
  - alpha6Rtilde/24 * phi^3 * del[del[phi,mu],mu] 
  - alpha6Rhat/2 * del[del[phi,mu],mu]  *  del[del[phi,mu2],mu2] 
  ];
\end{lstlisting}

Note that we have included WCs (denoted by {\tt alpha}) also for the kinetic and mass terms (squared), as well as for all operators that are redundant solely due to the equations of motion. 

In order for {\tt matchmakereft} to perform the reduction to the physical basis, we need to provide a set of relations that express the redundant WCs in terms of the irreducible ones, see Eq.~(\ref{reductiontogreeneqs}).  This is done at {\tt one\_scalar.red}: 
\begin{lstlisting}[language=Mathematica]
(* --- Contents of one_scalar.red --- *)
finalruleordered = {
	alpha6 -> - alpha6Rtilde * alpha4 *5 + alpha6Rhat * alpha4^2 * 10 + alpha6 ,
	alpha4 -> alpha4 - alpha6Rtilde * alpha2mass + 4 * alpha6Rhat * alpha2mass * alpha4 ,
	alpha4kin -> alpha4kin ,
	alpha2mass -> alpha2mass +  alpha6Rhat * alpha2mass^2
	}
\end{lstlisting}

Note that {\bf only} the  WCs corresponding to physical operators, among those appearing in the EFT Lagrangian and defined in the file \verb+one_scalar.fr+, must be present on the left hand side of the replacement rules in this file. The WCs corresponding to redundant operators appear only on the right hand side. When these rules are used, both redundant and non-redundant WCs have been matched and are known as functions of the parameters of the UV theory. The rules are therefore instructions on how to update the non-redundant WCs, to include the effect of the redundant ones.

With these files prepared we are ready to proceed with matching. In the matching directory, where {\tt two\_scalars.fr},{\tt one\_scalar.fr},{\tt one\_scalar.red} are present, we can run {\tt matchmakereft}: 
\begin{lstlisting}[numbers=none,backgroundcolor=\color{codegris}]
>matchmakereft
\end{lstlisting}
upon which we enter the python interface 
\begin{lstlisting}[numbers=none,backgroundcolor=\color{codegris}]
Checking for updates.
matchmakereft is up-to-date.

Welcome to matchmakereft v1.1.3
Please refer to SciPost Phys. 12, 198 (2022) arXiv:2112.10787 when using this code. 
For documentation please check the manual in /installationdirectory/docs/manual.pdf

matchmakereft> 
\end{lstlisting}
We first need to create the matchmaker models, i.e. the directories with all the necessary information for the UV and the EFT models. We do this by 
\begin{lstlisting}[numbers=none,backgroundcolor=\color{codegris}]
matchmakereft> create_model two_scalars.fr
\end{lstlisting}
which has the response
\begin{lstlisting}[numbers=none,backgroundcolor=\color{codegris}]
Creating model two_scalars_MM. This might take some time depending on the complexity of the model
Model two_scalars_MM created
It took 7 seconds to create it.
\end{lstlisting}
We can now observe that the directory {\tt two\_scalars\_MM} is created. We proceed with creating the EFT model
\begin{lstlisting}[numbers=none,backgroundcolor=\color{codegris}]
matchmakereft> create_model one_scalar.fr
Creating model one_scalar_MM. This might take some time depending on the complexity of the model
Model one_scalar_MM created
It took 7 seconds to create it.
\end{lstlisting}
The {\tt one\_scalar\_MM} directory is now created as well, and we are ready for the matching calculation. This is performed by the {\tt match\_model\_to\_eft} command:
\begin{lstlisting}[numbers=none,backgroundcolor=\color{codegris}]
matchmakereft> match_model_to_eft two_scalars_MM one_scalar_MM
\end{lstlisting}
Upon completion, the results of the matching are stored in the UV model directory, in this case {\tt two\_scalars\_MM}. The file {\tt two\_scalars\_MM/MatchingProblems.dat} contains troubleshooting information in case the matching procedure failed. In our case it is an empty list, indicating no problems:
\begin{lstlisting}
problist = {}
\end{lstlisting}

The result of the matching procedure is stored in {\tt two\_scalars\_MM/MatchingResults.dat}, a Mathematica file with a list of lists of replacement rules. These matching results can also be seen in printed form in Appendix C of the original \mme publication~\cite{Carmona:2021xtq}. In this example, the kinetic operator receives a one-loop matching correction and, therefore, $\phi$ is no longer canonically normalized. A field redefinition is needed to obtain a canonically normalized theory on which we can apply the corresponding redundancies to go to the physical basis. \Mme does these two processes (canonical normalization and going to the physical basis) automatically. The resulting WCs in the physical basis, up to one loop order and $O\left(\frac{\kappa^{2n}}{M_H^{2n}}\frac{m_L^2}{M_H^2}\right)$ can also be seen in printed form in Appendix C of Ref.~\cite{Carmona:2021xtq}.

As mentioned above, \mme can do many more things than just finite matching. One can for instance compute the RG equations for both models and check the consistency of the logarithmic terms from the RG equations and the finite matching. We refer to the manual for a detailed explanation.

We would like to finish this overview of \mme by mentioning two projects we are currently working on in the Granada group, that will either end up being part of \mme or strongly use it for their development. Current matching programs perform the matching off-shell, producing the effective Lagrangian in a so-called Green's basis, in which some redundant operators are not required to describe physical, on-shell amplitudes. The usual process is to reduce this Green's basis to a physical basis, either manually (as it is currently done in \texttt{matchmakereft}) or in an automated way, as currently done by \texttt{matchete}~\cite{Fuentes-Martin:2022jrf}. We are looking into side-stepping this reduction step by performing an on-shell matching. This has a number of technical complications that have been essentially solved for the tree-level matching. We are working on extending the on-shell matching to the one-loop level. More details can be found in Section~\ref{section:chala}.

Another project we are currently working on was mentioned by R. Fonseca in his talk but is not covered elsewhere in this document (it has a significant overlap with the content of Section~\ref{section:misiak} by other authors). Even with current codes that automate the process of one-loop matching, repeating the calculation for different models is re-iterative and, in the case of many fields, can be computationally very expensive. Our idea is to define a generic model, in which the gauge structure is not fixed a priori and the field multiplicity simply appears as a dummy flavor index. This general EFT can be built with a single multiplet of real scalars, Weyl fermions and gauge bosons. We have defined the most general EFT with this generic field content up to mass dimension six and we are computing its RG equations. All the loop integrals, tensor reduction and kinematic projections are performed in this generic EFT in such a way that the calculation of the RG equations for any specific EFT can be obtained by means of a straight-forward group-theoretic calculation. The next step will be to define a generic theory with light and heavy fields and perform the finite one-loop matching for it.

%%%%%%%%%%%%%%%%%%%%%%%%%%%%%%%%%%%%%%%%%%%%%%%%%%%%%%%%%%%%%%%%%%%%%%%%%%%%%%%%%%%%%%%%%%%%%%%%
\contribution{\texttt{Sym2Int}: Automatic generation of EFT operators}{Renato M. Fonseca}{}\label{sec:Sym2Int}
%%%%%%%%%%%%%%%%%%%%%%%%%%%%%%%%%%%%%%%%%%%%%%%%%%%%%%%%%%%%%%%%%%%%%%%%%%%%%%%%%%%%%%%%%%%%%%%%

EFTs are a powerful tool for probing potential new physics in a model-independent way. At a time when there is a lack of clarity on how to extend the SM, its related EFTs have been receiving an increasing amount of attention. For example, the number of SMEFT operators have been counted with several techniques in the last few years, up to high mass dimensions. Building an explicit basis of operators is more complicated, but here too there has been notable progress. I will go through my recent work on using the software packages \texttt{GroupMath} and \texttt{Sym2Int} to automatically build explicit bases of operators for EFTs, given their fields and symmetries.

\subsubsection{Using computers to generate Lagrangians}

With EFTs one can study the low energy consequences of a model without
having to handle or even know all of its intricacies; by integrating
out the heavy fields one can obtain a Lagrangian which describes well
the large distance behavior of the original theory. However, the reduction
in the number of fields comes at the cost of introducing a potentially
large set of local operators of high dimension. As such, the very
first step in the study of an effective field theory is to establish
a basis of operators which encodes all possible interactions between
the light fields. Put simply, one needs a Lagrangian. As the reader
is certainly aware, most theories are invariant under some group of
transformations --- for example the Lorentz group and/or a gauge
group --- therefore the task of finding all interactions is inseparable
from the problem of finding invariant combinations of products of
representations of some group.

There is a long history of using computers in particle physics, given
the complexity of the calculations one needs to perform. Indeed, there
are many codes specialized in various tasks, from calculating Feynman
rules all the way to generating events at colliders. However, at the
very beginning of such a stack of programs, it would be useful to
have one more code which, to some degree, alleviates the heavy burden
on the user of having to provide the full Lagrangian of the model.
To the best of my knowledge, \texttt{SARAH} \cite{Staub:2010jh}\texttt{
}and \texttt{Susyno} \cite{Fonseca:2011sy} were the first codes to
build symmetry-invariant Lagrangians (superpotentials, to be more
specific) with the user having only to specify the representation
of the (super)fields under some gauge group.

In the case of \texttt{Susyno}, it builds the most general renormalizable
supersymmetric (SUSY) Lagrangian allowed by the gauge symmetry, as
well as the soft SUSY breaking terms, and then applies known formulas
in order to derive the two-loop renormalization group equations for
all the free parameters (such as the gauge and Yukawa couplings).
The group theory code used to perform the first step grew over time
and was eventually released as the standalone \texttt{GroupMath} program
\cite{Fonseca:2020vke}. It is also used in other packages, such as
\texttt{SARAH 4} \cite{Staub:2013tta}, \texttt{Pyr@te 2+} \cite{Lyonnet:2016xiz},
\texttt{DRalgo} \cite{Ekstedt:2022bff} and \texttt{Sym2Int} \cite{Fonseca:2017lem}.
The aim of this last program, which will be the main topic from now
on, is to go from \uline{sym}metries \uline{to} \uline{int}eractions:
given some input fields (that is, representations of the Lorentz and
gauge groups) it computes all operators up to some mass dimension.
\texttt{Sym2Int} is currently being extended in order to be able to
provide explicit expressions for the operators.\footnote{See also the \texttt{AutoEFT} code mentioned in~\cite{Harlander:2023psl}.}

\subsubsection{The current \texttt{Sym2Int} program}

Details on how to use the program can be found in \cite{Fonseca:2017lem},
as well as on the program's webpage. For illustrative purposes, with
the following input one can obtain the list of SMEFT interactions
up to dimension 8:\\

\texttt{gaugeGroup{[}SM{]} \textasciicircum = \{SU3, SU2, U1\};}

\texttt{fld1 = \{\textquotedbl u\textquotedbl , \{3, 1, 2/3\}, \textquotedbl R\textquotedbl ,
\textquotedbl C\textquotedbl , 3\};}

\texttt{fld2 = \{\textquotedbl d\textquotedbl , \{3, 1, -1/3\},
\textquotedbl R\textquotedbl , \textquotedbl C\textquotedbl ,
3\};}

\texttt{fld3 = \{\textquotedbl Q\textquotedbl , \{3, 2, 1/6\}, \textquotedbl L\textquotedbl ,
\textquotedbl C\textquotedbl , 3\};}

\texttt{fld4 = \{\textquotedbl e\textquotedbl , \{1, 1, -1\}, \textquotedbl R\textquotedbl ,
\textquotedbl C\textquotedbl , 3\};}

\texttt{fld5 = \{\textquotedbl L\textquotedbl , \{1, 2, -1/2\},
\textquotedbl L\textquotedbl , \textquotedbl C\textquotedbl ,
3\};}

\texttt{fld6 = \{\textquotedbl H\textquotedbl , \{1, 2, 1/2\}, \textquotedbl S\textquotedbl ,
\textquotedbl C\textquotedbl , 1\};}

\texttt{fields{[}SM{]} \textasciicircum = \{fld1, fld2, fld3, fld4,
fld5, fld6\};}

\texttt{GenerateListOfCouplings{[}SM, MaxOrder -> 8{]};}

~

The number of interactions in this particular EFT --- up to dimension
15 and for an arbitrary number of flavors (which is set to 3 in the
code above) --- can be found in a couple of hours. As far as I know,
this is the only cross-check of the numbers provided for the first
time in \cite{Henning:2015alf} using the the Hilbert series.

It is worth noting that \texttt{Sym2Int} also computes some important
information on the symmetry of flavor indices. For example, $L_{i}L_{j}HH$
is found to be symmetric under the exchange of $i\leftrightarrow j$,
while $Q_{i}Q_{j}Q_{l}L_{k}$ has a more complicated symmetry. Let
us distinguish an \textit{operator}, where we expand flavor indices
(for $n$ flavors there are $n\left(n+1\right)/2$ operators of the
form $LLHH$) from a\textit{ Lagrangian term} which are tensors in
flavor space (there is just one term of the form $LLHH$). Then, the
information mentioned earlier can be used to infer the minimum number
of terms needed to write a model's Lagrangian. In the case of three
$Q$'s and one $L$, even though there are four ways of contracting
the various spinor and $SU(2)_{L}$ indices, it is possible to write
all of them as a single Lagrangian term $\omega_{ijkl}Q_{i}Q_{j}Q_{l}L_{k}$
(see \cite{Fonseca:2019yya} for a more thorough discussion of this
topic).

\subsubsection{An upgrade: building operators and terms explicitly}

Counting operators and terms is not the same as building a Lagrangian.
For the latter one needs to know the explicit form of each interaction,
which implies knowing how the various field indices are contracted.
It is also worth highlighting that neither the current version of
\texttt{Sym2Int} nor the Hilbert series method can be used to determine
how the derivatives --- if there is any --- are applied to the fields.

We desire a model's Lagrangian, but I would like to point out that
the Lagrangian often consists of a complicated function of the field
components with a very low information density. Consider for example
the interactions between a fermion transforming under the fundamental
representation of $SU(n)$ and the gauge bosons of this group. The
relevant Clebsch-Gordan factors coincide with the entries $T_{ij}^{A}$
of the matrices of the fundamental representation of $SU(n)$, containing
a total of $n^{2}\left(n^{2}-1\right)$ of entries. Given that fields
can be redefined, one has to wonder what is the information contained
in all these numbers: in principle, observable quantities should depend
on these Clebsch-Gordan factors through combinations of the tensor
$T$ with no open indices, such as $T_{ij}^{A}T_{ji}^{A}$ (see for
instance \cite{Cvitanovic:1976am}).

It is therefore conceivable that in the future we might find it unnecessary
to know a model's Lagrangian in full detail. With this cautionary
remark, the fact remains that at present we do need these complicated
expressions for the study of models in general, and EFTs in particular.
For this reason, the \texttt{Sym2Int} code is in the process of being
upgraded such that it not only counts but also computes explicitly
operators and terms. In the following I will make a few remarks concerning
this upgrade.

\paragraph{Output for humans vs output for other codes}

While designing a program which automatically generates lists of operators,
one must take into account whether the results are to be used directly
by a person, or fed into some other software package -- such as \texttt{FeynRules}
\cite{Alloul:2013bka} or \texttt{Matchmakereft} \cite{Carmona:2021xtq}.
In the second case, the readability of the result is not so important.
For example, the two possible ways of contracting three color octets
can be described by a $2\times8\times8\times8$-dimensional tensor
$c_{aijk}$ containing the relevant Clebsch-Gordan factors. But even
in this rather modest example, a human might find it hard to read
such data format.

A related problem is that these Clebsch-Gordan factors are not unique.
In the previous case, we are free to take any invertible combination
of the two contractions, $c_{aijk}\rightarrow X_{aa^{\prime}}c_{a^{\prime}ijk}$
with $\textrm{det}\left(X\right)\neq0$, and one can also make a rotation
in the eight-dimensional space of the octet, leading to the change
$c_{aijk}\rightarrow U_{ii^{\prime}}U_{jj^{\prime}}U_{kk^{\prime}}c_{ai^{\prime}j^{\prime}k^{\prime}}$
for some unitary matrix $U$.

Currently, the package \texttt{GroupMath} contains a function \texttt{Invariants}
which, subject to time and memory constraints, can compute the above
group theory data for any product of representations of a semi-simple
Lie algebra. However, due to the two issues discussed above, there
is room for improvement. Conveniently, the gauge symmetry of many
models is completely described by $SU(n)$ groups (and perhaps $U(1)$'s
which are easy to handle). Motivated by this, a future version of
\texttt{GroupMath }will include $SU(n)$-specific code capable of
providing the same information as \texttt{Invariants} but using instead
the tensor method which is familiar to physicists (see chapter 4 in
\cite{Cheng:1984vwu}). The numerical output will still consist of
large tensors with Clebsch-Gordan factors, but now it is also possible
to include a human readable string which identifies each of them.
For the product of three octets, writing them as $3\times3$ traceless
matrices $\Omega_{\;j}^{i}$, $\Omega_{\;j}^{\prime i}$ and $\Omega_{\;j}^{\prime\prime i}$,
the two invariant combinations alluded above would be $\Omega_{\;j}^{i}\Omega_{\;k}^{\prime j}\Omega_{\;i}^{\prime\prime k}$
and $\Omega_{\;j}^{i}\Omega_{\;k}^{\prime\prime j}\Omega_{\;i}^{\prime k}$;
the \texttt{GroupMath} program will be able to provide the value of
these two expressions as well as the corresponding formulas/strings.

The approach to Lorentz indices is similar: for each invariant expression
the program keeps track of a tensor (with the $SO(1,3)$ Clebsch-Gordan
factors) as well as a human readable string involving the familiar
scalars, spinors, gamma matrices, derivatives and field strength tensors.
If there are fermions in the interaction, with the $\gamma^{\mu}$
and $C$ matrices one can replace open spinor indices with vector
indices. Then, from an expression containing a set of open vector
indices only, one can construct all Lorentz invariants by contracting
it in all possible ways with the metric and Levi-Civita tensors, $\eta_{ab}$
and $\epsilon_{abcd}$. As for derivatives, they should be distributed
in all possible ways by the different fields in the operator. This
approach of contracting indices and applying derivatives in every
conceivable way leads to a highly redundant list of Lorentz invariant
expressions; such a list can easily be pruned by looking for linear
relations among the various polynomials.

\paragraph{Dealing with gauge indices and Lorentz indices separately}

Building operators explicitly implies handling potentially very large
polynomials of the numerous field components, thus calculations are
time consuming even for low dimensional interactions. In each step
of the design of a code that handles explicit operators, one must
therefore be aware of this problem and try to mitigate its impact.

In my opinion, an important step in reducing the computational and
memory requirements of handling operators is to segregate gauge indices
from the space-time indices of spinor and vector fields. If $c_{g_{1}g_{2}\cdots}$
and $\kappa_{l_{1}l_{2}\cdots}$ are the Clebsch-Gordan factors for
the two type of indices, instead of working with the full operator
$\mathcal{O}\equiv c_{g_{1}g_{2}\cdots}\kappa_{l_{1}l_{2}\cdots}\Phi_{g_{1},l_{1}}\Phi_{g_{2},l_{2}}\cdots$,
it is preferable to manipulate --- somehow --- the simpler polynomials
$\mathcal{O}_{G}\equiv c_{g_{1}g_{2}\cdots}\Phi_{g_{1}}\Phi_{g_{2}}\cdots$
and $\mathcal{O}_{L}\equiv c_{l_{1}l_{2}\cdots}\Phi_{l_{1}}\Phi_{l_{2}}\cdots$
involving only one type of indices.

Some may consider this to be an elementary observation but, as the
following example will show, it is not trivial to implement it. Consider
that both the gauge and the Lorentz groups are described by the $SU(2)$
group, and ignore for simplicity that some fields (fermions) anti-commute.
Now take some field $\Phi$ which is a doublet under both groups:
both $\mathcal{O}_{G}=\epsilon_{g_{1}g_{2}}\Phi_{g_{1}}\Phi_{g_{2}}$
and $\mathcal{O}_{L}=\epsilon_{l_{1}l_{2}}\Phi_{l_{1}}\Phi_{l_{2}}$
are identically zero, so whatever is our algorithm to handle these
two polynomials we would conclude that there is no $\Phi^{2}$ operator.
And yet, by considering the two indices together, we readily find
that $\mathcal{O}\equiv\epsilon_{g_{1}g_{2}}\epsilon_{l_{1}l_{2}}\Phi_{g_{1},l_{1}}\Phi_{g_{2},l_{2}}$
is not null. What is happening here is that both the Lorentz indices
$l_{i}$ and the gauge indices $g_{i}$ are contracted anti-symmetrically,
so by considering each set of indices separately, we get vanishing
polynomials, while the operator $\mathcal{O}$ is symmetric --- not
anti-symmetric --- under the change $\left(g_{1},l_{1}\right)\leftrightarrow\left(g_{2},l_{2}\right)$.

A possible solution is to distinguish equal fields in the simpler
polynomials $\mathcal{O}_{G}$ and $\mathcal{O}_{L}$; in the previous
example, $\mathcal{O}_{G}=\epsilon_{g_{1}g_{2}}\Phi_{g_{1}}\Phi_{g_{2}}^{\prime}$
and $\mathcal{O}_{L}=\epsilon_{l_{1}l_{2}}\Phi_{l_{1}}\Phi_{l_{2}}^{\prime}$
are no longer null, as long as we keep $\Phi^{\prime}\neq\Phi$. A
solution along this lines is viable, and indeed it has been successfully
tested in \texttt{Sym2Int}. Without going into details, I will simply
note that one must still account, in the end, for the fact that $\Phi^{\prime}$
is the same as as $\Phi$.\footnote{Setting back $\Phi=\Phi^{\prime}=\cdots$ at the end of the calculation
can have non-trivial consequences. To see this, using the same quantum
numbers as before, let us consider the quartic operators $\Phi^{4}$.
We start by differentiating the fields, thus considering instead interactions
of the type $\Phi\Phi^{\prime}\Phi^{\prime\prime}\Phi^{\prime\prime\prime}$.
In that case, there are two independent polynomials $\mathcal{O}_{G}$,
and well as two independent polynomials $\mathcal{O}_{L}$ (after all
the product $\boldsymbol{2}\times\boldsymbol{2}\times\boldsymbol{2}\times\boldsymbol{2}$
contains two singlets). There is therefore a total of four operators
$\Phi\Phi^{\prime}\Phi^{\prime\prime}\Phi^{\prime\prime\prime}$,
but once we require that $\Phi^{\prime\prime\prime}=\Phi^{\prime\prime}=\Phi^{\prime}=\Phi$
they all become proportional to a single expression.} Things become even more complicated when fields have flavor.

\paragraph{Flavor}

It turns out that in some models (for example in SMEFT), some representations
of the symmetry group are present more than once. We tend to account
for this mysterious multiplicity by adding a flavor index to the relevant
fields, $\Phi\rightarrow\Phi_{i}$, and write Lagrangians with flavored
tensors, such as the Yukawa matrices. Each of them is associated with
what I previously called a Lagrangian term, which may correspond to
many operators once the indices are expanded.

Flavor constitutes a significant complication. If all the fields in
an interaction are distinct, such as in $L_{i}^{*}L_{j}Q_{k}^{*}Q_{l}$,\footnote{The precise form of the spinor indices is irrelevant for the discussion.}
accounting for multiple flavors is trivial. The problem are those
cases containing repeated fields, such as $L_{i}^{*}L_{j}L_{k}^{*}L_{l}$,
as they will have some underlying symmetry under permutations of the
flavor indices. One approach to these troublesome cases might be to
consider each flavor combination at a time, effectively expanding
the flavor indices. Not only is such an approach very taxing computationally,
but it is also not clear how the results are to be presented ---
ideally one would like to undo the expansion of the indices and write
as few terms in the Lagrangian as possible.

As the operator dimension increases, so does the number of intervening
fields. At some point we are bound to find cases such as $Q_{i}Q_{j}Q_{k}L_{l}$
in SMEFT, where the same field appears three or more times. The relevant
permutation group is then no longer abelian, and we may have to consider
mixed (or multi-term) symmetries in the flavor indices (see \cite{Fonseca:2019yya}).

This problem is most acute in a model where only one index distinguishes
all scalars and all fermions; such model is therefore a good test-bed
for \texttt{Sym2Int}'s new code. Indeed, we may represent a generic
EFT as a theory with an arbitrary number of real scalars $\phi_{i}$
and Weyl fermions $\psi_{i}$ with covariant derivatives
\begin{equation}
D_{\mu}\psi_{i}=\partial_{\mu}\psi_{i}-\mathrm{i}gt_{ij}^{A}V_{\mu}^{A}\psi_{j}\textrm{ and }D_{\mu}\phi_{a}=\partial_{\mu}\phi_{a}-\mathrm{i}g\theta_{ab}^{A}V_{\mu}^{A}\phi_{b}
\end{equation}
where $t^{A}$ and $\theta^{A}$ are generic hermitian matrices (the
$\theta^{A}$ must also be anti-symmetric). It was precisely in this
general framework that the two-loop RG equations for dimension 4 operators were derived in \cite{Machacek:1983tz,Machacek:1983fi,Machacek:1984zw}.
We may however extend it to non-renormalizable operators, not just
to study complicated flavor symmetries, but also to derive important
results for a general EFT. Indeed we are currently in the process
of deriving the one-loop RG equations for this EFT (up to dimension-six interactions),
as well as the matching relations between it and a general renormalizable
UV model \cite{Fonseca-Olgoso-Santiago}. Once these are known, it
becomes unnecessary to go back to the computation of loops and amplitudes
for every single model; the task of computing RG equations and matching conditions
for a particular model is reduced to writing a Lagrangian, which involves
only some algebra and group theory.

\subsubsection{Outlook}

The \texttt{Sym2Int} code, as it exists now, lists and counts all
the possible operators in an effective field theory, up to some cutoff.
It is currently being extended to also build them explicitly, while
handling in a satisfactory way the fact that some fields have flavor.
With the approach being pursued, the presence of flavor does not significantly
affect the computational time, but it does pose some complicated problems
of a conceptual nature which still have to be addressed. Side-stepping
these for now, the code was already used to compute all Green operators
in SMEFT up to dimension 10, for an arbitrary number of fermion flavors,
with their counting matching the correct result.

%%%%%%%%%%%%%%%%%%%%%%%%%%%%%%%%%%%%%%%%%%%%%%%%%%%%%%%%%%%%%%%%%%%%%%%%%%%%%%%%%%%%%%%%%%%%%%%%
\contribution[{\includegraphics[width=0.35\textwidth]{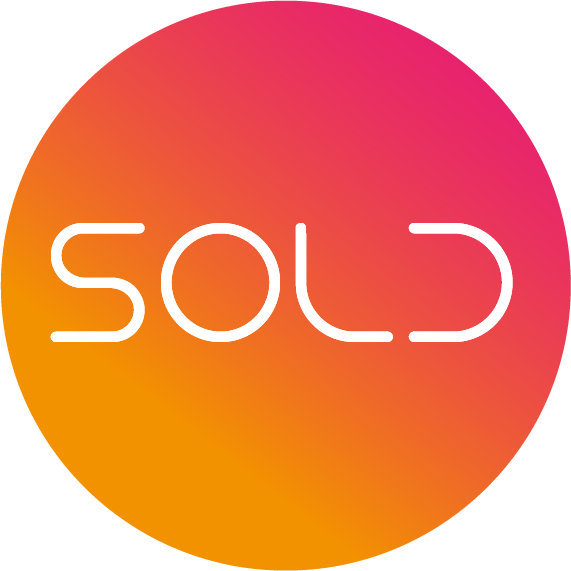}}]{\texttt{SOLD}: Towards the one-loop matching dictionary in the SMEFT}{José Santiago}{}\label{sec:SOLD}
%%%%%%%%%%%%%%%%%%%%%%%%%%%%%%%%%%%%%%%%%%%%%%%%%%%%%%%%%%%%%%%%%%%%%%%%%%%%%%%%%%%%%%%%%%%%%%%%

The benefit of using EFTs to compute experimental observables for different UV models is undeniable, but EFTs are much more powerful than that. They really shine when the top-down approach is combined with power-counting arguments, which allow for a complete classification of observable UV models in the form of IR/UV dictionaries. The EFT is a double expansion in the mass dimension of the operators and the loop order of the WCs, with operators of higher dimension being less relevant at low energies and WCs of higher loop order being smaller than lower loop ones. Eventually, contributions of high enough mass dimension and/or loop order are smaller than the experimental precision and can be disregarded as unobservable.

Given a finite order in mass dimension and loop order, the complete set of UV models that contribute to and EFT up to these orders can be exhaustively classified. Once the complete classification is achieved one can go one step further and compute the resulting WCs for all the UV models in the list. This way we obtain true IR/UV dictionaries that relate the WCs of the EFT (and therefore to experimental observables via the bottom-up approach) to all UV models that contribute to the EFT at the particular order the dictionary has been computed. These dictionaries can be used in an iterative way to obtain a complete map of the implications of experimental data on models of new physics. Indeed, given a particular experimental constraint or anomaly, one can list all models that are restricted by (or explain in the case of anomalies) that particular measurement. With this list, we can then obtain all other experimental implications of these models, that can be tested in a correlated way with different experimental data.

The tree-level, dimension-six dictionary for the SMEFT was computed a few years back~\cite{deBlas:2017xtg}, building on previous efforts~\cite{delAguila:2000rc,delAguila:2008pw,delAguila:2010mx,deBlas:2014mba}. This includes the most general extension of the SM with new scalars, fermions or vectors that contribute at tree level to the SMEFT operators up to mass-dimension six. While essential for the classification of large effects, this leading dictionary falls short when compared with the current precision of many experimental measurements, even more taking into account that certain WCs are only generated in weakly coupled extensions of the SM at the one-loop order. The first step towards the calculation of the one-loop, dimension-six IR/UV dictionary for the SMEFT has been recently published in~\cite{Guedes:2023azv}. 

\subsection*{Towards the one-loop, dimension-six IR/UV dictionary in the SMEFT}

Contrary to the tree-level dictionary, in which the list of new fields is finite (a total of 48 new scalars, fermions or vectors appear in the dictionary), at one-loop order the list is infinite, due to the fact that some contributions only constrain the quantum numbers of the product of multiples fields rather than each of them independently. Despite this infinite number of models, it is still possible to classify them in the form of a finite number of conditions on these models. Still, the calculation of the complete one-loop dictionary for the SMEFT at dimension six is a formidable task and, with the help of the computer tool \texttt{matchmakereft}~\cite{Carmona:2021xtq}, we have just finished the first step towards it~\cite{Guedes:2023azv}. In particular, we have considered the most general extension of the SM with an arbitrary number of scalar and fermionic fields\footnote{The best strategy to perform the matching in models with heavy gauge bosons is currently under study in collaboration with J. Fuentes-Martín, P. Olgoso and A.E. Thomsen.} that contribute at one-loop order to those operators in the Warsaw basis~\cite{Grzadkowski:2010es} which cannot be generated at tree level in any weakly-coupled extension of the SM. This includes all operators in the basis that contain at least one field-strength gauge tensor. Our results, that include the full classification of models, a partial list of the specific representations (up to a certain representation dimension, to be chosen by the user) and the actual value of the corresponding WCs are too long to be reported in print form, and we have published them in electronic form as a \texttt{Mathematica} package called \texttt{SOLD} (for Smeft One Loop Dictionary), available via its  \href{https://gitlab.com/jsantiago_ugr/sold}{\texttt{Gitlab} repository} (it can also trivially installed directly from a \texttt{Mathematica} notebook). Tools to create models suitable for the full one-loop matching using \texttt{matchmakereft} are also available within \texttt{SOLD} (see~\cite{Guedes:2023azv} for details). The fact that \texttt{matchmakereft} performs the matching calculation in a gauge-blind fashion for most of the computation has significantly helped the development of the dictionary.

%%%%%%%%%%%%%%%%%%%%%%%%%%%%%%%%%%%%%%%%%%%%%%%%%%%%%%%%%%%%%%%%%%%%%%%%%%%%%%%%%%%%%%%%%%%%%%%%
\contribution{MatchingDB: A format for matching dictionaries}{Juan Carlos Criado}{}\label{sec:MatchingDB}
%%%%%%%%%%%%%%%%%%%%%%%%%%%%%%%%%%%%%%%%%%%%%%%%%%%%%%%%%%%%%%%%%%%%%%%%%%%%%%%%%%%%%%%%%%%%%%%%

MatchingDB is a format for the storage, exchange and exploration of EFT matching results up to one-loop order.
Its specification, both in human- and machine-readable forms, together with a Python interface, is located at the MatchingDB GitLab repository:
\begin{center}
\href{https://gitlab.com/jccriado/matchingdb/}{gitlab.com/jccriado/matchingdb}
\end{center}
It aims to provide:
\begin{itemize}
  \item A unified language/tool-independent format for the communication of EFT matching results.
  \item An efficient workflow for the practical use of matching dictionaries.
\end{itemize}
The format is particularly useful to store and publish large matching dictionaries, whose size might make it impractical to provide them in the form of human-readable equations and tables.
The Python interface makes it easy to interact with them, and quickly obtain the relevant information for specific applications.
An example of such a dictionary is the complete tree-level dictionary~\cite{deBlas:2017xtg} between the dimension-six SMEFT and any of its UV completions, which is provided in MatchingDB format under the \href{https://gitlab.com/jccriado/matchingdb/-/tree/main/dictionaries}{dictionaries} directory in the MatchingDB repository.
Other matching databases will be made available at the same place.

MatchingDB can also be employed as a data exchange format between tools, allowing to compare results from different matching codes, and providing an interface to connect them to packages for RG running and observable calculations.
It will be implemented as an output format in \texttt{MatchingTools}~\cite{Criado:2017khh} and \texttt{Matchmakereft}~\cite{Carmona:2021xtq}.

Some of the features currently offered by the combination of the MatchingDB format and the accompanying Python package are: listing the heavy fields and UV couplings that generate a given WC; listing the EFT contributions of a given set of heavy fields; providing LaTeX output, both for the UV Lagrangian and for matching corrections; and providing numerical output that matches WCxf~\cite{Aebischer:2017ugx} for the SMEFT.

\begin{figure*}
  \centering
  \includegraphics[width=\textwidth]{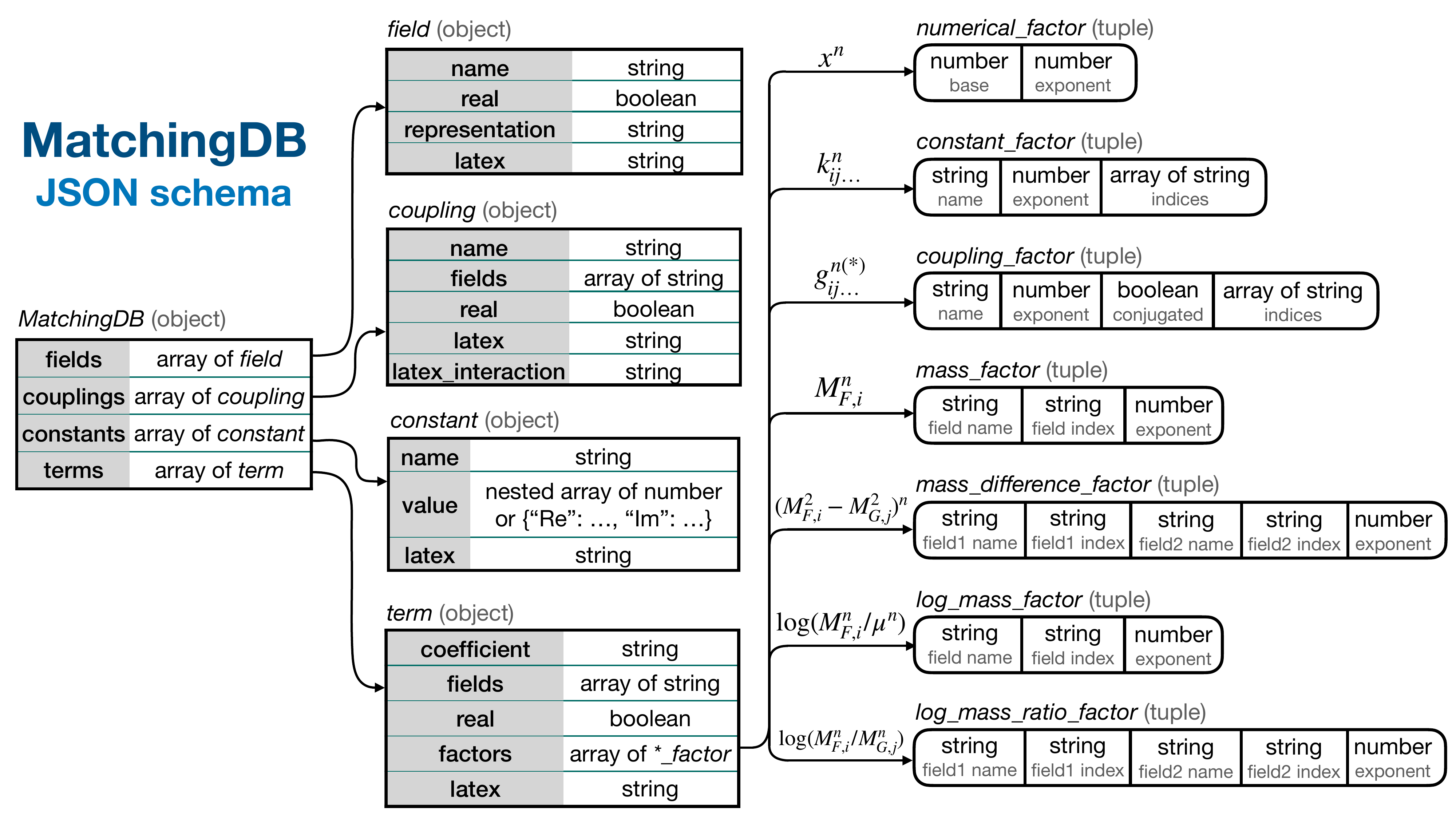}
  \caption{Diagram summarizing the MatchingDB JSON schema. Rectangles represent json objects, with the property names on the left column and the corresponding value types on the right. Capsule-shaped items represent a tuple, with the type of each of its items given in a larger font, and an short explanation of their meaning on a smaller one directly below. }
  \label{fig:matchingdb-schema}
\end{figure*}

\subsubsection{Format definition}

MatchingDB data can be stored either as a plain-text JSON~\cite{json} file or as an SQLite~\cite{sqlite} database.
The format is defined by its JSON schema, which is given in the
\href{https://gitlab.com/jccriado/matchingdb/-/raw/main/matchingdb.json}{matchingdb.json}
file at the root of the MatchingDB repository, using the JSON Schema language~\cite{JsonSchema}.
A MatchingDB JSON file must comply with this schema, which can be checked using any of the standard tools for this purpose, such as the \texttt{jsonschema} Python package~\cite{python-jsonschema}.
Alternatively, MatchingDB data can be stored as an SQLite database, with a structure based on the JSON schema.
The SQLite representation may provide faster access to the information in larger databases.
Below, I describe the format informally, starting with the JSON representation.
A diagram summarizing it is provided in Fig.~\ref{fig:matchingdb-schema}.

The root value of the data must be an object with 4 name/value pairs, with names \texttt{"fields"}, \texttt{"couplings"}, \texttt{"constants"} and \texttt{"terms"}. The corresponding values should be arrays whose values are objects with the following structures:
\begin{description}
  \item[\texttt{field}] Represents a heavy field in the UV theory that has been integrated out. It contains the following key/value pairs:
        \begin{itemize}
          \item \texttt{"name"} \emph{(string)}: a name identifying the field.
          \item \texttt{"real"} \emph{(boolean)}: determines whether the field is real or complex.
          \item \texttt{"representation"} \emph{(string)}: the group theory representation. The format for this is free in principle, but intended to be self-consistent in each database.
          \item \texttt{"latex"} \emph{(string)}: math-mode LaTeX code representing the field.
        \end{itemize}

  \item[\texttt{coupling}] Represents a coupling present in the UV theory. Its name/value pairs are:
        \begin{itemize}
          \item \texttt{"name"} \emph{(string)}: a name identifying the coupling constant.
          \item \texttt{"fields"} \emph{(array of string)}: a \textbf{sorted} list of the heavy fields that appear in the interaction.
          \item \texttt{"real"} \emph{(boolean)}: determines whether the coupling constant is real or complex.
          \item \texttt{"latex"} \emph{(string)}: math-mode LaTeX code representing the coupling constant.
          \item \texttt{"latex\_interaction"} \emph{(string)}: math-mode LaTeX code representing the full interaction term in the UV theory, including the coupling constant.
        \end{itemize}

  \item[\texttt{constant}] Represents a scalar constant such as $\pi$ or a constant tensor such as the Kronecker delta, which appears in the matching results. Its name/value pairs are
        \begin{itemize}
          \item \texttt{"name"} \emph{(string)}: a name identifying the constant.
          \item \texttt{"value"} \emph{(nested array of numbers or object)}: the numerical value of the constant. If the constant is a scalar, it should be a number. If it is a tensor, it should be provided as a nested array of numbers. Finally, if the constant is complex, it should be given as an object of the form: \texttt{\{"Re": ..., "Im": ...\}}, with the values having the real scalar o tensor type.
          \item \texttt{"latex"} \emph{(string)}: math-mode LaTeX code representing the constant.
        \end{itemize}
  \item[\texttt{term}] Represents a term appearing in the matching corrections to some WCs in the EFT:
        \begin{itemize}
          \item \texttt{"coefficient"} \emph{(string)}: a name identifying the WC in which the term appears.
          \item \texttt{"fields"} \emph{(array of string)}: a \textbf{sorted} list of the heavy fields that contribute to the term.
          \item \texttt{"factors"} \emph{(array of *\_factor)}: a list of the factors that appear in the term, as described below.
          \item \texttt{"free\_indices"} \emph{(array of string)}: a list of the free indices of the term, which must coincide with the ones of the corresponding coefficient, and appear at least once in the \texttt{"factors"} list.
        \end{itemize}
\end{description}
The \texttt{"fields"} array in both couplings and terms should be sorted with the lexicographic order.
This allows to compare them efficiently to a set of heavy fields when querying the database.

The full analytical formulas for the matching corrections to the WCs in the EFT are stored in the \texttt{"factors"} property of the \texttt{term} objects.
The matching correction to any WC $\mathcal{C}$ is a sum of terms $\mathcal{T}^{(N)}$, with each term being a product of factors $\mathcal{F}^{(N)}_A$:
\begin{equation}
  \mathcal{C} = \sum_N \mathcal{T^{(N)}}, \qquad
  \mathcal{T}^{(N)} = \prod_A \mathcal{F}^{(N)}_A.
\end{equation}
Each factor is assumed to be of one of 7 possible forms.
Every form has an associated JSON type, all of them being tuples, that is, inhomogeneous arrays with a fixed type for each of its items:
\begin{description}
  \item[$\mathcal{F} = b^n$] (\texttt{numerical\_factor}): a number $b$ to some power $n$. Represented as a tuple $[b, n]$ of type:\\ \texttt{[number, number]}.

  \item[$\mathcal{F} = k^n_{ij\dots}$] (\texttt{constant\_factor}): a constant $k$, to some power $n$, with some flavor indices $i$, $j$, \ldots Represented as a tuple $[g, n,\allowbreak [i, j, \ldots]]$ of type:\\ \texttt{[string, number, array of string]}.

  \item[$\mathcal{F} = g^{n(*)}_{ij\dots}$] (\texttt{coupling\_factor}): a coupling constant $g$, to some power $n$, possibly complex conjugated ($c = \rm True/False$), with some flavor indices $i$, $j$, \ldots \\ Represented as a tuple $[g, n, c,\allowbreak [i, j, \ldots]]$ of type:\\ \texttt{[string, number, boolean, array of string]}.

  \item[$\mathcal{F} = M^n_{F,i}$] (\texttt{mass\_factor}): the mass of a field $F$, to some power $n$, with a flavor index $i$. Represented as a tuple $[F, i, n]$ of type:\\ \texttt{[string, string, number]}.

  \item[$\mathcal{F} = (M^2_{F,i} - M^2_{G,j})^n$] (\texttt{mass\_difference\_factor}): the difference between the masses of two fields. Represented as a tuple $[F, i, G, j, n]$ of type:\\ \texttt{[string, string, string, string, number]}.

  \item[$\mathcal{F} = \log(M^n_{F,i}/\mu^n)$] (\texttt{log\_mass\_factor}): the log of the mass of a field $F$, to some power $n$, with a flavor index $i$. Represented as a tuple $[F, i, n]$ of type:\\ \texttt{[string, string, number]}.

  \item[$\mathcal{F} = \log(M^n_{F,i}/M^n_{G,j})$] (\texttt{log\_mass\_difference\_factor}):\\ the log of the ratio between the masses of two fields. Represented as a tuple $[F, i, G, j, n]$ of type:\\ \texttt{[string, string, string, string, number]}.
\end{description}
This completes the specification of the MatchingDB format in JSON form.

MatchingDB data can also be stored as an SQLite database.
The SQLite representation consists of 4 tables, named \texttt{fields}, \texttt{couplings}, \texttt{constants} and \texttt{terms}.
Their columns take their names from the keys of the associated objects.
Every object is stored as a row, with each of its values encoded as a string containing the corresponding JSON code.

\subsubsection{Python interface}

The \texttt{matchingdb} Python package is provided under the \href{https://gitlab.com/jccriado/matchingdb/-/tree/main/python}{\texttt{python}} directory of the MatchingDB repository.
It can be installed by cloning the repository, moving into the \texttt{python} directory and running
\begin{verbatim}
  > pip install .
\end{verbatim}
The package exposes two classes: \texttt{JsonDB} and \texttt{SQLiteDB}, for creating and querying MatchingDB dictionaries, in the JSON and the SQLite representations, respectively.
Both classes have the same methods, with the same arguments and the same behaviour.
An existing database can be loaded as:
\begin{lstlisting}[language=ppython]
db = JsonDB.load("db.json")
# or SQLiteDB.load("db.sqlite")
\end{lstlisting}
A new one can be created through:
\begin{lstlisting}[language=ppython]
db = JsonDB.new("db.json", data=my_data)
# or SQLiteDB.new("db.sqlite", data=my_data)
\end{lstlisting}
where \texttt{my\_data} is the data to be included, as a JSON value that validates against the MatchingDB schema, represented as a Python object through the mapping displayed in Table~\ref{tab:matchingdb-json-python-mapping}.

\begin{table}
  \centering
  \begin{tabular}{ll}
    \hline\noalign{\smallskip}
    JSON & Python \\
    \noalign{\smallskip}\hline\noalign{\smallskip}
    object & dict \\
    array & list or tuple \\
    string & str \\
    number & int or float \\
    boolean & bool \\
    \noalign{\smallskip}\hline
  \end{tabular}
  \caption{Mapping between JSON and Python types.}
  \label{tab:matchingdb-json-python-mapping}
\end{table}

New items can be inserted into a database through:
\begin{lstlisting}[language=ppython]
db.insert(item, table)
\end{lstlisting}
where table is one of \texttt{"fields"}, \texttt{"couplings"}, \texttt{"const\\ants"}, or \texttt{"terms"}, and item is a dict complying with the corresponding sub-schema, which can be found at \texttt{\$defs/<table>/items} in the full schema.
Any changes made to a database must be saved with
\begin{lstlisting}[language=ppython]
db.save()
\end{lstlisting}
in order for them to persist.

There are 4 methods to query a database:
\begin{itemize}
  \item \texttt{select\_fields()}
  \item \texttt{select\_couplings()}
  \item \texttt{select\_constants()}
  \item \texttt{select\_terms()}
\end{itemize}
They filter the items of each of the corresponding arrays according to certain conditions, and prepare the selected items in the desired output state.
A summary of their arguments is provided in Table~\ref{tab:matchingdb-python}.
All of them are optional.
If provided, the \texttt{name}, \texttt{fields} and \texttt{coeffi\\cients} arguments select only those items for which the corresponding property coincides with the given one.
\texttt{fields\_criterion} further configures the behaviour of the \texttt{fields} argument, following Table~\ref{tab:matchingdb-fields-criterion}.

\begin{table*}
  \footnotesize
  \hspace{-25pt}
  \begin{tabular}{lll}
    \hline\noalign{\smallskip}
    Method
    & Arguments
    & Description
    \\
    \noalign{\smallskip}\hline\noalign{\smallskip}
    \multirow{2}{*}{\texttt{select\_fields()}}
    & \texttt{name: str} 
    & If provided, select the field with the given name \\
    & \texttt{output\_format: str} 
    & One of \{\texttt{"raw"} (default), \texttt{"pandas"}\}
    \\
    \noalign{\smallskip}\hline\noalign{\smallskip}
    \multirow{4}{*}{\texttt{select\_couplings()}}
    & \texttt{name: str} & If provided, select the coupling with the given name \\
    & \texttt{fields: [str]} & If provided, select the couplings with the given set of fields \\
    & \texttt{fields\_criterion: str} & One of \{\texttt{"equals"} (default), \texttt{"subset"}, \texttt{"supset"}\} \\
    & \texttt{output\_format: str} & One of \{\texttt{"raw"} (default), \texttt{"pandas"}, \texttt{"latex"}\} \\
    \noalign{\smallskip}\hline\noalign{\smallskip}
    \multirow{2}{*}{\texttt{select\_constants()}}
    & \texttt{name: str} & If provided, select the constant with the given name \\
    & \texttt{output\_format: str} & One of \{\texttt{"raw"} (default), \texttt{"pandas"}\}
    \\
    \noalign{\smallskip}\hline\noalign{\smallskip}
    \multirow{5}{*}{\texttt{select\_terms()}}
    & \texttt{coefficient: str} & If provided, select the terms contributing to a given coefficient \\
    & \texttt{fields: [str]} & If provided, select the terms with the given set of fields \\
    & \texttt{fields\_criterion: str} & One of \{\texttt{"equals"} (default), \texttt{"subset"}, \texttt{"supset"}\} \\
    & \texttt{output\_format: str} & One of \{\texttt{"raw"} (default), \texttt{"pandas"}, \texttt{"latex"}, \texttt{"numeric"}\} \\
    & \texttt{parameters: Iterable} & The non-vanishing parameters for numerical output \\
    \noalign{\smallskip}\hline
  \end{tabular}
  \caption{Summary of the querying methods of the \texttt{JsonDB} and \texttt{SQLiteDB} classes of the \texttt{matchingdb} Python package.}
  \label{tab:matchingdb-python}
\end{table*}

\begin{table}
  \centering
  \begin{tabular}{ll}
    \hline\noalign{\smallskip}
    Value & Select an \texttt{item} if: \\
    \noalign{\smallskip}\hline\noalign{\smallskip}
    \texttt{"equals"} & \texttt{item["fields"] == sorted(fields)} \\
    \texttt{"subset"} & \texttt{set(item["fields"]) <= set(fields)} \\
    \texttt{"supset"} & \texttt{set(item["fields"]) >= set(fields)} \\
    \noalign{\smallskip}\hline
  \end{tabular}
  \caption{Possible values and behavior of the \texttt{fields\_criterion} argument to the \texttt{select\_couplings()} and \texttt{select\_terms()} methods of the \texttt{JsonDB}, and \texttt{SQLiteDB} classes. In the right column, \texttt{fields} refers to the argument of these methods with the same name.}
  \label{tab:matchingdb-fields-criterion}
\end{table}

The \texttt{output\_format} argument must be one of the following strings:
\begin{description}
  \item[\texttt{"raw"} (default).] The method returns a list of the selected items, represented as Python values, following Table~\ref{tab:matchingdb-json-python-mapping}.
  \item[\texttt{"pandas"}.] The method returns a Pandas~\cite{reback2020pandas} dataframe with a simplified version of the output. Provides an easy way to visually explore the data.
  \item[\texttt{"latex"}.] Returns math-mode LaTeX code representing the output. \texttt{select\_couplings()} returns a single string with the formula for the selected sector of the UV Lagrangian. \texttt{select\_terms()} returns a dict with coefficient names as keys and strings with their selected terms as values. The other 2 methods do not accept this option.
  \item[\texttt{"numeric"}.] Available in \texttt{select\_terms()} only. Returns a function for the numerical evaluation of WCs. The additional \texttt{parameters} argument of \texttt{select\_terms()} must be set to an iterable containing all the names of all the parameters that will be set to non-vanishing values. This allows to prepare the output function to be efficiently evaluated many times.
\end{description}
The function returned by \texttt{select\_terms()} when \texttt{output\\\_format="numeric"} takes two arguments:
\begin{description}
  \item[\texttt{parameters: dict}.] A dict whose keys are the UV parameters (couplings, masses and matching scale), and whose values are NumPy~\cite{harris2020array} arrays with one axis for each of the indices of the corresponding parameter.
        Masses are named \texttt{"M\_<field>"} where \texttt{"<field>"} is the name of the field. The matching scale is \texttt{"mu"}. The values of constants in the \texttt{"constants"} array of the database are included automatically.

  \item[\texttt{expand\_flavor: bool} (optional).]\phantom{.}
        \begin{itemize}
          \item If \texttt{False} (default), the output of the function is a dictionary with WC names as keys, and NumPy arrays as values, with one axis per EFT flavor index of the coefficient.
          \item If \texttt{True}, flavor indices are expanded, and the output becomes a dictionary with keys of the form \texttt{"<coeff>\_<flavor\_indices>"}, and values being either float (if real) or dictionaries with keys  \texttt{"Re"}, \texttt{"Im"} and floats as values (if complex). If the names given to the coefficients in the database follow the conventions of WCxf, the output will be compatible with the \texttt{values} section of a WCxf WC file.
        \end{itemize}
\end{description}

\subsubsection{Example}

The \href{https://gitlab.com/jccriado/matchingdb/-/tree/main/python/examples}{python/examples} directory contains examples showcasing several features of the \texttt{matchingdb} package.
Here, I will present a brief example on how to extract different types of information from the tree-level dimension-six SMEFT matching dictionary~\cite{deBlas:2017xtg} (given in MatchingDB format at \href{https://gitlab.com/jccriado/matchingdb/-/raw/main/dictionaries/smeft_dim6_tree.json}{dictionaries/smeft\_dim6\_tree\\.json}). To load this dictionary, one can do:
\begin{lstlisting}[language=ppython]
from matchingdb import JsonDB
db = JsonDB.load("smeft_dim6_tree.json")
\end{lstlisting}
One can then get a summary view of all terms that appear in the WC for the $\mathcal{O}_{ll}$ Warsaw operator through:
\begin{lstlisting}[language=ppython]
db.select_terms(
  coefficient="ll", output_format="pandas"
)
\end{lstlisting}
\begin{lstlisting}[language=poutput]
Output:
  coefficient fields     couplings
0          ll   [S1]    [yS1, yS1]
1          ll  [Xi1]  [yXi1, yXi1]
2          ll    [B]    [glB, glB]
3          ll    [W]    [glW, glW]
4          ll    [W]    [glW, glW]
\end{lstlisting}
From this table, one can see which UV fields and couplings generate this operator at tree level.
Information on these fields can be obtained as:
\begin{lstlisting}[language=ppython]
db.select_fields(
  name="S1", output_format="pandas"
)
\end{lstlisting}
\begin{lstlisting}[language=poutput]
    name   real representation
0   S1    False       S(1,1,1)
\end{lstlisting}
All the matching corrections to any WC induced by the $\mathcal{S}_1$ field can be found via:
\begin{lstlisting}[language=ppython]
db.select_terms(
  fields=["Xi"], output_format="pandas"
)
\end{lstlisting}
\begin{lstlisting}[language=poutput]
  coefficient fields   couplings
0          ll   [S1]  [yS1, yS1]
\end{lstlisting}
This implies that $\mathcal{S}_1$ only contributes to $\mathcal{O}_{ll}$.
The formula for this contribution in LaTeX code can be obtained through:
\begin{lstlisting}[language=ppython]
db.select_terms(
  fields=["Xi"], output_format="latex"
)
\end{lstlisting}
\begin{lstlisting}[language=poutput]
{'ll': ' + \\frac{  \\left(y_{\\mathcal...
\end{lstlisting}
which renders as:
$
  + \frac{  \left(y_{\mathcal{S}_1}\right)_{ajl}^{*} \left(y_{\mathcal{S}_1}\right)_{aik}}{  M_{\mathcal{S}_1,a}^{2}}
$.

This formula can also be numerically evaluated given the values of the UV parameters: the coupling $y_{\mathcal{S}_1}$ and the mass $M_{\mathcal{S}_1}$. This is done as:
\begin{lstlisting}[language=ppython]
evaluator = db.select_terms(
  fields=["S1"],
  output_format="numeric",
  parameters={"yS1", "M_S1"},
)

import numpy as np
n = 2  # number of S1 flavors
parameters = {
    "yS1": np.random.random(size=(n, 3, 3)),
    "M_S1": np.random.random(size=(n,)),
  }

evaluator(parameters, expand_flavor=True)
\end{lstlisting}
\begin{lstlisting}[language=poutput]
{'ll_0000': ..., 'll_0001': ...,
 'll_0002': ...,  ...}
\end{lstlisting}
The final output here has the format of the \texttt{values} field of the WCxf format, and is thus suitable for interfacing with numerical tools for running and the calculation of observables.
The \texttt{evaluator()} function is optimized for multiple evaluations, with the corresponding database lookup being performed once, in the \texttt{select\_terms()} call.

%%%%%%%%%%%%%%%%%%%%%%%%%%%%%%%%%%%%%%%%%%%%%%%%%%%%%%%%%%%%%%%%%%%%%%%%%%%%%%%%%%%%%%%%%%%%%%%%
\contribution{RG equations in generic EFTs\label{section:misiak}}{Miko{\l}aj Misiak and Ignacy Na{\l\c{e}}cz}{}
%%%%%%%%%%%%%%%%%%%%%%%%%%%%%%%%%%%%%%%%%%%%%%%%%%%%%%%%%%%%%%%%%%%%%%%%%%%%%%%%%%%%%%%%%%%%%%%%

The SMEFT RG equations at one loop were determined in
Refs.~\cite{Jenkins:2013zja,Jenkins:2013wua,Alonso:2013hga}, and the RG equations for the LEFT WCs have been determined in the past, dependently on phenomenological needs, sometimes up to the four-loop level~\cite{Czakon:2006ss}. However, the two-loop SMEFT RG equations remain unknown. Instead of deriving the RG equations separately in various Effective Field
Theories (EFTs), one can consider a generic case, as done for
renormalizable models (see below).  Particular results are then found
by substitutions. Our goal in the current (ongoing) project is to
evaluate one-loop RG equations for all the dimension-six operators in the generic case.

\subsubsection{Operator classification}\label{class}

We shall consider EFTs of the LEFT and SMEFT type, where the gauge
group is an arbitrary finite product of finite-dimensional Lie
groups. Real scalars $\phi_a$ and left-handed spin-$\frac12$ fermions
$\psi_k$ are going to be the matter fields. Obviously, any complex
scalar can always be written in terms of two real ones, while
right-handed spin-$\frac12$ fermions can always be described as
charge-conjugated left-handed ones.

To simplify our calculation in its initial steps, we assume a discrete
symmetry $\{\phi \to -\phi, ~\psi \to i\psi\}$. It turns out to forbid
all odd-dimensional operators. However, it gives no restriction on
even-dimensional ones when they have already been required to
be Lorentz-invariant. In more generic EFTs, where no such discrete
symmetry is imposed, RG equations for odd-dimensional operators can be
obtained from the even-dimensional case by treating one of the scalar
fields as an auxiliary gauge-singlet that takes a fixed vacuum
expectation value.

The generic EFT Lagrangian we are going to consider reads
\begin{align}
{\mathcal L} &= -\frac14 F^A_{\mu\nu} F^{A\,\mu\nu} + \frac12 (D_\mu\phi)_a (D^\mu\phi)_a + i \bar\psi_j (\not\!\!D \psi)_j- \frac12 m^2_{ab} \phi_a\phi_b -\frac{1}{4!} \lambda_{abcd} \phi_a \phi_b \phi_c \phi_d\nonumber\\
&\quad-\frac12 \left(Y^a_{jk} \phi_a \psi^T_j C \psi_k + {\rm h.c.}\right) + {\mathcal L}_{\rm g.f.} + {\mathcal L}_{\rm FP}+ \frac{1}{\Lambda^2} \sum Q_N + {\mathcal O}\left(\frac{1}{\Lambda^4}\right), \label{Leff}
\end{align}
where $Q_N$ stand for linear combinations of dimension-six operators
multiplied by their WCs.

Let us absorb the gauge couplings into the structure constants and generators. Then
$F^A_{\mu\nu}  = \partial_\mu V^A_\nu - \partial_\nu V^A_\mu - f^{ABC} V^B_\mu V^C_\nu$,
$(D_\rho F_{\mu\nu})^A  = \partial_\rho F^A_{\mu\nu} - f^{ABC} V_\rho^B F^C_{\mu\nu}$,
$(D_\mu\phi)_a = \left(\delta_{ab}\partial_\mu + i\theta^A_{ab} V^A_\mu\right) \phi_b$, and
$(D_\mu\psi)_j = \left(\delta_{jk}\partial_\mu + i t^A_{jk} V^A_\mu\right) \psi_k$.

RG equations for couplings at the dimension-four interactions in
Eq.~\eqref{Leff} were calculated up to two loops in a series of papers
by Machacek and Vaughn almost 40 years
ago~\cite{Machacek:1983tz,Machacek:1983fi,Machacek:1984zw}. Some
corrections to their results were found more recently in
Refs.~\cite{Luo:2002ti,Schienbein:2018fsw}. Even at the one-loop
level, it is only the latter paper~\cite{Schienbein:2018fsw} that we
fully agree with. Generic RG equations for the gauge and Yukawa couplings at
the four- and three-loop levels, respectively, were recently determined
in Ref.~\cite{Bednyakov:2021qxa,Davies:2021mnc} by combining information on results in various specific
models. Earlier three-loop results for the gauge coupling beta functions
can be found in Refs.~\cite{Pickering:2001aq,Poole:2019kcm}.

We perform our one-loop calculation off shell, using the
background-field gauge method. Therefore, we need to begin with
classifying all the dimension-six operators in the off-shell
basis. Such operators are gauge invariant but many linear combinations
of them vanish by the Equations of Motion (EOM). Once the RG equations in the
off-shell basis are found, one needs to pass to the on-shell basis
where no linear combination of operators vanishes by the EOM. Only in
the latter case are the RG equations gauge-parameter independent.

The off-shell basis we use consists of the following 22 terms\footnote{
Very similar off-shell and on-shell bases of dimension-six operators
were presented by R.~Fonseca and J.~Santiago at the SMEFT-Tools workshop. We
have verified that all the differences between our setups are due to
convention choices only.}
\begin{align}
Q_1    &= \frac{1}{6!} W^{( 1)}_{abcdef}\, \phi_a \phi_b \phi_c \phi_d \phi_e \phi_f,&
Q_2    &= \frac{1}{4} W^{( 2)}_{abcd}\, (D_\mu\phi)_a (D^\mu\phi)_b \phi_c \phi_d,\nonumber\\
Q_3    &= \frac12 W^{( 3)}_{ab}\, (D^\mu D_\mu\phi)_a (D^\nu D_\nu\phi)_b,&
Q_4    &= \frac12 W^{( 4)A}_{ab}\, (D^\mu\phi)_a (D^\nu\phi)_b F^A_{\mu\nu},\nonumber\\
Q_5    &= \frac14 W^{( 5)AB}_{ab}\, \phi_a \phi_b F^A_{\mu\nu}F^{B\,\mu\nu},&
Q_6    &= \frac14 W^{( 6)AB}_{ab}\, \phi_a \phi_b F^A_{\mu\nu}\widetilde F^{B\,\mu\nu},\nonumber\\
Q_7    &= \frac12 W^{( 7)AB}\, \left(D^\mu F_{\mu\nu}\right)^A \left( D_\rho F^{\rho\nu}\right)^B,&
Q_8    &= \frac{1}{3!} W^{( 8)ABC}\, F^{A\,\mu}_{~~~\;\nu}F^{B\,\nu}_{~~~\;\rho}F^{C\,\rho}_{~~~\;\mu},\nonumber\\
Q_9    &= \frac{1}{3!} W^{( 9)ABC}\, F^{A\,\mu}_{~~~\;\nu}F^{B\,\nu}_{~~~\;\rho}\widetilde F^{C\,\rho}_{~~~\;\mu},&
Q_{10} &= \frac18 W^{(10)}_{jkln}\, (\psi_j^T C \psi_k)(\psi_l^T C \psi_n) + {\rm h.c.},\nonumber\\
Q_{11} &= \frac14 W^{(11)}_{jkln}\, (\bar\psi_j \gamma_\mu \psi_k)(\bar\psi_l \gamma^\mu \psi_n),&
Q_{12} &= i W^{(12)}_{jk}\, \bar\psi_j \left(\not\!\! D \not\!\! D \not\!\! D\psi\right)_k,\nonumber\\
Q_{13} &= \frac12 W^{(13)}_{a,jk}\, \phi_a (D_\mu\psi)_j^T C (D^\mu\psi)_k + {\rm h.c.},&
Q_{14} &= W^{(14)}_{a,jk}\, \phi_a \psi_j^T C (D_\mu D^\mu\psi)_k + {\rm h.c.},\nonumber\\
Q_{15} &= \frac12 W^{(15)}_{a,jk}\, \phi_a (D_\mu\psi)_j^T C \sigma^{\mu\nu} (D_\nu\psi)_k + {\rm h.c.},&
Q_{16} &= \frac{i}{2} W^{(16)}_{ab,jk}\, \phi_a\phi_b\, \left[ (\bar\psi\,\raisebox{3.5mm}{}^{\leftarrow}\hspace{-4mm} \not\!\! D)_j \psi_k- \bar\psi_j (\not\!\! D\psi)_k \right],\nonumber\\
Q_{17} &= W^{(17)}_{ab,jk}\, \phi_a(D_\mu \phi)_b\, \bar\psi_j \gamma^\mu \psi_k,&
Q_{18} &= \frac{1}{12} W^{(18)}_{abc,jk}\, \phi_a\phi_b\phi_c\, \psi_j^T C \psi_k + {\rm h.c.},\nonumber\\
Q_{19} &= \frac12 W^{(19)A}_{a,jk}\, \phi_a F^A_{\mu\nu}\, \psi_j^T C \sigma^{\mu\nu} \psi_k + {\rm h.c.},&
Q_{20} &= i W^{(20)A}_{jk}\, F^A_{\mu\nu}\, \left[ (\bar\psi\,\raisebox{3.5mm}{}^{\leftarrow}\hspace{-3.3mm} D^\nu)_j \gamma^\mu \psi_k - \bar\psi_j \gamma^\mu (D^\nu \psi)_k\right],\nonumber\\
Q_{21} &= i W^{(21)A}_{jk}\, \widetilde F^A_{\mu\nu}\, \bar\psi_j\gamma^\mu (D^\nu \psi)_k,&
Q_{22} &= W^{(22)A}_{jk}\, \left(D^\mu F_{\mu\nu}\right)^A \bar\psi_j\gamma^\nu \psi_k,
\end{align}
where $W^{(N)}$ contain both the WCs and the necessary
Clebsch-Gordan coefficients that select singlets from various tensor
products of the gauge group representations.  In general, each
$W^{(N)}$ contains many independent WCs, and 
many gauge-singlet operators are present in each $Q_N$. 

After applying the EOM, we find an on-shell
basis that consists of 11 operators only. They are conveniently chosen as
$\{Q_1,Q_2,Q_5,Q_6,Q_8,Q_9,Q_{10},Q_{11},Q_{17},Q_{18},Q_{19}\}$. There
is a subtlety for $Q_2$ whose $W$-coefficient has more symmetries in the
on-shell basis, namely $W^{(2)}_{abcd} = W^{(2)}_{cdab}$ and
$W^{(2)}_{(abcd)} = 0$, apart from just $W^{(2)}_{abcd} =
W^{(2)}_{(ab)(cd)}$ in the off-shell case.

\subsubsection{Sample Off-Shell Results}\label{sample.off}
As a sample off-shell result, let us quote the RG equation we have obtained
for $W^{(1)}$ in the Feynman-'t Hooft gauge. Terms that are due to the presence of fermions $\psi$ are going to be denoted by $(\ldots)_\psi$ in what follows. The RG equation
reads
\begin{align}\label{W1RGE}
\mu \frac{d W^{(1)}_{abcdef}}{d\mu} &= \frac{1}{16\pi^2} \left( 2 X^{(1)} + X^{(2)} + X^{(3)} - 6 X^{(4)} + 2 X^{(5)} + 2 X^{(6)} + 2 X^{(7)}\right. \nonumber\\
&\quad \left. -\, 12 X^{(8)} + 6 X^{(9)} + (\ldots)_\psi \right)_{abcdef}
\end{align}
where
\begin{align}
X^{(1)}_{abcdef} &= \frac{1}{48} \sum \theta^A_{ag} \theta^A_{bh} W^{(1)}_{cdefgh},&
X^{(2)}_{abcdef} &= \frac{2\pi^2}{15} \sum (\gamma_\phi)_{ag} W^{(1)}_{bcdefg},\nonumber\\
X^{(3)}_{abcdef} &= \frac{1}{48} \sum \lambda_{abgh} W^{(1)}_{cdefgh},&
X^{(4)}_{abcdef} &= \frac{1}{4} \sum \theta^A_{ag} \theta^A_{bh} \theta^B_{cg} \theta^B_{di} W^{(2)}_{hief},\nonumber\\
X^{(5)}_{abcdef} &= \frac{1}{16} \sum \lambda_{adhi} \lambda_{bcgi} W^{(2)}_{ghef},&
X^{(6)}_{abcdef} &= \frac{1}{8} \sum \theta^A_{ei} \theta^A_{fj} \lambda_{adhi} \lambda_{bcgj} W^{(3)}_{gh},\nonumber\\
X^{(7)}_{abcdef} &= \frac{1}{16} \sum \lambda_{aeij} \lambda_{bfhj} \lambda_{cdgi} W^{(3)}_{gh},&
X^{(8)}_{abcdef} &= \frac{1}{4} \sum \theta^A_{cg} \theta^A_{dh} \theta^B_{bh} \theta^C_{ag} W^{(5)BC}_{ef},\nonumber\\
X^{(9)}_{abcdef} &= \frac{1}{2} \sum \theta^A_{fi} \theta^B_{eh} \theta^C_{cg} \theta^C_{di} \theta^D_{ag} \theta^D_{bh} W^{(7)AB}.
\end{align}
The sums go over such permutations of uncontracted indices that make
each $X^{(N)}_{abcdef}$ totally symmetric. The scalar field anomalous
dimensions in $X^{(2)}$ are given by
\begin{equation}
(\gamma_\phi)_{ab} = \frac{1}{32\pi^2} \left[ Y^a_{ij} Y^{b*}_{ij} + Y^b_{ij} Y^{a*}_{ij}
                     -4 \theta^A_{ac} \theta^A_{cb} \right].
\end{equation}

\subsubsection{Automatic Computations}\label{autom}

Our calculation begins with generating the Feynman rules from the
Lagrangian~\eqref{Leff} with the help of {\tt
FeynRules}~\cite{Alloul:2013bka}. Next, {\tt
FeynArts}~\cite{Hahn:2000kx} is used to construct expressions for all
the necessary one-loop diagrams. Calculation of their divergent parts
is very simple, most efficiently achieved with the help of a
self-written code. Simplification of the evaluated results requires
applying various identities that stem from gauge invariance and/or EOM
(see the next section). For this purpose, the code {\tt
xTensor}~\cite{xTensor:2016xx} is very helpful, as it allows us to
impose all the relevant symmetries of the considered tensors in a
straightforward manner. However, full automation of the necessary
simplifications has not yet been achieved, which is the main reason why
our project is still quite far from getting completed. New ideas are
currently being tested.

\subsubsection{Simplification Methods}\label{simpl}

Gauge invariance of the theory imposes some identities on the couplings and $W$-coefficients.
To derive such an identity for the Yukawa couplings, one considers an infinitesimal gauge transformation
\begin{align}
Y^a_{jk}\phi_a(\psi_j)^T CP_L\psi_k \rightarrow Y^a_{jk}(\delta_{ab}-i\epsilon^A \theta^A_{ab})\phi_b \left[(\delta_{jl}-i\epsilon^B  t^B_{jl})\psi_l\right]^T CP_L(\delta_{kn}-i\epsilon^C t^C_{kn})\psi_n.
\end{align}
Since the Yukawa term is gauge invariant,
\begin{equation}
\phi_a(\psi_j)^T CP_L\psi_k \epsilon^A [ -\theta^A_{ab}Y^b_{jk}+(t^A)^T_{jl} Y^a_{lk}+Y^a_{jl}t^A_{lk} ] = 0,
\end{equation}
it follows that
\begin{equation}
(t^A)^T_{jl} Y^a_{lk}+Y^a_{jl}t^A_{lk}-\theta^A_{ab}Y^b_{jk}=0.
\end{equation}
A generic, purely fermionic operator can be written as
\begin{align}
W^{(n)}_{j_1j_2...k_1k_2...l_1l_2...}\psi_{k_1}^T\omega C\psi_{k_2}\ldots \overline{\psi}_{l_1}\omega C\overline{\psi}_{l_2}^T\ldots \overline{\psi}_{j_1}\gamma\psi_{j_2}\ldots,
\end{align}
where $\omega$ that contracts spinor indices is either the identity or
the $\sigma_{\mu\nu}$ matrix. Some of the spinor fields may be
replaced by their covariant derivatives of arbitrary degree.  For such
operators, the quantity that must vanish due to gauge invariance reads
\begin{align}
&t^{E}_{m k_1} W^{(N)}_{j_1 j_2...m k_2...l_1l_2...}\,+\,t^{E}_{m k_2 }W^{(N)}_{j_1 j_2...k_1m...l_1l_2...}-t^{E*}_{m l_1} W^{(N)}_{j_1 j_2...k_1 k_2...ml_2...}\,-\,t^{E*}_{m l_2 }W^{(N)}_{j_1 j_2...k_1k_2...l_1m...}\nonumber\\
&\quad-t^{E*}_{m j_1} W^{(N)}_{m j_2...k_1k_2...l_1l_2...}\,+\,t^{E}_{m j_2 }W^{(N)}_{j_1 m...k_1k_2...l_1l_2...}+\,\ldots\,.
\end{align}
Analogously, for the $W$-coefficients of operators with bosonic fields only, the quantity that must vanish reads
\begin{align}
if^{B E A_1} W^{(N) B A_2...A_k}_{a_1...a_m}\,+\,\ldots\,+\,if^{B E A_k}W^{(N) A_1...B}_{a_1...a_m}
+\,\theta^{E}_{b a_1} W^{(N) A_1...A_k}_{b a_2...a_m}\,+\,\ldots\,+\,\theta^{E}_{b a_m}W^{(N) A_1...A_k}_{a_1...b}.
\end{align}
Both types of terms arise on the r.h.s.\ for operators that involve both the fermionic and bosonic fields.

Once the RG equations in the off-shell basis are found, we should pass to the
on-shell ones by using the EOM. Let us illustrate this using the
$W$-coefficient of $Q_5$. We start from the observation that $Q_7$ is
reducible by the gauge-field EOM
\begin{equation}
(D_{\mu}F^{\mu \nu})^A= - i \theta^A_{ab}\phi_b (D^{\nu}\phi)_a+ (\ldots)_\psi+ \mathcal{O}(\frac{1}{\Lambda}).
\end{equation}
An operator $\widetilde{Q_7}$ that vanishes on-shell is obtained by a simple redefinition
\begin{equation}
\widetilde{Q}_{7} := Q_7+\frac{1}{2} Q_4^{\prime} +\frac{1}{4} Q_5^{\prime} + (\ldots)_\psi,
\end{equation}
with
\begin{align}
Q_4^\prime    &:= i W^{(7)\;AC}\theta^{C}_{ab}(D^{\mu}\phi)_a(D^{\nu}\phi)_b F^A_{\mu \nu},&
Q_5^{\prime}  &:= \frac{1}{4}  \left( \sum  W^{(7)\;AC} \theta^{C}_{ac}\theta^{B}_{bc} \right) \phi_a \phi_b F^A_{\mu \nu} F^{B\;\mu\nu}.
\nonumber\\
\end{align}
Next, $Q_4^\prime$ and $Q_5^\prime$ are absorbed into $Q_4$ and $Q_5$:
\begin{align}
\overline{W}^{(4)\;}{}^{A}_{ab}  &:= W^{(4)\;}{}^{A}_{ab}-i W^{(7)\;AC}\theta^{C}{}_{ab},&
\overline{W}^{(5)\;}{}^{AB}_{ab} &:= W^{(5)\;}{}^{AB}_{ab}- \frac{1}{4}  W^{(7)\;AC} \theta^{C}_{ac}\theta^{B}_{bc}\,.
\end{align}
To get an on-shell expression for the $W$-coefficient of $Q_5$, another redefinition is necessary:
\begin{equation}
\widetilde{Q_{4}} := Q_4+\tfrac{1}{4} Q^{\prime\prime}_5+(\ldots),
\end{equation}
with
\begin{equation}
Q^{\prime\prime}_5:=\tfrac{i}{4}\sum\overline{W}^{(4)}_{ac}{}^A\theta^B_{cb}\phi_a \phi_b F^A_{\mu \nu} F^{B\;\mu\nu}.
\end{equation}
It yields
\begin{align} 
\widetilde{W}^{(5)\;}{}^{AB}_{ab} &= \overline{W}^{(5)\;}{}^{AB}_{ab}+\frac{i}{4} \sum \overline{W}^{(4)\;}{}^{A}_{ac}\theta^{B}_{bc}= W^{(5)\;}{}^{AB}_{ab}+\frac{i}{4} \sum W^{(4)\;}{}^{A}_{ac}\theta^{B}_{bc}\,.
\end{align}
Finally, applying $\mu\frac{d}{d\mu}$ to both sides of the above equation, one obtains the on-shell RG equation for $\widetilde{W}^{(5)}$:
\begin{eqnarray}
\mu\frac{d\widetilde{W}^{(5)\;}{}^{AB}_{ab}}{d\mu} &=& \mu\frac{dW^{(5)\;}{}^{AB}_{ab}}{d\mu}
+ \frac{i}{4} \sum\left( \mu\frac{d W^{(4)\;}{}^{A}_{ac}}{d\mu}\theta^{B}_{bc} + W^{(4)\;}{}^{A}_{ac} \theta^{\underline{B}}_{bc} \gamma_{\underline{B}}  \right),
\end{eqnarray}
where
\begin{equation}
\gamma_{B}=\frac{1}{48\pi^2} \left[ -11 C_2(G_B)
+\frac{1}{2}\text{tr}(\theta^A_{\underline{B}}\theta^A_{\underline{B}})
+2\text{tr}(t^A_{\underline{B}} t^A_{\underline{B}}) \right]
\end{equation}
and~ $C_2(G_{\underline{B}}) \delta^{\underline{B}C} = f^{BDE} f^{CDE}$.

\subsubsection{Sample On-Shell Results}\label{sample.on}

Three out of six on-shell-irreducible bosonic operators, namely $Q_6$,
$Q_8$ and $Q_9$, transform trivially to the on-shell basis.  The
corresponding RG equations that we find in their case take the form
\begin{align}\label{rge.on}
\mu\frac{dW^{(6)}{}^{AB}_{ab}}{d\mu} &= \frac{1}{16\pi^2}
\left( 2 Z^{(1)}\!+\!2 Z^{(2)}\!+\!8 Z^{(3)}\!-\!8 Z^{(4)} + \!2 Z^{(5)}\!+\!Z^{(6)}\!+\!Z^{(7)}\!+\!2 Z^{(8)}\!+\!6 Z^{(9)}\right)^{AB}_{ab},\nonumber\\
\mu \frac{dW^{(8)}{}^{ABC}}{d\mu} &= \frac{1}{16\pi^2}
[ 12\,C_2(G_{\underline{B}})+48\pi^2\,\gamma_{\underline{B}} ]\,W^{(8)}{}^{A\underline{B}C},\nonumber\\
\mu \frac{dW^{(9)}{}^{ABC}}{d\mu} &= \frac{1}{16\pi^2}
[ 12\,C_2(G_{\underline{B}})+48\pi^2\,\gamma_{\underline{B}}]\,W^{(9)}{}^{A\underline{B}C},
\end{align}
where
\begin{align}
Z^{(1)AB}_{ab} &= W^{(6)}{}^{AB}_{cd}\, \theta ^{C}{}_{ac}\, \theta ^{C}{}_{bd}\,,&
Z^{(2)AB}_{ab} &= \sum W^{(6)}{}^{BC}_{bd}\, \theta ^{A}{}_{cd} \,\theta ^{C}{}_{ac}\,,\nonumber\\
Z^{(3)AB}_{ab} &= C_2(G_{\underline{B}}) W^{(6)}{}^{A\underline{B}}_{ab}\,,&
Z^{(4)AB}_{ab} &= f^{ACE} f^{BDE} W^{(6)}{}^{CD}_{ab}\,,\nonumber\\
Z^{(5)AB}_{ab} &= 16\pi^2 W^{(6)}{}{}^{A\underline{B}}_{ab}\, \gamma_{\underline{B}}\,,&
Z^{(6)AB}_{ab} &= 8\pi^2\sum W^{(6)}{}^{AB}_{bc} (\gamma_{\phi}){} _{ac}\,,\nonumber\\
Z^{(7)AB}_{ab} &= W^{(6)}{}^{AB}_{cd} \lambda _{abcd}\,,&
Z^{(8)AB}_{ab} &= i\sum W^{(9)}{}^{BCD} \,\theta ^{A}{}_{ac}\, \theta ^{C}{}_{bd} \,\theta ^{D}{}_{cd}\,,\nonumber\\
Z^{(9)AB}_{ab} &= \tfrac{i}{2}\sum W^{(9)}{}^{BCD} \,\theta ^{A}{}_{cd} \,\theta ^{C}{}_{ac} \,\theta ^{D}{}_{bd}\,.
\end{align}
The RG equations for $Q_8$ and $Q_9$ in Eq.~\eqref{rge.on} agree with those in
Refs.~\cite{Braaten:1990gq,Braaten:1990zt}. As far as $Q_6$ in the
generic case is concerned, we are not aware of any published one-loop
RG equation so far. However, we have checked that in the SMEFT case we reproduce
the RG equations found in
Refs.~\cite{Jenkins:2013zja,Jenkins:2013wua,Alonso:2013hga}. Such a
comparison tests the sum $2Z^{(5)}+Z^{(6)}+Z^{(7)}$ in Eq.~\eqref{rge.on}.

\subsubsection{Summary}\label{summ}

Our goal is to evaluate one-loop RG equations for dimension-six operators in a
generic class of EFTs. In the absence of fermions, the calculation has
been completed~\cite{Nalecz:2021xx} in the off-shell basis, with
partial reduction to the on-shell one. As far as the operators with
fermions are concerned, only partial off-shell results have been
obtained so far~\cite{Mieszkalski:2021xx}. The main issue that remains
to be resolved is automatization of tensor expression simplifications
that must be performed after evaluation of the necessary Feynman
diagrams. Eventually, once our project is completed, one-loop RG equations for
many practically relevant specific EFTs will be possible to determine
via straightforward substitutions.

%%%%%%%%%%%%%%%%%%%%%%%%%%%%%%%%%%%%%%%%%%%%%%%%%%%%%%%%%%%%%%%%%%%%%%%%%%%%%%%%%%%%%%%%%%%%%%%%
\contribution{Two-loop Renormalization for \texorpdfstring{$\chi$QED}{chiral-QED} in the BMHV Scheme
}{\hspace{75pt} Herm\`{e}s B\'{e}lusca-Ma\"{i}to, Amon Ilakovac,\newline\phantom{a}\hspace{75pt} Marija Ma\dj{}or-Bo\v{z}inovi\'{c}, Paul K\"{u}hler,\newline Dominik St\"{o}ckinger, and Matthias Wei\ss{}wange}{}
%%%%%%%%%%%%%%%%%%%%%%%%%%%%%%%%%%%%%%%%%%%%%%%%%%%%%%%%%%%%%%%%%%%%%%%%%%%%%%%%%%%%%%%%%%%%%%%%

Dimensional regularization (DReg) is an indispensable tool for practical calculations at the (multi-)loop level. Its popularity is not least of all due to manifest preservation of symmetries of vector-like theories of the classical action to all loop orders, which aids greatly both in renormalizability proofs, as well as in the practical determination of counterterms \cite{Jegerlehner:2000dz}. Many powerful theorems such as the quantum action principle can be rigorously derived in this framework.
However, it is known from experiment that the world is described by chiral gauge theories such as the electroweak sector of the SM. For such theories no invariant regulator is known, and dimensional schemes like DReg clash with their chiral nature. Technically,
this is reflected in the definition of $\gamma_5$ (or, equivalently, $\varepsilon^{\mu\nu\rho\sigma}$) in DReg, where inconsistencies arise from relying on the simultaneous validity of customary $4$-dimensional relations in the dimensionally regularized setting.

Therefore, one cannot literally apply the familiar relations involving $\gamma_5$ but must define an appropriate scheme.
In the following we summarize the main results presented in Refs.~\cite{Belusca-Maito:2020ala,Belusca-Maito:2021lnk,Belusca-Maito:2023wah}.
We adopt the Breitenlohner--Maison--'t Hooft--Veltman (BMHV)~\cite{tHooft:1972tcz,Breitenlohner:1977hr} scheme, which treats Lorentz covariants as being comprised of a $4$-dimensional (barred) and $-2\epsilon$-dimensional evanescent (hatted) part,
\begin{equation}
    g^{\mu\nu}=\bar{g}^{\mu\nu}+\hat{g}^{\mu\nu}.
\end{equation}
Inconsistencies are avoided by giving up the anti-commutativity of $\gamma_5$, 
\begin{align}
    \{\gamma_5,\gamma^{\mu}\} = 2\gamma_5\hat{\gamma}^{\mu}
    \, , &&
    \{\gamma_5,\bar{\gamma}^{\mu}\} = 0
    \, , &&
    [\gamma_5,\hat{\gamma}^{\mu}] = 0
    \, .
\end{align}
Its distinguishing feature is its consistency, and hence, reliability at the multi-loop level, but it introduces spurious symmetry breakings. There are a number of alternative schemes (cf. \cite{Jegerlehner:2000dz}, also references in \cite{Belusca-Maito:2020ala,Belusca-Maito:2021lnk}) like the naive scheme \cite{Chanowitz:1979zu} or reading-point prescriptions \cite{Kreimer:1993bh,Korner:1991sx}, which are computationally simpler, but their consistency at higher loop orders is generally not ensured. 

At the level of the full quantum theory, expressed in terms of
the 1-particle-irreducible (1-PI) quantum effective action $\Gamma$,
%% (generating functional of the 1-PI Green's functions),
the Becchi--Rouet--Stora--Tyutin (BRST) symmetry is formulated by the Slavnov--Taylor identity,
\begin{equation}
\label{eq:FullSTI}
    \mathcal{S}(\Gamma) = \int \dInt[4]{x} \frac{\delta \Gamma}{\delta \phi_i(x)}\frac{\delta \Gamma}{\delta K_{\phi_i}(x)}=0 \, ,
\end{equation}
with generic quantum fields $\phi$ and corresponding BRST sources $K_\phi$.
Its validity to all orders is an essential ingredient in ensuring unitarity and physicality of the $S$-matrix. Hence we require that our full quantum theory obeys the Slavnov--Taylor identity.
It turns out that any violation of BRST symmetry, and hence \cref{eq:FullSTI}, can be related to the insertion of a local operator into the effective action by the regularized quantum action principle \cite{Lam:1972mb,Lowenstein:1971jk,Breitenlohner:1977hr},
\begin{equation}
\label{eq:QAP}
    \mathcal{S}(\Gamma_\text{DRen}) = \Delta\cdot\Gamma_\text{DRen} \, .
\end{equation}
Evaluating the r.h.s. of \cref{eq:QAP} determines the finite symmetry restoring counterterms without the need to compute products of Green's functions including higher-order terms from the l.h.s. of \cref{eq:FullSTI}.

We, therefore, aim for the systematic determination of the full counterterm structure---comprised of non-symmetric, singular counterterms needed for consistency at higher orders, and finite, symmetry-restoring counterterms---for various toy models of increasing complexity and loop orders and, eventually, the SM.
So far we have studied the scheme at the one-loop level for a generic Yang-Mills theory \cite{Belusca-Maito:2020ala} (see also \cite{Cornella:2022hkc,Martin:1999cc} for related works) and at the two-loop level for an abelian model \cite{Belusca-Maito:2021lnk}. The latter will serve to illustrate our methods in this article. For an extensive review on the topic of chiral gauge theories and renormalization see \cite{Belusca-Maito:2023wah}.

\subsubsection{Application to \ChiQED/}

We consider an Abelian gauge theory with a family of $N_f$ right-handed fermions \cite{Belusca-Maito:2021lnk}, which we denote as \ChiQED/. The $d$-dimensional treatment only affects the fermionic sector non-trivially, where the kinetic term is kept $d$-dimensional while the interaction term is purely $4$-dimensional:
\begin{equation}
\label{eq:FermionicPart}
    \int \dInt[d]{x} \left( \imath \overline{\psi}_i \overline{\slashed{\partial}} \psi_i + \imath \overline{\psi}_i \widehat{\slashed{\partial}} \psi_i + e {\mathcal{Y}_R}_{ij} \overline{\psi}_i \Proj{L} \slashed{A} \Proj{R} \psi_j \right)\equiv
    \overline{S_{\overline{\psi} \psi}} + \widehat{S_{\overline{\psi} \psi}} + \overline{S_{\overline{\psi} A \psi_R}}
    \,,
\end{equation}
where $\mathcal{Y}_{Rij}=(\mathrm{diag}(\mathcal{Y}_R^{1},\dots,\mathcal{Y}_R^{N_f}))_{ij}$ is the hypercharge matrix for the abelian model.
The full regularized action at tree-level becomes
\begin{equation}
\label{eq:S0_dD_ChiQED}
\begin{split}
\hspace*{-1pt}
    S_0 &= \eqref{eq:FermionicPart} +
    \int \dInt[d]{x} \left(
        - \frac{1}{4} F_{\mu\nu} F^{\mu\nu}
        - \frac{1}{2 \xi} (\partial_\mu A^\mu)^2
        - \bar{c}\partial^2 c
        + K_\phi s_d\phi
        \right)
    \\
    &\equiv
    \eqref{eq:FermionicPart} + S_{AA} + S_\text{g-fix} + S_{\bar{c}c} + S_{\rho c} + S_{\bar{R} c \psi} + S_{\overline{\psi} c R}
    \,,
\end{split}
\end{equation}with fields $\phi\in\{A^{\mu},\overline{\psi}^i,\psi^i\}$ and sources $K_\phi\in\{\rho^{\mu},R^i,\overline{R}^i\}$.
We can see that the symmetry is violated for the regularized action at tree level,
giving rise to the following breaking vertex:
\vspace*{-5pt}
\begin{align}
    %% \widetilde{\mathcal{S}_d(S_0)} =
    \mathcal{S}_d(S_0) = \int \dInt[d]{x} \widehat{\Delta}(x)
    \leadsto
    &
    \raisebox{-35pt}{\includegraphics[scale=0.5]{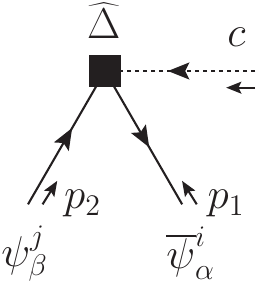}}
    &= e \, {\mathcal{Y}_R}_{ij} \left(\widehat{\slashed{p_1}} \Proj{R} + \widehat{\slashed{p_2}} \Proj{L} \right)_{\alpha\beta}
    \, .
\end{align}

The standard UV-renormalization of the model at one loop order leads to the singular counterterm action,
\begin{equation}
\label{eq:SingularCT1Loop}
    S_\text{sct}^{(1)} =
        \bigl(\textcolor{black}{\text{symmetric}}\bigr)
        - \frac{\hbar \, e^2}{16 \pi^2 \epsilon} \frac{\Tr[\mathcal{Y}_R^2]}{3} \int \dInt[d]{x} \frac{1}{2} \bar{A}_\mu \widehat{\partial}^2 \bar{A}^\mu
    \, ,
\end{equation}
where the unspecified terms correspond to $4$-dimensional multiplicative renormalization. The last term, evanescent and non--gauge invariant, is necessary for canceling the divergence in \eqref{fig:1LoopDeltaDiag}.
At one-loop order the symmetry restoration is rather straightforward, with \cref{eq:QAP} boiling down to%
\footnote{By $\Gamma^{(n)}_{\text{subren}}$ we refer to the $n$-loop effective action including counterterms up to order $n-1$. In the case of $n=1$, this corresponds to the unrenormalized and dimensionally regularized action.}
\begin{subequations}
\begin{equation}
    \mathcal{S}_d(\Gamma_\text{subren})^{(1)} =
    \widehat{\Delta}\cdot\Gamma^{(1)}_\text{subren} \, ,
    \qquad\text{cf. \eqref{fig:1LoopDeltaDiag}} \, .
\end{equation}
\vspace*{-10pt}
\begin{align}
\label{fig:1LoopDeltaDiag}
    \includegraphics[scale=0.45]{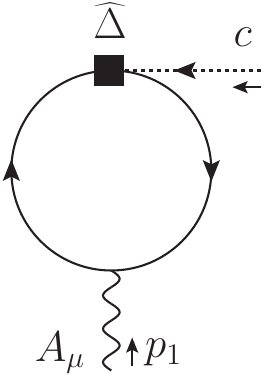}
    %&&
    %\raisebox{+5pt}{\includegraphics[scale=0.45]{Figures/BIMSp1_EvsctDeGG.pdf}}
    &&
    \raisebox{-10pt}{\includegraphics[scale=0.45]{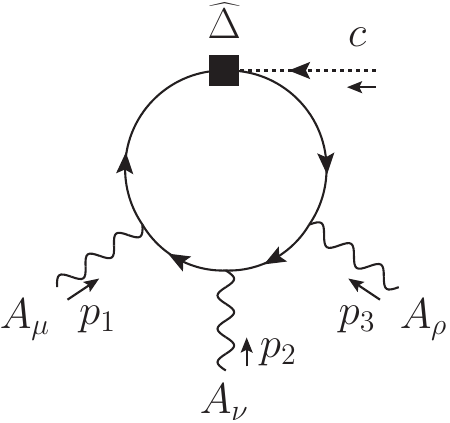}}
    &&
    \includegraphics[scale=0.45]{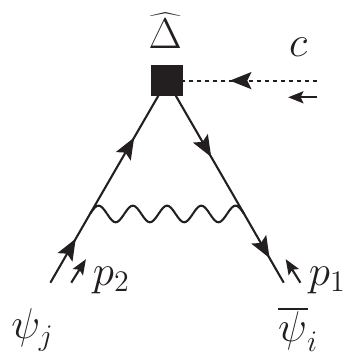}
\end{align}
\end{subequations}
These are the only%
\footnote{The triple photon triangle variant ($\widehat{\Delta}\cdot\Gamma)_{cAA}$ must vanish by the anomaly cancellation requirement $\mathrm{Tr}[\mathcal{Y}^3]=0$.}
(divergent) contributing diagrams.
The finite part of \eqref{fig:1LoopDeltaDiag} leads to the finite, non-invariant BRST-restoring counterterm action $S^{(1)}_\text{fct}$, whose structure \cite{Belusca-Maito:2020ala} is the same at two loops and will be highlighted below.

At order $\hbar^{>1}$ (using $ \hbar $ as the loop-counting parameter), \cref{eq:QAP} implies
\begin{equation}
    \mathcal{S}_d(S_0+S_\text{ct}) = (\widehat{\Delta}+\Delta_\text{ct})\cdot\Gamma_\text{DRen} \, ,
\end{equation}
and more explicitly at two-loop order:
\begin{subequations}
\begin{align}
\label{eq:CheckDeltasct2L}
    \big(\widehat{\Delta}\cdot\Gamma_\text{subren}^{(2)} + \Delta_\text{sct}^{(1)}\cdot\Gamma_\text{subren}^{(1)}+ \Delta_\text{fct}^{(1)}\cdot\Gamma_\text{subren}^{(1)}
    + \Delta_\text{sct}^{(2)}\big)_\text{div} &= 0 \, ,
    \\
\label{eqs:ThirdEq2LConditions}
    \mathop{\text{LIM}}_{d \to 4} \big(\widehat{\Delta}\cdot\Gamma_\text{subren}^{(2)} + \Delta_\text{sct}^{(1)}\cdot\Gamma_\text{subren}^{(1)}+\Delta_\text{fct}^{(1)}\cdot\Gamma_\text{subren}^{(1)}
    + \Delta_\text{fct}^{(2)}\big)_\text{fin} &= 0 \,.
\end{align}
\end{subequations}
The singular counterterms have the same structure as at one-loop (including the evanescent $\int \dInt[d]{x} \frac{1}{2} \bar{A}_\mu \widehat{\partial}^2 \bar{A}^\mu$), except for a novel non-gauge invariant, $4$-dimensional piece,
\begin{equation}
\label{eq:SingularCT2Hbar2Loop} %% SingularCT2Loop
    S_\text{sct}^{(2,\,2)} \supset
    -\left(\frac{\hbar \, e^2}{16 \pi^2}\right)^2 \frac{1}{3 \epsilon} \sum_j (\mathcal{Y}_R^j)^2
    \left( \frac{5}{2} (\mathcal{Y}_R^j)^2 - \frac{2}{3} \Tr[\mathcal{Y}_R^2] \right) \overline{S^{j}_{\overline{\psi}\psi_R}}
    \, .
\end{equation}
There exist additional divergent one-loop diagrams containing one insertion of a finite counterterm $\overline{S_\text{fct}^{(1)}}$ (similar to diagrams \cref{fig:Evsct_SAA_insertions,fig:Evsct_SPsiPsi_insertions}), whose divergent parts define new counterterms, $S_\text{sct}^{(2,\,1)}$:
\begin{equation}
\label{eq:SingularCT2Hbar1Loop} %% See also eq:Sfct1_Insert
    S_\text{sct}^{(2,\,1)} =
    -\left( \overline{S_\text{fct}^{(1)}} \cdot \Gamma^{(1)} \right)^\text{div}
    \, .
\end{equation}
They possess a genuine one-loop structure, even though they are of order $\hbar^2$.

\begin{figure}
    \begin{align*}
        \includegraphics[scale=0.45]{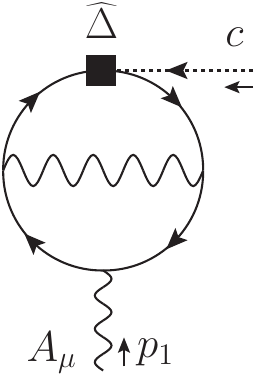}
        &&
        \includegraphics[scale=0.45]{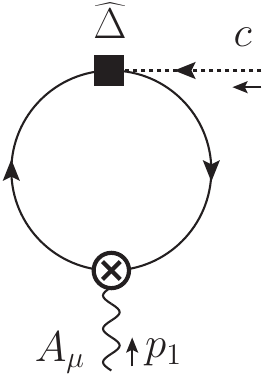}
        &&
        \includegraphics[scale=0.45]{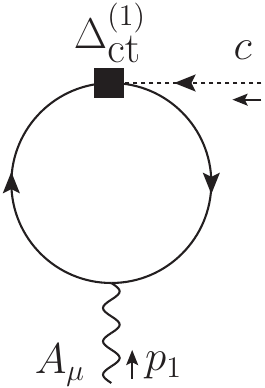}
        \\
        \includegraphics[scale=0.45]{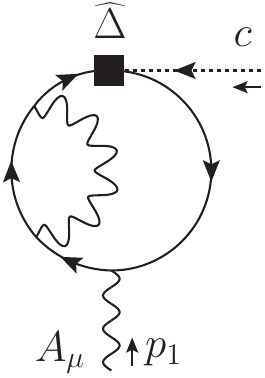}
        &&
        \includegraphics[scale=0.45]{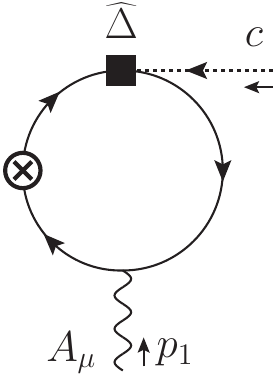}
        &&
        \includegraphics[scale=0.45]{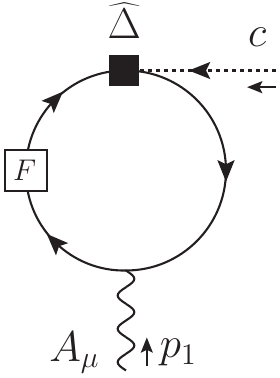}\nonumber
    \end{align*}
    \caption{The two-loop diagrams with one insertion of the tree-level $\widehat{\Delta}$-vertex and the relevant counterterms for the subdivergences. 
    \label{fig:TwoLoopcA}
    }
\end{figure}

For the r.h.s. of \cref{eq:QAP} three structures arise: the proper two-loop diagrams with one insertion of the tree-level $\widehat{\Delta}$-vertex (first diagram in \cref{fig:TwoLoopcA}),
a new insertion of BRST-transformed non-invariant one-loop counterterms into one-loop diagrams (last diagram in the first line of \cref{fig:TwoLoopcA}), and the last term in \cref{eq:CheckDeltasct2L,eqs:ThirdEq2LConditions} that determines the finite counterterms and provides a consistency check for the divergent ones, respectively.
The complete order-$\hbar^2$ finite counterterms are, in \emph{Feynman gauge} $\xi = 1$,
\begin{multline}
\label{eq:Sfct2L}
    %% S_\text{fct}^{(2)} \equiv
    \overline{S_\text{fct}^{(2)}} =
    \left(\frac{\hbar \, e^2}{16\pi^2}\right)^2
    \left\{
    \int \dInt[4]{x} \left(
        \frac{11 \Tr[\mathcal{Y}_R^4]}{24} \frac{1}{2} \bar{A}_\mu \overline{\partial}^2 \bar{A}^\mu
        + \frac{3 e^2 \Tr[\mathcal{Y}_R^6]}{2} \frac{1}{4} (\bar{A}^2)^2
    \right)
    \right.
    \\
    \left.
    - \sum_j (\mathcal{Y}_R^j)^2 \,
    \left( \frac{127}{36} (\mathcal{Y}_R^j)^2 - \frac{1}{27} \Tr[\mathcal{Y}_R^2] \right)
    \overline{S^{j}_{\overline{\psi}\psi_R}}
    \right\}
    %% \\
    %% + \text{any BRST-symmetric term}
    \, .
\end{multline}
Remarkably, they have the same compact structure as at one-loop and directly correspond to the restoration of three well-known QED Ward identities, i.e., transversality of the photon two- and four-point function, as well as the relation between electron self energy and electron-photon interaction. Indeed, we can explicitly check for the restoration of the symmetry by evaluating the relevant identities and confirming that only after adding those counterterms, are they satisfied in our model \cite{Belusca-Maito:2021lnk}.

\subsubsection{RG Equation in Dimensional Renormalization}

The RG equation \cite{Gell-Mann:1954yli,Iliopoulos:1974ur} describes the \emph{invariance} of bare correlation (Green's) functions under a change of the arbitrary renormalization scale: in DReg it is the \emph{``unit of mass\footnotemark''}%
\footnotetext{In off-shell schemes this is instead an arbitrary energy scale.}
$\mu$ \cite{tHooft:1973mfk,Bonneau:1980zp},
that is associated to each loop of any diagram: $\mu^\epsilon \int \dInt[d]{x}$.
%%%%
For example, the 1-PI quantum effective action $\Gamma$
depends on $\mu$ both explicitly and implicitly via the $\mu$-dependence of the field renormalizations $Z_\phi^{1/2}$ and renormalized parameters:
$\Gamma[\{\phi(\mu)\}; e(\mu), \xi(\mu), \mu]$.
Its invariance under a total $\mu$-variation is represented by the RG equation
(summation over fields $\phi \in \text{\ChiQED/}$ is implied%
\footnote{%
Here and in the following, we use the shorthand notations: $\mu \partial_\mu := \mu \partial/{\partial \mu}$, $\partial_e := \partial/{\partial e}$, etc.
}%
):
\begin{subequations}
\begin{equation}
\label{eq:Generic_RGE}
    \mu \frac{\operatorname{d}\Gamma}{\operatorname{d}\mu} = 0
    = \mu \partial_\mu \Gamma
    + \left( \beta_e e \partial_e + \beta_\xi \partial_\xi
        - %% \sum_{\phi \in \text{\ChiQED/}}
        \gamma_\phi N_\phi \right) \Gamma
    \, .
\end{equation}

In \cref{eq:Generic_RGE}, $\mu \partial_\mu$ is the \emph{RG differential operator}.
The $N_\phi$ are field-numbering (``leg-counting'') differential operators, defined
by %% \label{eq:FieldCount_ops}
$N_\phi \equiv \int \dInt[d]{x} \phi(x) {\delta}/{\delta \phi(x)}$,
for bosonic fields, ghosts, and for right-handed (and left-handed anti-) fermions $\phi := \Proj{R} \psi$, $\phi := \overline{\psi} \Proj{L}$.
The coefficient functions $\beta_{e,\xi}$ are the beta-functions for the coupling constant $e$ and the gauge parameter $\xi$, and $\gamma_\phi$ are the anomalous dimensions for the fields $\phi$, defined by
\begin{align}
    \beta_e = \frac{1}{e} \mu \frac{\operatorname{d}e}{\operatorname{d}\mu} \, , &&
    \beta_\xi = \mu \frac{\operatorname{d}\xi}{\operatorname{d}\mu} \, , &&
    \gamma_\phi = \frac{1}{2} \mu \frac{\operatorname{d}\ln{Z_\phi}}{\operatorname{d}\mu} \, .
\end{align}
\end{subequations}

\paragraph{``Modified'' multiplicative renormalization (MultRen)}
\label{subsect:MultRen}
Standard renormalization transformation \cite{tHooft:1973mfk,Machacek:1983tz,Machacek:1983fi,Machacek:1984zw}
consists in renormalizing fields multiplicatively, while couplings are usually renormalized additively:
\begin{equation}\begin{gathered}
    e \to e + \delta e
    \, , \qquad
    \xi \to Z_A \xi
    \, , \qquad
    A_\mu \to \sqrt{Z_A} A_\mu
    \, , \\
    ({\psi_R}_i , \overline{\psi_R}_i) \to \sqrt{Z_\psi} ({\psi_R}_i , \overline{\psi_R}_i)
    \, , \qquad
    ({\psi_L}_i , \overline{\psi_L}_i) \to ({\psi_L}_i , \overline{\psi_L}_i)
    \, ,
\end{gathered}\end{equation}
(BRST sources renormalize the inverse way from their corresponding dynamical fields).
Beta-functions and anomalous dimensions can be found from the $1/\epsilon$ poles of the renormalizations $\delta e$ and $Z_\phi$.

The situation becomes more involved when new evanescent singular and finite symmetry-restoring counterterms are generated during renormalization.
One way to proceed is to extend~\cite{Bos:1987fb,Schubert:1988ke} (also Section~8 in~\cite{Belusca-Maito:2020ala}) the original tree-level action $S_0$ with those new generated operators, associated with new \emph{auxiliary couplings} $\rho_\mathcal{O} := \sigma_i, \rho_i$.
A new tree-level action $S_0^*$ is thus defined,
% \begin{equation}
%     S_0^* = S_0 + \int \dInt[d]{x} \rho_\mathcal{O} \mathcal{O}(x) \, ,
% \end{equation}
which, in the case of \ChiQED/, can take the following form:%
\footnote{%
The operator $\int \dInt[d]{x} \frac{1}{2} \bar{A}_\mu \overline{\partial}^2 \bar{A}^\mu$ generated in the finite counterterms %% is not independent \emph{per se}, but
is equal to the combination $\overline{S_{AA}} + \xi \overline{S_\text{g-fix}}$ of the 4-dimensional photon kinetic and gauge-fixing terms.
The operator $\int \dInt[d]{x} {e^2}/{4} (\bar{A}^2)^2$, associated to $\rho_3$, does not actually contribute to the RG evolution at the $\hbar^2$ order.
}
\begin{equation}
\label{eq:S0_dD_ChiQED_AuxCoupls}
\begin{split}
    S_0^* =\;& S_0
        + \rho_1 \delta\text{fct}_\psi \overline{S_{\overline{\psi} \psi}}
        + \sigma_1 \widehat{S_{\overline{\psi} \psi}}
        + \rho_2 \delta\text{fct}_A (\overline{S_{AA}} + \xi \overline{S_\text{g-fix}})
        \\
       &+ \sigma_2 \widehat{S_{AA}}
        + \int \dInt[d]{x} \left( \sigma_3 \frac{1}{2} \bar{A}_\mu \widehat{\partial}^2 \bar{A}^\mu + \rho_3 \frac{e^2}{4} (\bar{A}^2)^2 \right)
    \, .
\end{split}
\end{equation}
The coefficients $\delta\text{fct}_\psi$ and $\delta\text{fct}_A$ arise from the finite BRST-restoring counterterms $S_\text{fct}$.
The generated modified effective action $\Gamma^*_\text{DReg}[\phi, \rho_\mathcal{O}]$ and counterterms can be expanded in $\rho_\mathcal{O}$, whose lowest-order terms correspond to the quantities evaluated in the original theory.

One obtains an RG equation for $\Gamma^*_\text{DReg}[e, \xi, \{\sigma_i\}, \{\rho_i\}]$, with beta-functions $\widetilde{\beta}$ for $e, \xi$ and auxiliary couplings $\sigma_i, \rho_i$, and anomalous dimensions $\widetilde{\gamma_\phi}$:
\begin{equation}
\label{eq:RGE_MultRen}
    \mu \partial_\mu \Gamma^*_\text{DReg} =
    \left( - \widetilde{\beta_e} e \partial_e - \widetilde{\beta_\xi} \partial_\xi - \widetilde{\beta_{\sigma_i}} \partial_{\sigma_i} - \widetilde{\beta_{\rho_i}} \partial_{\rho_i} + \widetilde{\gamma_\phi} N_\phi \right) \Gamma^*_\text{DReg}
    \, .
\end{equation}
The genuine renormalized theory generated by the original $S_0$ can be recovered in the limit $\sigma_i, \rho_i \to 0$, since $\sigma_i, \rho_i$ are unphysical and are absent in $S_0$.
The true $\beta$ and $\gamma$ functions for $\Gamma$ will depend on $\widetilde{\beta}$ and $\widetilde{\gamma_\phi}$ and are obtained for the \emph{4-dimensional renormalized} effective action $\Gamma$, defined by
\begin{equation}
    \Gamma[e, \xi] = \mathop{\text{LIM}}_{d \to 4} \lim_{\sigma_i,\rho_i \to 0} \Gamma^*_\text{DReg}[e, \xi, \{\sigma_i\}, \{\rho_i\}]
    \, ,
\end{equation}
where: {\it (i)} divergences are MS-subtracted from $\Gamma$ with suitable singular counterterms, and {\it (ii)} $d \to 4$, with {\it (iii)} remaining finite evanescent quantities set to zero.
%%%%
The corresponding RG equation, obtained from \cref{eq:RGE_MultRen} when both sides are taken under those same limits, has the final structure:
\begin{equation}
\label{eq:RGE_No_evsct}
    \mu \partial_\mu \Gamma
    =
    \left( - \beta_e e \partial_e - \beta_\xi \partial_\xi + \gamma_\phi N_\phi \right) \Gamma
    \quad \sim \quad
    \mathop{\text{LIM}}_{d \to 4} \lim_{\sigma_i,\rho_i \to 0} \mu \partial_\mu \Gamma^*_\text{DReg}
    \, .
\end{equation}

The procedure \cite{Bos:1987fb,Schubert:1988ke} then consists in evaluating the effects of the evanescent and non-symmetric operators, that dilute into the non-evanescent ones, via the following terms in the limit $\sigma_i, \rho_i \to 0$ and the renormalized limit $d \to 4$:
\begin{align}
\label{eq:MultRen_AuxOpsEffects}
    %\left.
    - \widetilde{\beta_{\sigma_i}} \partial_{\sigma_i} \Gamma^*_\text{DReg}
    %\right|_{\sigma_i,\rho_i \to 0}
    \, ,
    &&
    %\left.
    - \widetilde{\beta_{\rho_i}} \partial_{\rho_i} \Gamma^*_\text{DReg}
    %\right|_{\sigma_i,\rho_i \to 0}
    \, .
\end{align}
They correspond, from the Regularized Action Principle \cite{Breitenlohner:1977hr,Breitenlohner:1975hg,Breitenlohner:1976te}, to diagrammatic vertex insertions of their associated operators:
${\partial \Gamma^*_\text{DReg}}/{\partial \rho_\mathcal{O}} = \left( \mathcal{O} + {\partial S_\text{ct}^*}/{\partial \rho_\mathcal{O}} \right) \cdot \Gamma_\text{DReg}$,
%%\begin{equation}
%%    \left. \frac{\partial \Gamma^*_\text{DReg}}{\partial \rho_\mathcal{O}} \right|_{\rho_\mathcal{O} \to 0}
%%    % = \left. \frac{\partial (S_0^* + S_\text{ct}^*)}{\partial \rho_\mathcal{O}} \cdot \Gamma^*_\text{DReg} \right|_{\rho_\mathcal{O} \to 0}
%%    = \left. \left( \mathcal{O} + \frac{\partial S_\text{ct}^*}{\partial \rho_\mathcal{O}} \right) \cdot \Gamma^*_\text{DReg} \right|_{\rho_\mathcal{O} \to 0}
%%    \, ,
%%\end{equation}
where ${\partial S_\text{ct}^*}/{\partial \rho_\mathcal{O}}$ removes the divergences from $\mathcal{O} \cdot \Gamma_\text{DReg}$.
Finally, they are re-cast as new contributions to $\beta_e e \partial_e \Gamma$, $\beta_\xi \xi \partial_\xi \Gamma$ and $\gamma_\phi N_\phi \Gamma$, from which shifts to the $\beta_e$ and $\gamma_\phi$ are obtained.

\paragraph{RG equation in Algebraic Renormalization (AlgRen)}
\label{subsect:AlgRen}
The other and more streamlined method for obtaining the RG equation is that of the ``Algebraic Renormalization'' framework~\cite{Piguet:1995er,Martin:1999cc}.
It is based on the properties of the theory and of the RG evolution regarding the BRST symmetry (see, e.g., Section~7 of \cite{Belusca-Maito:2020ala}).
It applies at the level of the BRST-restored \emph{4-dimensional renormalized} effective action $\Gamma$.

After symmetry restoration, $\Gamma$ is now BRST invariant, and the RG operator inherits the symmetries from $\Gamma$:
{\it (i)} the RG evolution is BRST invariant,
%%: $\mathcal{S}_\Gamma (\mu \partial_\mu \Gamma) = 0$,
{\it (ii)} it satisfies the \emph{gauge-fixing condition},
%%: ${\delta (\mu \partial_\mu \Gamma)}/{\delta B} = 0$,
{\it (iii)} and the \emph{ghost equation}.
%%: $\mathcal{G} \, \mu \partial_\mu \Gamma = 0$,
%% with $\mathcal{G} = \delta/{\delta \bar{c}} + \partial^\mu \delta/{\delta \rho^\mu}$.
%%%%
The RG equation for $\Gamma$ is thus an expansion in a basis of 4-dimensional operators, with ghost number $= 0$, satisfying these same constraints ($\phi = A, \psi, c$):
%%\begin{subequations}
\begin{equation}
\label{eq:RGE_Beta_Gamma_def}
    \underbrace{\mu \partial_\mu \Gamma}_{= \mathfrak{R}}
    =
    \underbrace{\left( -\beta_e e \partial_e + \gamma_\phi \mathcal{N}_\phi \right) \Gamma}_{= \mathfrak{W}}
    \, .
\end{equation}
%% The operator $e \partial_e$ already verifies such conditions, and
The $\mathcal{N}_\phi$ are BRST-invariant field-counting operators \cite{Belusca-Maito:2020ala,Belusca-Maito:2022wem} that are \emph{linear combinations} of the basic $N_\phi$ operators previously introduced:
\begin{align*}
    \mathcal{N}_A = (N_A + 2 \xi \partial_\xi - \cdots)
    \, , &&
    \mathcal{N}_\psi = (N_\psi^R + N_{\overline{\psi}}^L - \cdots)
    \, , &&
    \mathcal{N}_c = %% (N_c - N_\zeta) S_0 \equiv
        N_c
    \, .
\end{align*}
%%\end{subequations}

The Quantum Action Principle (QAP) \cite{Lowenstein:1971jk,Lam:1972mb,Lam:1973qa,Clark:1976ym,Breitenlohner:1977hr,Breitenlohner:1975hg,Breitenlohner:1976te,Piguet:1980nr,Piguet:1995er}, asserts that
the variations of $\Gamma$ (terms $\mathfrak{W}$) %% in \cref{eq:RGE_Beta_Gamma_def}
with respect to parameters and fields naturally present in $S_0$, are equivalent to a renormalized insertion of local \emph{$d$-dimensional} operators in $\Gamma$,
derived from the \emph{finite dimensional-regularized action}, $S_0 + S_\text{fct}$,
%% , constituted by the full $d$-dimensional tree-level action $S_0$ and the finite counterterms $S_\text{fct}$
\begin{equation}
\label{eq:diffOps_as_Insertions}
    \mathcal{D} \Gamma =
        N[\mathcal{D} (S_0 + S_\text{fct})] \cdot \Gamma
    \, ,
    \qquad\text{with}\qquad
        \mathcal{D} = e \partial_e \; ; \; \mathcal{N}_\phi
    \, .
\end{equation}
%%%%
Because $\mu$ is \emph{not} a parameter of $S_0$, but is a modification of the loop integration, the QAP does not directly apply to $\mu \partial_\mu \Gamma$ itself (term $\mathfrak{R}$). Nonetheless, it can also be expressed as a renormalized insertion (Bonneau \cite{Bonneau:1980zp}):
\begin{equation}
\label{eq:RGE_Bonneau}
    \mathfrak{R} \equiv
    \mu \partial_\mu \Gamma =
    \sum_{N_l \geq 1}
    N_l \, N[\text{r.s.p.}\, \Gamma_\text{DReg}^\text{$N_l$ loops}] \cdot \Gamma
    \, .
\end{equation}
%%%%
In this equation, $\text{r.s.p.}\, \Gamma_\text{DReg}^\text{$N_l$ loops}$ designates the \emph{residue of simple $1/(4-d)$ pole} of the \emph{$N_l$-loop} 1-PI diagrams, made from Feynman rules derived from the action $(S_0 + S_\text{fct})$,
and \emph{sub-renormalized} using lower-order singular counterterms $S_\text{sct}$.

The procedure then consists in re-expressing \cref{eq:RGE_Beta_Gamma_def} using \labelcref{eq:diffOps_as_Insertions,eq:RGE_Bonneau}, and all the operator insertions into a basis of (independent) 4-dimensional ones.
Note that the inserted evanescent $\widehat{\mathcal{M}_j}$ operators manifesting there are not linearly independent quantities, and need to be expanded into independent insertions of 4-dimensional operators $\overline{\mathcal{M}_i}$ (Bonneau identities \cite{Bonneau:1980zp,Bonneau:1979jx}):
$N[\widehat{\mathcal{M}_j}] \cdot \Gamma = {\textstyle \sum_i} c_{ji} N[\overline{\mathcal{M}_i}] \cdot \Gamma$,
with $c_{ji} \sim \mathcal{O}(\hbar)$.
%%%%
Grouping all terms together, one obtains the final form of the RG equation, as a system of equations for the $\beta_e$ and the $\gamma_\phi$ functions,
%%(with coefficients $r_i$, $w_{e,i}$ and $w_{\phi,i}$ to be determined)
ensuring their self-consistency:
\begin{equation}
\label{eq:RGE_Beta_Gamma_system}
    \mu \partial_\mu \Gamma
    =
    \underbrace{{\textstyle \sum_i} r_i N[\overline{\mathcal{M}_i}] \cdot \Gamma}_{= \mathfrak{R}}
    =
    \underbrace{{\textstyle \sum_i} \left( - \beta_e w_{e,i} + \gamma_\phi w_{\phi,i} \right) N[\overline{\mathcal{M}_i}] \cdot \Gamma}_{= \mathfrak{W}}
    \, .
\end{equation}

\subsubsection{Results for the two-loop RG evolution}

Mainly focusing on the AlgRen method, we now describe how all the $\hbar^2$-order terms entering \cref{eq:RGE_Beta_Gamma_system} can be explicitly evaluated in order to determine the two-loop $\beta_e^{(2)}$ and $\gamma_\phi^{(2)}$ functions. For all details we refer to \cite{Belusca-Maito:2022wem}.

\begin{figure}[t]
    \centering
    %% 1loop_Evsct_SAA.png
    \includegraphics[scale=0.4]{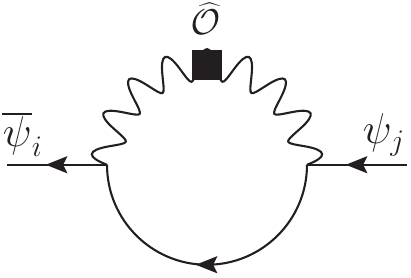} \hspace{1cm}
    \includegraphics[scale=0.4]{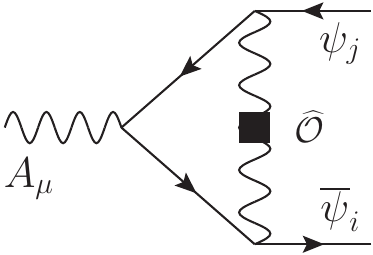}
    \caption{Diagrams with insertion of $\widehat{\mathcal{O}} \equiv \widehat{S_{AA}}$ or $-\text{r.s.p.}\, \widehat{S_\text{sct}^{(1)}} \propto \int \dInt[d]{x} \frac{1}{2} \bar{A}_\mu \widehat{\partial}^2 \bar{A}^\mu$.}
\label{fig:Evsct_SAA_insertions}
\end{figure}

\begin{figure}[t]
    \centering
    %% 1loop_Evsct_SPsiPsi.png
    \begin{tabularx}{\columnwidth}{*{3}{>{\centering\arraybackslash}X}}
        \includegraphics[scale=0.45]{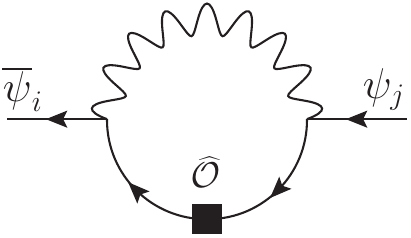}
        &
        \includegraphics[scale=0.4]{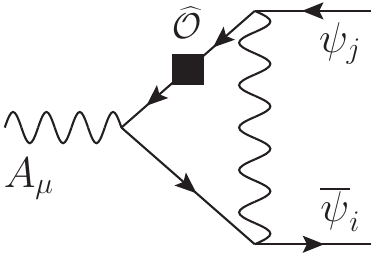} \newline + perm.
        &
        \includegraphics[scale=0.4]{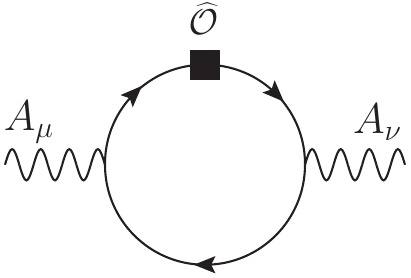}   \newline + perm.
        % \\
        % \includegraphics[scale=0.4]{Figures/EvsctAAA.pdf}  \newline + 5 perms.
        % &
        % \includegraphics[scale=0.4]{Figures/EvsctAAAA.pdf} \newline + 23 perms.
    \end{tabularx}
    \caption{\emph{Non-vanishing} diagrams with insertion of $\widehat{\mathcal{O}} \equiv \widehat{S^{i}_{\overline{\psi} \psi}}$.}
\label{fig:Evsct_SPsiPsi_insertions}
\end{figure}

\noindent
The left-hand side of \cref{eq:RGE_Beta_Gamma_system} can be grouped into four different terms,
$\mathfrak{R} = \mathfrak{R_1} + \mathfrak{R_2} + \mathfrak{R_3} + \mathfrak{R_4}$.
\begin{itemize}
    \item $\mathfrak{R_1}$ corresponds to insertions of one-loop 4-dimensional singular counterterms into one-loop diagrams,
    $\mathfrak{R_1} = N[-\text{r.s.p.}\, \overline{S_\text{sct}^{(1)}}] \cdot \Gamma^{(1)}$.
%%%%
    \item $\mathfrak{R_2}$ corresponds to insertions of one-loop evanescent singular counterterms into one-loop diagrams, \cref{fig:Evsct_SAA_insertions,fig:Evsct_SPsiPsi_insertions}:
    $\mathfrak{R_2} = N[-\text{r.s.p.}\, \widehat{S_\text{sct}^{(1)}}] \cdot \Gamma^{(1)}$.
%%%%
    \item $\mathfrak{R_3}$ corresponds to two-loop singular counterterms, obtained from the $1/(4-d)$ pole of one-loop diagrams involving insertions of one-loop finite counterterms (see discussion around \cref{eq:SingularCT2Hbar1Loop}),
    $\mathfrak{R_3} = - \text{r.s.p.}\, \overline{ S_\text{sct}^{(2,\,1)} }$.
    Those diagrams are similar to those of \cref{fig:Evsct_SAA_insertions,fig:Evsct_SPsiPsi_insertions},
    but with $\widehat{\mathcal{O}} \rightarrow \overline{S_\text{fct}^{(1)}}$.
%%%%
    \item $\mathfrak{R_4}$ corresponds to the genuine two-loop $\hbar^2$ singular counterterms (see discussion around \cref{eq:SingularCT2Hbar2Loop}). Note that in the language of \cref{eq:RGE_Bonneau}, it is only these $\mathfrak{R_4}$ terms that receive a factor $N_l = 2$, whereas all other contributions $\mathfrak{R_{1,2,3}}$ receive a factor $N_l = 1$. This subtlety is not present in the case of manifest symmetry preservation.
\end{itemize}

\noindent
Similarly, the right-hand side of \cref{eq:RGE_Beta_Gamma_system} can be grouped into four terms,
$\mathfrak{W} = \mathfrak{W_1} + \mathfrak{W_2} + \mathfrak{W_3} + \mathfrak{W_4}$.
\begin{itemize} 
    \item $\mathfrak{W_1}$ corresponds to contributions from the one-loop RG coefficients combined with the insertions of the respective differential operators, i.e.
    $\mathfrak{W_1} =
        -\beta_e^{(1)} N[e \partial_e \overline{S_0}] \cdot \Gamma^{(1)}
        + \gamma_\phi^{(1)} N[\mathcal{N}_\phi \overline{S_0}] \cdot \Gamma^{(1)}$.
    Note that there is an automatic agreement $\mathfrak{R_1} = \mathfrak{W_1}$, in accord with the one-loop RG coefficients.
%%%%
    \item $\mathfrak{W_2}$ corresponds to the contributions from one-loop RG coefficients combined with insertions of tree-level evanescent operators,
    $\mathfrak{W_2} =
        2 \gamma_A^{(1)} N[\widehat{S_{AA}}] \cdot \Gamma^{(1)}
        + %\sum_i
        \gamma_{\psi_i}^{(1)} N[\widehat{S^{i}_{\overline{\psi} \psi}}] \cdot \Gamma^{(1)}$,
    corresponding to \cref{fig:Evsct_SAA_insertions,fig:Evsct_SPsiPsi_insertions}.
%%%%
    \item $\mathfrak{W_3}$ corresponds to contributions from one-loop RG coefficients combined with finite one-loop counterterms,
    $\mathfrak{W_3} =
        %\sum_i
        \left( \gamma_{\psi_i}^{(1)}
        + \gamma_A^{(1)} \xi \frac{\partial}{\partial \xi}
        - \beta_e^{(1)} \right) \mathcal{N}_{\psi_i} \overline{S_\text{fct}^{(1)}}$.
%%%%
    \item $\mathfrak{W_4}$ contains the genuine ``two-loop'' $\hbar^2$-order $\beta$-functions and anomalous dimensions of \ChiQED/ to be determined:
    $\mathfrak{W_4} = -\beta_e^{(2)} e \partial_e \overline{S_0}
        + \gamma_\phi^{(2)} \mathcal{N}_\phi \overline{S_0}$.
\end{itemize}

%%%%

In the MultRen method, there exists a one-to-one correspondence with the terms obtained in the AlgRen method.
The singular counterterms $\mathfrak{R_{3,4}}$ generate contributions $\widetilde{\beta_{e,\xi}}$ and $\widetilde{\gamma_{A,\psi_i}}$ to the $\beta$ and $\gamma$ functions.
The terms $- \widetilde{\beta_{\sigma_i}} \partial_{\sigma_i} \Gamma^*_\text{DReg}$
for $i = 1,2,3$, evaluated following \cref{eq:MultRen_AuxOpsEffects} in \cref{subsect:MultRen}, correspond to $\mathfrak{W_2}$ and $\mathfrak{R_2}$.
Those are evaluated with the very same diagrammatic calculations as in AlgRen, \cref{fig:Evsct_SAA_insertions,fig:Evsct_SPsiPsi_insertions}.
Likewise, one evaluates the terms $- \widetilde{\beta_{\rho_i}} \partial_{\rho_i} \Gamma^*_\text{DReg}$
($i = 1,2$), that correspond to $\mathfrak{W_3}$ in the AlgRen method.

All these quantities, except for the unknown two-loop RG coefficients in $\mathfrak{W_4}$, are known or calculable from one-loop diagrams.
The equation $\mathfrak{R} = \mathfrak{W}$ can therefore be solved to obtain these coefficients.
%% $\mathfrak{W_4}$ satisfies an overdetermined system of equations:
%% \begin{equation}
%%     \mathfrak{W_4}
%%     = \mathfrak{R_2} + \mathfrak{R_3} + \mathfrak{R_4} - \mathfrak{W_2} - \mathfrak{W_3}
%%     \, .
%% \end{equation}
%% one for each independent 4-dimensional operator that belongs to the operator basis.
The resulting $\hbar^2$-order $\beta$ and $\gamma$ functions of \ChiQED/ are (in Feynman gauge $\xi = 1$):
\begin{subequations}
\label{eq:2loopRGE}
\begin{align}
    \beta_e^{(2)} &= \gamma_A^{(2)} = \gamma_c^{(2)}
        = \left(\frac{\hbar \, e^2}{16 \pi^2}\right)^2 2 \Tr[\mathcal{Y}_R^4]
    \, ,
    \\
    \gamma_{\psi_i}^{(2)} &=
        -\left(\frac{\hbar \, e^2}{16 \pi^2}\right)^2 \left( \frac{2}{9} \Tr[\mathcal{Y}_R^2] (\mathcal{Y}_R^i)^2 + \frac{3}{2} (\mathcal{Y}_R^i)^4 \right)
    \, .
\end{align}
\end{subequations}
The two compared AlgRen and MultRen methods agree in the obtained results.

\subsubsection{Summary and Outlook}
We have demonstrated the practical renormalization of a chiral Abelian toy model up to two-loop order. The main result consists in the full set of non-invariant, singular counterterms as well as the finite, non-invariant symmetry-restoring counterterms which implement the Slavnov--Taylor identity at the two-loop level. These counterterms are found to be rather compact and of a similar structure at one- and two-loop order. Importantly, it is verified that they ensure the validity of the usual Ward identities.
The beta-functions and anomalous dimensions of the renormalization group equation of the model have been derived using two approaches: in the Algebraic Renormalization framework and in a modified version of the more customary multiplicative renormalization method. The methods are equivalent and provide the same final results. However, the application of algebraic renormalization is more straightforward, as it does not require any ``auxiliary couplings''.
We are currently working on the three-loop renormalization as well as the two-loop study of the non-Abelian case. All of this is in preparation for the application to the SM.

%%%%%%%%%%%%%%%%%%%%%%%%%%%%%%%%%%%%%%%%%%%%%%%%%%%%%%
%%%%%%%%%%%%%%%%%%%%%%%%%%%%%%%%%%%%%%%%%%%%%%%%%%%%%%
\section{Phenomenological Studies and Applications\label{section:pheno}}
%%%%%%%%%%%%%%%%%%%%%%%%%%%%%%%%%%%%%%%%%%%%%%%%%%%%%%
%%%%%%%%%%%%%%%%%%%%%%%%%%%%%%%%%%%%%%%%%%%%%%%%%%%%%%

Another important use case for computer tools has to do with phenomenological analyses. Typical tasks performed by such codes include the extraction of theory parameters from data, the prediction of observables in terms of NP parameters, or setting bounds on the underlying parameter space. Tools are for instance used to determine the WCs of higher-dimensional operators, by extracting them from observables in an automated global analyses. Typical fitting tools that allow for such analyses are \texttt{SMEFiT}~\cite{Hartland:2019bjb}, as well as \texttt{smelli}, \texttt{HighPT}, and \texttt{HEPfit}, which are all discussed further in this section. Common observable calculators, that consist of a large data base of predefined observables are \texttt{flavio}~\cite{Straub:2018kue}, \SuperIso~\cite{Mahmoudi:2007vz,Mahmoudi:2008tp}, \FlavBit~\cite{Workgroup:2017myk}, as well as the package \texttt{EOS}, which is further discussed below. Furthermore, there are several clustering tools and Montecarlo enablers on the market, such as \texttt{ClusterKing}~\cite{Aebischer:2019zoe} and \texttt{Pandemonium}~\cite{Laa:2021dlg}, the package \texttt{SMEFTsim} \cite{Brivio:2017btx}, as well as \sfr which can be used for Montecarlo simulations including Dim8 SMEFT operators, and which is discussed in the last subsection.

%%%%%%%%%%%%%%%%%%%%%%%%%%%%%%%%%%%%%%%%%%%%%%%%%%%%%%%%%%%%%%%%%%%%%%%%%%%%%%%%%%%%%%%%%%%%%%%%
\contribution[{\includegraphics[width=2.5cm]{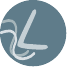}}]{\texttt{smelli:} Towards a global SMEFT likelihood}{Peter Stangl}
%%%%%%%%%%%%%%%%%%%%%%%%%%%%%%%%%%%%%%%%%%%%%%%%%%%%%%%%%%%%%%%%%%%%%%%%%%%%%%%%%%%%%%%%%%%%%%%%

The Python package \texttt{smelli} is a powerful tool for constraining SMEFT WCs and parameters of UV models matched to the SMEFT. Its goal is to provide a likelihood that is as global as possible while being fast enough to allow comprehensive fits and parameter scans.

NP extensions of the SM aim to resolve certain theoretical issues or tensions with experimental data. Typically, however, they have effects on many observables beyond their original purpose.
It is therefore crucial to carry out global phenomenological analyses of NP models in order to assess their viability and to show their actual superiority over the SM.
This is a challenging task as it involves computing predictions for a large number of observables and doing so for each model.
Fortunately, this problem can be tremendously simplified by using the SMEFT in an intermediate step.
In particular, a global likelihood function that yields the probability of observing the experimental data given SMEFT WCs\footnote{%
In the following we will focus on the WCs of operators up to and including dimension-six.
} can also be used as a likelihood function for model parameters of all NP models that can be matched to the SMEFT.
For such models, a global phenomenological analysis can be divided into two parts.
\begin{enumerate}

\item\label{smelli:enum:match} The NP model has to be matched to the SMEFT in order to express the SMEFT WCs at the matching scale $\Lambda_\text{NP}$, $\vec C(\Lambda_\text{NP})$, in terms of model parameters $\vec\xi$, i.e.\
 \begin{equation}\label{smelli:eq:matching}
  \vec C(\Lambda_\text{NP}) = f_\text{match}(\vec\xi)\,,
 \end{equation}
where the matching function $f_\text{match}$ and the model parameters $\vec\xi$ depend on the specific NP model.
 It might be necessary to include one-loop effects in this step, in particular if the leading contribution to relevant WCs is not generated by the tree-level matching.

 \item\label{smelli:enum:pheno} The SMEFT WCs at the scale $\Lambda_\text{NP}$, $\vec C(\Lambda_\text{NP})$, have to be constrained by experimental data.
 This requires the computation of theory predictions for a large number of observables at various scales, both in the SMEFT and in the LEFT.
 Importantly, WCs at different scales and in different EFTs are connected by RG running and matching.
 The one-loop contributions introduced by the RG running have been shown to be crucial in constraining NP models (see e.g.\ Refs.~\cite{Feruglio:2017rjo,Crivellin:2018yvo}).
 Theoretical predictions and experimental measurements of all relevant observables can then be used to construct a global likelihood function for the SMEFT WCs at the scale $\Lambda_\text{NP}$,
 \begin{equation}
  L_\text{SMEFT}\left(\vec C(\Lambda_\text{NP})\right)\,.
 \end{equation}
 Through Eq.~\eqref{smelli:eq:matching}, this also directly provides a likelihood function for the parameters~$\vec\xi$ of a NP model,
 \begin{equation}
  L_\text{NP}\left(\vec\xi\right) = L_\text{SMEFT}\left(f_\text{match}(\vec\xi)\right).
 \end{equation}

\end{enumerate}

The matching in step \ref{smelli:enum:match} depends only on the NP model, but is independent of both the experimental data and the theoretical predictions of the observables.
Full tree-level matching of generic models to SMEFT has been performed in Ref.~\cite{deBlas:2017xtg} and several tools are being developed to fully automate generic one-loop matching~\cite{Fuentes-Martin:2022jrf,Carmona:2021xtq,DasBakshi:2018vni}.

The phenomenological part in step \ref{smelli:enum:pheno} is independent of the NP model, so that a SMEFT likelihood function, once constructed, can be be used for generic phenomenological analyses of NP models.
It is important to stress that different sectors of observables should not be considered separately, since RG effects mix all sectors, and matching a NP model to the SMEFT will generally lead to effects in many sectors. It is therefore crucial to consider a \textit{global} SMEFT likelihood function that encompasses as many sectors as possible.

\subsubsection{\texttt{smelli} -- the \underline{SME}FT \underline{l}ike\underline{li}hood}

To establish a comprehensive global likelihood function in the space of dimension-six SMEFT WCs, the open source Python package \texttt{smelli} -- the \underline{SME}FT \underline{l}ike\underline{li}hood -- was introduced in Ref.~\cite{Aebischer:2018iyb}.
It builds on several other open-source projects that provide key components:
\begin{itemize}
 \item \texttt{wilson}~\cite{Aebischer:2018bkb} -- running and matching beyond the SM

\texttt{wilson} is a Python package for the running and matching of WCs in the LEFT and the SMEFT. It implements the one-loop running of all dimension-six operators in the SMEFT~\cite{Jenkins:2013zja,Jenkins:2013wua,Alonso:2013hga}, matching to the LEFT at the electroweak scale~\cite{Jenkins:2017jig}, and one-loop running of all dimension-six LEFT operators in QCD and QED~\cite{Aebischer:2017gaw,Jenkins:2017dyc}. Furthermore, it takes into account effects from rediagonalization of Yukawa matrices after running above the EW scale \cite{Aebischer:2020mkv,Aebischer:2020lsx}.

 \item \texttt{flavio}~\cite{Straub:2018kue} -- A Python package for flavour and precision physics in and beyond the SM

 The Python package \texttt{flavio} can compute theoretical predictions for a wide range of observables from different sectors, including flavour physics, electroweak precision tests, Higgs physics, and other precision tests of the SM.
 NP contributions are taken into account in terms of WCs of dimension-six operators in the SMEFT and the LEFT.
\texttt{flavio} also comes with an extensive database of experimental measurements and allows the construction of likelihoods based on these measurements and their corresponding theoretical predictions.

\item \texttt{WCxf}~\cite{Aebischer:2017ugx} -- the Wilson coefficient exchange format

\texttt{smelli}, \texttt{wilson}, and \texttt{flavio} all use the Wilson coefficient exchange format (WCxf) to represent WCs, which makes it easy to interface these codes with each other and with any other code that supports the WCxf standard.

\end{itemize}
In order to achieve a reasonably fast evaluation of the likelihood function in \texttt{smelli}, two simplifying approximations are used to deal with nuisance parameters $\vec\theta$ that enter the theory predictions $\vec{O}_\text{th}(\vec C, \vec \theta)$:
\begin{itemize}
 \item
 For observables with negligible theoretical uncertainties compared to the experimental uncertainties, each likelihood $L_\text{exp}^i$ from a given experimental measurement is evaluated with nuisance parameters fixed to their central values $\vec \theta_0$,
 \begin{equation}
  L_\text{exp}^i\left(\vec{O}_\text{th}(\vec C, \vec \theta_0)\right)\,.
 \end{equation}

 \item
 For observables with significant theoretical uncertainties,\footnote{All theoretical uncertainties under consideration are \emph{parametric} uncertainties that are due to the uncertainties of the nuisance parameters $\vec\theta$.} both the theoretical and experimental uncertainties are approximated as multivariate Gaussian and a combined likelihood is constructed for all correlated observables.
 The experimental covariance matrix $\Sigma_\text{exp}$ and the central experimental values $\vec O_\text{exp}$ are extracted from the original experimental likelihoods.
 The theoretical covariance matrix $\Sigma_\text{th}$ is obtained by sampling the nuisance parameters $\vec\theta$ from their respective likelihood distributions, while their central values $\vec \theta_0$ are used for the theoretical predictions $\vec{O}_\text{th}(\vec C, \vec \theta_0)$.
 Both covariance matrices enter the combined likelihood $\tilde L_\text{exp}$ defined by
 \begin{equation}
  -2 \ln \tilde L_\text{exp}\left(\vec{O}_\text{th}(\vec C, \vec \theta_0)\right) = \vec D^T (\Sigma_\text{exp}+\Sigma_\text{th})^{-1} \vec D\,,
  \qquad
  \vec D = \vec{O}_\text{th}(\vec C, \vec \theta_0) - \vec O_\text{exp}\,.
 \end{equation}

\end{itemize}
The global likelihood is then constructed by combining the individual approximated likelihood functions,
\begin{equation}
  L_\text{SMEFT}\left(\vec C\right)\approx
  \tilde L_\text{exp}\left(\vec{O}_\text{th}(\vec C, \vec \theta_0)\right)
  \times
  \prod_i L_\text{exp}^i\left(\vec{O}_\text{th}(\vec C, \vec \theta_0)\right)\,.
\end{equation}
The \texttt{smelli} Python package that provides this global likelihood function is available in the Python package manager \texttt{pip} and can be installed using
\begin{itemize}
 \item[] \texttt{python3 -m pip install smelli {-}{-}user}
\end{itemize}
which will download \texttt{smelli} with all dependencies from the Python package archive (PyPI) and install it in the user's home directory.
The source code of the package and more information about using it can be found in
\begin{itemize}
 \item the \texttt{smelli} GitHub repository~\url{https://github.com/smelli/smelli},
\item the \texttt{smelli}  API documentation~\url{https://smelli.github.io/smelli},
 \item the introductory tutorial in Ref.~\cite{Stangl:2020lbh}.
\end{itemize}

\subsubsection{Status and prospects of \texttt{smelli}}
The \texttt{smelli} project is under active development and has been extended several times in recent years, in particular also since the SMEFT-Tools 2019 workshop~\cite{Proceedings:2019rnh} where \texttt{smelli v1.3} was presented.

The first version of \texttt{smelli} focused on flavour and electroweak precision observables.
These included flavour-changing neutral and charged current $B$ and $K$ decays, meson-antimeson mixing observables in the $B$, $K$ and $D$ systems, charged-lepton flavour violating $B$, $K$, $\tau$, $\mu$ and $Z$ decays, as well as $Z$ and $W$ electroweak precision observables and the anomalous magnetic moments of the charged leptons.

In the context of Ref.~\cite{Falkowski:2019hvp}, \texttt{smelli} has been extended to Higgs physics, and the signal strengths of various decay ($h\to \gamma\gamma$, $Z\gamma$, $ZZ$, $WW$, $bb$, $cc$, $\tau\tau$, $\mu\mu$) and production channels ($gg$, VBF, $Zh$, $Wh$, $t\bar{t}h$) have been implemented.

With \texttt{smelli v2.0} more new observables and features have been introduced.
Beta decays were implemented following Ref.~\cite{Gonzalez-Alonso:2018omy}, adding the lifetimes and correlation coefficients of neutron beta decay as well as super-allowed nuclear beta decays.
Furthermore, additional $K$ decays and the total and differential cross sections for $e^+ e^-\to W^+ W^-$ pair production, as measured at LEP-2, have been added.
Apart from new observables and some minor innovations, one of the most important new features of \texttt{smelli v2.0} is a proper treatment of the Cabibbo--Kobayashi--Maskawa (CKM) matrix in SMEFT.
Inspired by Ref.~\cite{Descotes-Genon:2018foz}, \texttt{smelli} uses a CKM input scheme that takes four observables as proxies for the four CKM parameters.
The default CKM input scheme uses $R_{K\pi}=\Gamma(K^+\to\mu^+\nu)/\Gamma(\pi^+\to\mu^+\nu)$ (mostly fixing $V_{us}$), $BR(B^+\to\tau \nu)$ (fixing $V_{ub}$), $BR(B\to X_c e \nu)$ (fixing $V_{cb}$), and $\Delta M_d/\Delta M_s$ (mostly fixing the CKM phase $\delta$).
The CKM elements are then expressed in terms of the four CKM input observables and the SMEFT WCs that enter the predictions of these observables.
This removes a major limitation of \texttt{smelli} and allows semi-leptonic charged current meson decays to be included in the likelihood.

Since \texttt{smelli v2.0} there have been several new developments that will be incorporated in future versions of \texttt{smelli}.
A new numerical method has been developed in the context of Ref.~\cite{Altmannshofer:2021qrr}, which allows a numerically efficient implementation of the NP-dependence of the theory covariance matrix.
This will remove another major limitation of \texttt{smelli} and will enable the inclusion of observables whose theoretical uncertainties have a strong NP dependence, as e.g.\ the neutron Electric Dipole Moment (EDM).
In addition, the new method of Ref.~\cite{Altmannshofer:2021qrr} increases the computational speed by orders of magnitude, resulting in a significantly shorter evaluation time of the global likelihood function and allowing for much more comprehensive analyses.
These new features have already been successfully applied in Ref.~\cite{Greljo:2022jac}, where a global likelihood was constructed that includes neutral and charged current Drell-Yan tails, which will be implemented in a future version of \texttt{smelli}.

%%%%%%%%%%%%%%%%%%%%%%%%%%%%%%%%%%%%%%%%%%%%%%%%%%%%%%%%%%%%%%%%%%%%%%%%%%%%%%%%%%%%%%%%%%%%%%%%
\contribution[{\includegraphics[width=2.8cm]{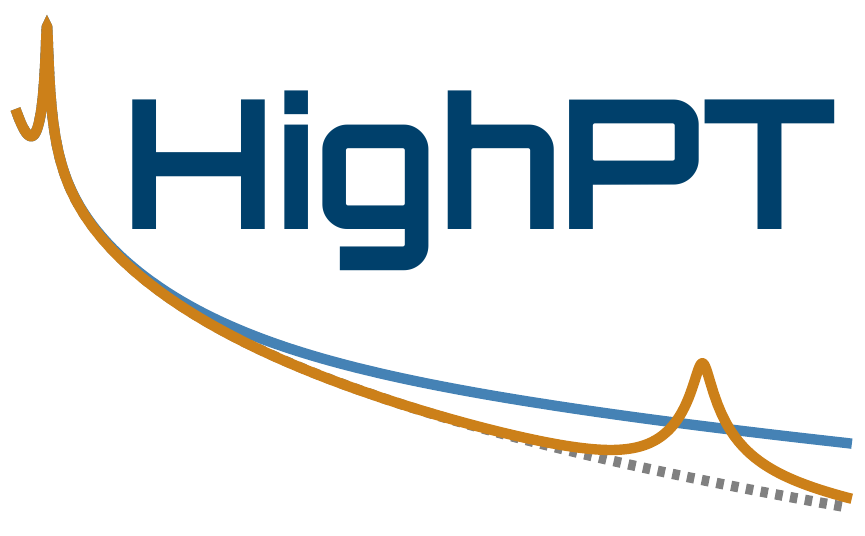}}]{{\tt HighPT}: A tool for Drell-Yan tails beyond the Standard Model}{Lukas Allwicher}
%%%%%%%%%%%%%%%%%%%%%%%%%%%%%%%%%%%%%%%%%%%%%%%%%%%%%%%%%%%%%%%%%%%%%%%%%%%%%%%%%%%%%%%%%%%%%%%%

High-$p_T$ tails in Drell-Yan processes can provide useful complementary information to low-energy and electroweak observables when investigating the flavor structure beyond the SM. The {\tt Mathematica} package {\tt HighPT} allows to compute Drell-Yan cross sections for dilepton and monolepton final states at the LHC. The observables can be computed at tree-level in the SMEFT, including the relevant operators up to dimension-eight, with a consistent expansion up to $\mathcal{O}(\Lambda^{-4})$. Furthermore, hypothetical TeV-scale bosonic mediators can be included at tree level in the computation of the cross-sections, thus allowing to account for their propagation effects. Using the Run-2 searches by ATLAS and CMS, the LHC likelihood for all possible leptonic final states can be constructed within the package, which therefore provides a simple framework for high-$p_T$ Drell-Yan analyses. We illustrate the main features of {\tt HighPT} with a simple example.

%\lipsum[1-2]
Semi-leptonic interactions have received a lot of attention in the literature in recent years, driven mainly by interesting data in $B$ meson decays.
In this context, it has been stressed several times that not only low-energy observables can contribute to constrain the new physics scenarios, but high-$p_T$ observables, especially Drell-Yan tails, can give complementary and independent information and, sometimes even more stringent bounds \cite{Fuentes-Martin:2020lea,deBlas:2013qqa,Angelescu:2020uug,Dawson:2018dxp,Marzocca:2020ueu}.
A comprehensive analysis of these effects has been implemented for the first time in {\tt HighPT} \cite{Allwicher:2022mcg,Allwicher:2022gkm}, a {\tt Mathematica} package that allows to compute hadronic cross-sections, event yields and the likelihoods from different LHC searches involving leptonic final states.
The aim is to provide an easy-to-use integrated framework to directly obtain a likelihood, in order to easily extract bounds on new physics parameters (both WCs in the SMEFT and couplings of TeV-scale mediators), to be then juxtaposed with low-energy experiments.

\subsubsection{Drell-Yan cross-section}
%\lipsum[1-2]
The most general Drell-Yan process can be written, at parton level, as the scattering
\begin{align}
  \bar q_i q_j' \to \ell_\alpha \bar\ell_\beta' \,,
\end{align}
where $i,j$ ($\alpha,\beta$) are quark (lepton) flavour indices, and $q,q'$ indicate either up- or down-type quarks, while $\ell,\ell'$ generically stand for either a charged lepton or a neutrino \footnote{However, we do not consider for the moment the process $pp\to\nu\bar\nu$.}.
The amplitude can be expressed in terms of form factors as\\
\resizebox{0.95\textwidth}{!}{
  \begin{minipage}{\linewidth}
    \begin{align}
      \begin{split}\label{eq:dilep-amp}
        \mathcal{A}(\bar{q}_i q_j' \to {\ell}_{\alpha} \bar\ell_{\beta}')\, =\, \frac{1}{v^2}\,\sum_{XY}&\,\Big\lbrace\,\,
        \left(\bar \ell_\alpha\gamma^\mu   \mathbb{P}_X \ell_\beta'\right)\left(\bar q_i\gamma_\mu  \mathbb{P}_Y q_j'\right)\, [\mathcal{F}^{XY,\,qq'}_{V}(\hat{s},\hat{t})]_{\alpha\beta ij}\\[0.15em]
        & + \left(\bar \ell_\alpha  \mathbb{P}_X\ell_\beta'\right)\left(\bar q_i  \mathbb{P}_Y q_j'\right)\, [\mathcal{F}^{XY,\,qq'}_{S}(\hat{s},\hat{t})]_{\alpha\beta ij}\\[0.25em]
        & + \left(\bar \ell_\alpha\sigma_{\mu\nu} \mathbb{P}_X\ell_\beta'\right)\left(\bar q_i \sigma^{\mu\nu}  \mathbb{P}_Y q_j'\right) \, \delta^{XY} \, [\mathcal{F}^{XY,\,qq'}_T(\hat{s},\hat{t})]_{\alpha\beta ij}\\[0.25em]
        & +  \left(\bar \ell_\alpha\gamma_{\mu} \mathbb{P}_X\ell_\beta'\right)\left(\bar q_i \sigma^{\mu\nu}  \mathbb{P}_Y q_j'\right)\,\frac{ik_\nu}{v}  \, [\mathcal{F}^{XY,\,qq'}_{D_q}(\hat{s},\hat{t})]_{\alpha\beta ij} \\[0.25em]
        & + \left(\bar \ell_\alpha\sigma^{\mu\nu}   \mathbb{P}_X \ell_\beta'\right) \left(\bar q_i \gamma_{\mu}   \mathbb{P}_Y q_j'\right)\,\frac{ik_\nu}{v} \, [\mathcal{F}^{XY,\,qq'}_{D_\ell}(\hat{s},\hat{t})]_{\alpha\beta ij}\,\Big\rbrace\,,
      \end{split}
    \end{align}
  \end{minipage}
}\\
which captures all possible $SU(3)_c\times U(1)_{\rm em}$ and Lorentz-invariant structures.
The sum over $X,Y=L,R$ extends over left- and right-handed chiralities, and we have defined the Mandelstam variables $\hat s = k^2 = (p_\ell + p_{\ell'})^2$, and $\hat t = (p_\ell - p_{q'})^2$.
The form factors $\cF_I$ can be decomposed as
\begin{align}
  \mathcal{F}_{I}(\hat s, \hat t) = \mathcal{F}_{I,\text{Reg}}(\hat s, \hat t) + \mathcal{F}_{I,\text{Poles}}(\hat s, \hat t) \,,
\end{align}
where
\begin{align}
  \cF_{I,\,\rm Reg}(\hat  s,\hat{t})\ =\ \sum_{n,m=0}^\infty \cF_{I \,(n,m)}\,\left(\frac{\hat s}{v^2}\right)^{\!n}\left(\frac{\hat t}{v^2}\right)^{\!m}\,,
\end{align}
is an analytic function in $\hat s$ and $\hat t$, describing local interactions (i.e. effective operators of $d\geq 6$), while $\cF_{I,\text{Poles}}$ captures the effect of simple poles in the $s$, $t$ or $u$ channel, due to some TeV-scale mediator.
The differential cross-section at parton level then is\\
\resizebox{0.92\textwidth}{!}{
  \begin{minipage}{\linewidth}
    \begin{align}
      \displaystyle\dfrac{{\rm d}\hat\sigma}{\rm{d}\hat t}(\bar{q}_i q^\prime_j \to {\ell}_{\alpha} \overline{\ell^{\prime}}_{\beta})  = \frac{1} {48\pi\,v^4}\,\sum_{XY} \sum_{IJ} M^{XY}_{IJ}(\hat s,\hat t)\,{\left[\mathcal{F}^{XY,\,qq^\prime}_I (\hat s,\hat t) \right]}_{\alpha\beta ij} {\left[\mathcal{F}^{XY,\,qq^\prime}_J(\hat s,\hat t)\right]}^{\ast}_{\alpha\beta ij}\,,
    \end{align}
  \end{minipage}
}\\
where $M_{IJ}^{XY}$ decribes the interference between different form factors.
This cross-section needs to be convoluted with the parton luminosity functions and integrated over the appropriate region in order to match the experimental searches (see \cite{Allwicher:2022gkm} for further details).

\subsubsection{Drell-Yan in the SMEFT}
%\lipsum[1-2]
When working in the context of the SMEFT, the WCs can be mapped to the form factor description of the scattering process by suitable matching conditions \cite{Allwicher:2022gkm}. Writing the SMEFT Lagrangian as
\begin{align}
  \cL_\mathrm{SMEFT}\ =\ \cL_\mathrm{SM}\ +\ \sum_{d,k} \dfrac{\cC_k^{(d)}}{\Lambda^{d-4}}\,\cO^{(d)}_k\ +\ \sum_{d,k} \bigg[\dfrac{\widetilde{\cC}_k^{(d)}}{\Lambda^{d-4}}\,\widetilde{\cO}^{(d)}_k\,+\,\rm{h.c.} \bigg] \,,
\end{align}
the cross-section, up to $\mathcal{O}(\Lambda^{-4})$, can be schematically written as
\begin{align}
  \hat{\sigma}\, \sim\,  \int[\mathrm{d}\Phi]\,\Bigg{\lbrace} \vert \cA_{\rm SM} \vert^2&+ \frac{v^2}{\Lambda^2}\sum_i 2\,\mathrm{Re}\Big{(}\cA_i^{(6)}\,\cA_{\rm SM}^\ast\Big{)}\\
  \ &+ \frac{v^4}{\Lambda^4}\bigg{[}\sum_{ij}2\,\mathrm{Re}\Big{(} \cA_i^{(6)}\,\cA_j^{(6)\,\ast} \Big{)} + \sum_{i} 2\,\mathrm{Re}\Big{(}\cA_i^{(8)}\,\cA_{\rm SM}^\ast  \Big{)}\bigg{]}+ \ \dots\Bigg{\rbrace}
\end{align}
where $\cA_i^{(6)}$ ($\cA_i^{(8)}$) indicates the contribution from dimension-six (dimension-eight) operators. The classes of operators contributing to Drell-Yan up to this order are summarized in Table \ref{tab:classes}.
%\begin{align}
%  \cF_{V} &=\ \cF_{V\,(0,0)} + \cF_{V\,(1,0)}\,\frac{\hat s}{v^2} + \cF_{V\,(0,1)}\,\frac{\hat t}{v^2}+
%    \sum_{a} \frac{v^2\left[\cS_{(a,\,\rm SM)}+\delta \cS_{(a)}\right]}{\hat s-m^2_a+im_a\Gamma_a}
%\end{align}
%\begin{align}
  %
%  \cF_{V\,(0,0)} \ &=\ \frac{v^2}{\Lambda^2}\,\cC^{\,(6)}_{\psi^4}\ +\ \frac{v^4}{\Lambda^4}\,\cC^{\,(8)}_{\psi^4H^2}\ +\ \frac{v^2 m_a^2}{\Lambda^4}\,\cC^{\,(8)}_{\psi^2H^2D^3}\ + \ \cdots \,, \\
  %
%  \cF_{V\,(1,0)} \ &=\ \frac{v^4}{\Lambda^4}\,\cC^{\,(8)}_{\psi^4D^2}\ + \ \cdots \,, \\
  %
%  \cF_{V\,(0,1)} \ &=\ \frac{v^4}{\Lambda^4}\,\cC^{\,(8)}_{\psi^4D^2}\ + \ \cdots \,, \\
%  \delta \cS_{(a)} \ &=\ \frac{m_a^2}{\Lambda^2}\,\cC^{\,(6)}_{\psi^2H^2D}\,+\frac{v^2m_a^2}{\Lambda^4}\,\left(\left[\cC_{\psi^2H^2D}^{\,(6)}\right]^2+\cC_{\psi^2H^4D}^{\,(8)}\right)\  +\ \frac{m_a^4}{\Lambda^4}\,\cC_{\psi^2H^2D^3}^{\,(8)}\ + \ \cdots\,,
%\end{align}

\begin{table}[t]
  \centering
  \resizebox{0.9\textwidth}{!}{
    \begin{minipage}{\linewidth}
      {\renewcommand{\arraystretch}{1.5}
        \begin{tabular}{lc|c c c | c c c c}
          \multicolumn{2}{c|}{Dimension} &     \multicolumn{3}{c|}{$d=6$} & \multicolumn{4}{c}{$d=8$}  
          \\\hline\hline
          \multicolumn{2}{l|}{Operator classes}  &  $\psi^4$ & $\psi^2 H^2 D$ & $\psi^2 X H$ & $\psi^4 D^2$ & $\psi^4 H^2$ & $\psi^2 H^4 D$ & $\psi^2 H^2 D^3$
          \\\hline
          \multicolumn{2}{l|}{Amplitude  scaling}  & $E^2/\Lambda^2$ & $v^2/\Lambda^2$ & $v E/\Lambda^2$ & $E^4/\Lambda^4$ & $v^2 E^2/\Lambda^4$ & $v^4/\Lambda^4$ & $v^2 E^2/\Lambda^4$\\\hline
          \multirow{2}{*}{Parameters} & \# $\mathbb{Re}$  & 456 & 45 & 48 & 168 & 171 & 44 & 52 
          \\
          & \# $\mathbb{I}$ & 399 & 25 & 48 & 54 & 63 & 12 & 12\\
        \end{tabular}
      }
    \end{minipage}
  }
  \caption{\sl\small Classes of SMEFT operators relevant to the high-$p_T$ observables and corresponding energy scaling of the amplitude. The parameter counting takes into account all structures entering the Drell-Yan cross-section up to  $\cO(1/\Lambda^{4})$. The operator bases for dimension-six and dimension-eight operators are taken from \cite{Grzadkowski:2010es} and \cite{Murphy:2020rsh}, respectively.}
  \label{tab:classes}
\end{table}

\subsubsection{Collider limits}

In order to compare the theory prediction for the cross-section with the searches performed by the experimental collaborations, detector effects, such as limited resolution or acceptance, must be taken into account. For binned distributions, this is done by introducing a response matrix $K$, such that
\begin{align}
  \sigma_q(x_{\rm obs}) = \sum_{p=1}^M K_{pq} \sigma_p (x) \,,
\end{align}
where $x$ indicates a generic particle-level observable, divided into $M$ bins, and $x_{\rm obs}$ is the experiment-level observable. $\sigma_q$ here indicates the cross-section for bin $q$.
The matrix $K$ needs to be extracted from Monte Carlo simulation for each independent combination of form factors.
With all the elements described so far, one can define a $\chi^2$ likelihood as
\begin{align}\label{eq:chi2}
    \chi^2(\theta)=\sum_{A\in\cA}\left(\frac{\cN_A(\theta)+\cN^{b}_A-\cN^{\rm obs}_A}{\Delta_A}\right)^2\,,
\end{align}
where $\cN_A^b$ is the number of background events and $\cN_A^{\rm obs}$ the number of observed events in bin $A$, both provided by the experimental collaborations.
The uncertainty $\Delta_A$ is obtained by adding in quadrature the background and observed uncertainties, $\Delta_A^2=(\delta\cN^b_A)^2+ \cN_A^{\rm obs}$, where the last term corresponds to the Poissonian uncertainty of the data.
$\cN_A(\theta)$, on the other hand, is the predicted number of events in bin $A$, depending on the new physics parameters $\theta$.
{\tt HighPT} includes recasts from ATLAS and CMS searches for all possible dilepton ($ee$, $\mu\mu$, $\tau\tau$, $e\mu$, $e\tau$, $\mu\tau$) and monolepton ($e\nu$, $\mu\nu$, $\tau\nu$) final states \cite{Allwicher:2022mcg}, such that a likelihood, written as a polynomial in the WCs, can be obtained for each of them.

\subsubsection{Using {\tt HighPT}: an example}
\begin{figure}[t]
  \centering
  \includegraphics[width=0.5\textwidth]{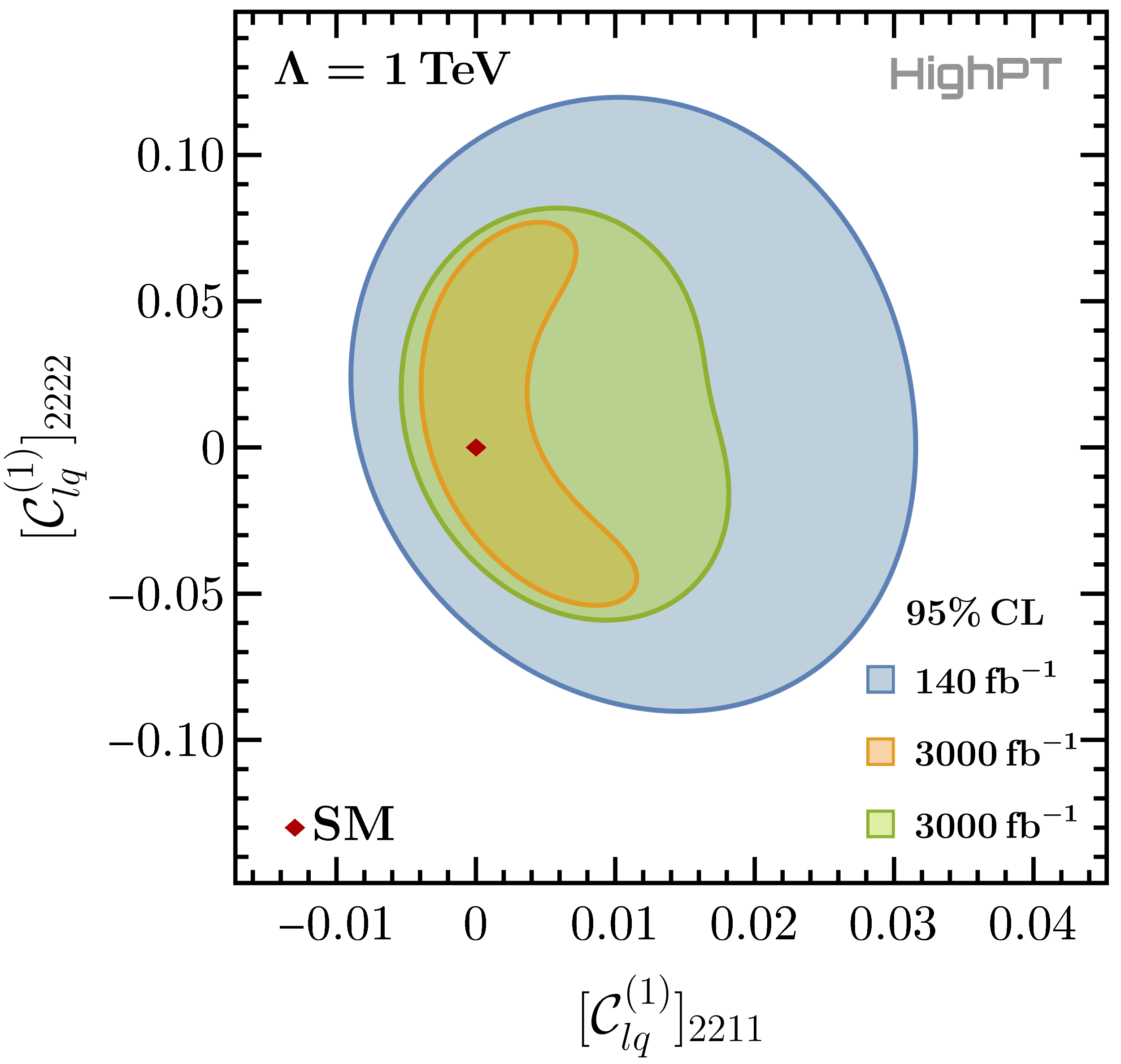}
  \caption{$95\%$\,CL regions for a fit of the given WCs to the dimuon search~\cite{CMS:2021ctt}. The blue region corresponds to the current constraints, whereas the orange and green regions correspond to projections for HL-LHC.}
  \label{fig:exampleplot}
\end{figure}

In order to briefly illustrate the main features of {\tt HighPT}, we show here an explicit example.
For a detailed review of all the functionalities see \cite{Allwicher:2022mcg}.
The main routine of the package is the function {\tt ChiSquareLHC}, yielding the $\chi^2$ likelihood as a list, with each element corresponding to a bin e.g. in $m_{\ell\ell}^2$ \footnote{The default binning is chosen such that each bin contains at least 10 events \cite{Allwicher:2022mcg}.}.
Consider the dimuon search by CMS \cite{CMS:2021ctt} and the dimension-six coefficients $[\cC_{lq}^{(1)}]_{2211}$, $[\cC_{lq}^{(1)}]_{2222}$, as in \cite{Grzadkowski:2010es}.
The likelihood can be extracted as
\vspace{3pt}
\begin{mmaCell}{Input}
    \mmaDef{L\(\mu\mu\)Binned} = ChiSquareLHC["di-muon-CMS",
      Coefficients->\{WC["lq1",\{2,2,1,1\}], WC["lq1",\{2,2,2,2\}]\}];
    \mmaDef{L\(\mu\mu\)} = Total[\mmaDef{L\(\mu\mu\)Binned}];
    
\end{mmaCell}
which computes the $\chi^2$ keeping only the specified operators, and to $\cO(\Lambda^{-4})$. The default setting is $\Lambda = 1$ TeV, but this can be changed at any time, together with the order of the EFT truncation \cite{Allwicher:2022mcg}.
Within the same framework, one can compute the projected likelihood for the HL-LHC by
\begin{mmaCell}{Input}
    \mmaDef{L\(\mu\mu\)3000} = Total[ChiSquareLHC["di-muon-CMS", 
      Coefficients-> \{WC["lq1",\{2,2,1,1\}],WC["lq1",\{2,2,2,2\}]\},
      Luminosity->3000, RescaleError->True]];
      
\end{mmaCell}
\begin{mmaCell}{Input}
    \mmaDef{L\(\mu\mu\)3000Const} = Total[ChiSquareLHC["di-muon-CMS",
       Coefficients-> \{WC["lq1",\{2,2,1,1\}],WC["lq1",\{2,2,2,2\}]\}, 
       Luminosity->3000, RescaleError->False]];
       
\end{mmaCell}
where the first option corresponds to a rescaling of the background uncertainty by $\Delta\mathcal{N}_A^b \rightarrow (L_\mathrm{projected}/L_\mathrm{current})^{1/2} \,\Delta\mathcal{N}_A^b=\sqrt{3000/140}\,\Delta\mathcal{N}_A^b$, while the second is the likelihood computed assuming that the ratio of background error over background is constant, i.e. $\Delta\mathcal{N}_A^b/\mathcal{N}_A^b=\text{const}$.
Minimizing these likelihoods, one can plot for example the 95\% C.L. contours as in Fig. \ref{fig:exampleplot}.

\subsubsection{Summary and outlook}
%\lipsum[1]
We have introduced {\tt HighPT}, a {\tt Mathematica} package designed to translate the data from Drell-Yan searches at the LHC into a likelihood function in terms of WCs.
It is worth stressing that, despite the focus in this brief overview has been on the SMEFT, {\tt HighPT} currently includes also a set of leptoquark mediators, allowing to include possible propagation effects of such new states in the computation of the cross-section \cite{Allwicher:2022mcg}.
We have shown in a short example how the $\chi^2$ can be computed, including also an option for HL-LHC projections.
Future directions of development for the package include the implementation of electroweak and low-energy observables, in order to be able to get a global likelihood for combined analyses in a unified framework.
Another possible extension is the inclusion of more high-$p_T$ observables related to semi-leptonic interactions, such as processes with a jet in the final state, and the inclusion of processes mediated by four-quark operators.

%%%%%%%%%%%%%%%%%%%%%%%%%%%%%%%%%%%%%%%%%%%%%%%%%%%%%%%%%%%%%%%%%%%%%%%%%%%%%%%%%%%%%%%%%%%%%%
\contribution[{\includegraphics[width=2.4cm]{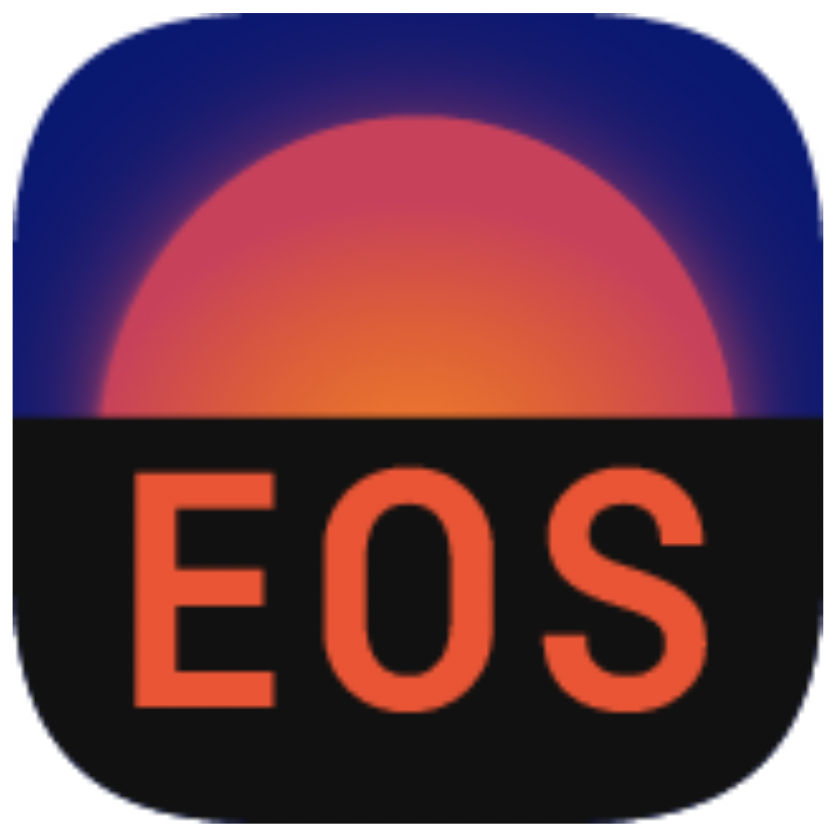}}]{\texttt{EOS:} Flavor Phenomenology with the \EOS Software}{M\'eril Reboud}{}
%%%%%%%%%%%%%%%%%%%%%%%%%%%%%%%%%%%%%%%%%%%%%%%%%%%%%%%%%%%%%%%%%%%%%%%%%%%%%%%%%%%%%%%%%%%%%%

Recent studies in flavor physics have revealed a consistent numerical pattern where large amounts of experimental data are analyzed to infer theory parameters of and BSM. Constraining the WCs of effective field theories have been proven particularly useful as they provide a model-independent framework to study scenarios BSM. In this context, performing the flavor analyses in a separated software and exporting the resulting constraints in terms of likelihoods in the WCs space is crucial for model building.

\EOS~\cite{vanDyk:2012zla,EOS:repo} is such a software.\footnote{\EOS developers welcome new contributors, feedback, questions and wishes on \url{https://github.com/eos/eos}.} It is an open source flavor software dedicated to the calculation of observables and the inference of theory parameters from an extendable list of models and constraints. It is particularly suited to the extraction of constraints on the parameters of effective field theories in the context of global analysis. It is written for three main use cases:
\begin{itemize}
    \item the numerical prediction of experimental observables with a wide range of theoretical and statistical techniques;
    \item the inference of theory parameters from an extensible database of experimental and theoretical likelihoods;
    \item and the production of Monte Carlo samples that can e.g. be used to study the experimental sensitivity to a specific observable.
\end{itemize}
\EOS is written in \cpp but offers a rich \Python interface meant to be used e.g. within a \Jupyter notebook.

\subsubsection{Installation and documentation}

\EOS can be installed using \Python package installer:
\begin{verbatim}
    python3 -m pip install --user eoshep
\end{verbatim}
and the \Python module can be accessed using
\begin{verbatim}
    import eos    
\end{verbatim}
\EOS documentation \cite{EOS:doc, EOSAuthors:2021xpv} contains further installation instructions, basic tutorials, as well as detailed examples for advanced usage.

\subsubsection{How to derive flavor constraints in the SMEFT}
The two main objects in \EOS are \texttt{Observable} and \texttt{Analysis}.
The former allows to compute any build-in (pseudo-)observable by specifying a set of parameters, options and kinematics. \EOS pre-built observables are classified by their \texttt{QualifiedNames} and can be listed using the \texttt{Observables} command. An updated list can also be found together with the documentation. Experimental measurements and theory constraints are expressed in terms of likelihoods with the \texttt{Constraint} class.

The \texttt{Analysis} class allows to evaluate a set of constraints within ranges of parameters provided by the user.
Once the analysis object is defined, it can be \texttt{optimized} to identify the best-fit point(s) and it accepts sampling routines. 

A SMEFT analysis therefore consists of the following steps:
\begin{enumerate}
    \item List the experimental and theory constraints relevant for the analysis.
    New constraints can be added using \texttt{manual\_constraints}.

    \item List the relevant nuisance parameters.
    The parameters of interest are the WCs of the effective theory relevant to the observables (e.g. \texttt{"ubmunumu::Re\{cVL\}"} for a study of $B\to\pi\mu\nu_\mu$).
    The matching from the low-energy effective theory to the SMEFT is performed at a later stage.

    \item Create an \texttt{Analysis} object, specifying \texttt{model:\ LEFT} as a global option.
    This analysis can be optimized to find the best-fit point and the corresponding goodness-of-fit information.

    \item Create posterior-predictive samples of the analysis, using one of the sampling routines: \texttt{sample\_mcmc}, \texttt{sample\_pmc} or \texttt{sample\_nested} (\EOS \texttt{>\ v1.0.5}).

    \item After marginalizing over the nuisance parameters, the samples can be exported to any matching software (e.g. \texttt{wilson}~\cite{Aebischer:2018bkb}) and converted to SMEFT parameters using the \EOS basis of the \texttt{WCxf} format \cite{wcxf:EOS-basis}.
\end{enumerate}

Alternatively, WCs can be imported from \texttt{wilson} directly into a \texttt{Parameters} object using the \texttt{FromWCxf} routine.

\subsubsection{\EOS vs. other flavor software}
\EOS is developed since 2011~\cite{vanDyk:2012zla} and was used in many phenomenological studies (see e.g.~\cite{Gubernari:2020eft,Bruggisser:2021duo,Leljak:2021vte,Bobeth:2021lya,Blake:2022vfl, Gubernari:2022hxn, Amhis:2022vcd} for the most recent ones). It is however not the only openly available flavour software and competes, among others, with \flavio~\cite{Straub:2018kue}, \SuperIso~\cite{Mahmoudi:2007vz,Mahmoudi:2008tp}, \HEPfit~\cite{DeBlas:2019ehy} and \FlavBit~\cite{Workgroup:2017myk}.
The unique features of \EOS are described below.
\begin{itemize}
    \item \EOS is particularly suited to study and compare different models of hadronic matrix elements (theory calculations, parameterizations...). It thus implements the possibility to select from various hadronic models at run time. As far as theory calculations are concerned, \EOS implements all the necessary tools for the evaluation of these elements using QCD sum rules.

    \item The careful implementation of hadronic matrix elements makes it the primary tool to a simultaneous inference of hadronic and new physics parameters.
    The underlying correlations are of primary importance when combining many experimental results.

    \item \EOS also offers the possibility of producing pseudo-events from an extensible set of PDFs. These events can then be used, e.g. for sensitivity studies and in preparation for experimental measurements.
\end{itemize}

\subsubsection{Recent and future developments}
\EOS development is done via its GitHub page, where the issue tracker allows the user to ask for new features, observables or constraints. The long-term plans are also discussed via the discussion panel.

In parallel to the implementations of new observables, parameterizations and recent experimental results, a considerable work has been done on the improvement of statistical tools. This development was performed in preparation to new phenomenology analyses which now usually reach $\mathcal{O}(100)$ nuisance parameters. Such large numbers make the approach based on basic Monte-Carlo techniques inefficient if not impossible. \EOS now offers an interface to the \texttt{dynesty}~\cite{Speagle:2020,dynesty:v2.0.3} package to make full use of nested sampling algorithm.

In the long-term, \EOS will contain ``pre-packaged'' low-energy analyses.
The idea is to simplify the use of low-energy constraints in the conception of new physics models. In particular, following the steps described above can be particularly time- and CPU-demanding when the number of nuisance parameters is large. This is typically the case in flavor physics due to involved parameterizations of the hadronic form factors. For example, the extraction of the WCs of the $b\to s\mu\mu$ weak effective theory using $B\to K\mu\mu$, $B\to K^*\mu\mu$ and $B_s\to \phi\mu\mu$ requires at least $130$ parameters for a consistent description of the hadronic transitions~\cite{Gubernari:2022hxn}. Provided that these nuisance parameters are uncorrelated to the other parameters entering a global SMEFT analysis, repeating this analysis in its entirety would be pointless and computationally challenging.

We therefore propose to simplify the publication of likelihoods containing only the parameters of interests (the WCs in this case). The posterior densities can be fitted with a Gaussian Mixture Model and used in \EOS or other flavor software.

%%%%%%%%%%%%%%%%%%%%%%%%%%%%%%%%%%%%%%%%%%%%%%%%%%%%%%%%%%%%%%%%%%%%%%%%%%%%%%%%%%%%%%%%%%%%%%%%
\contribution[{\includegraphics[width=.5\textwidth]{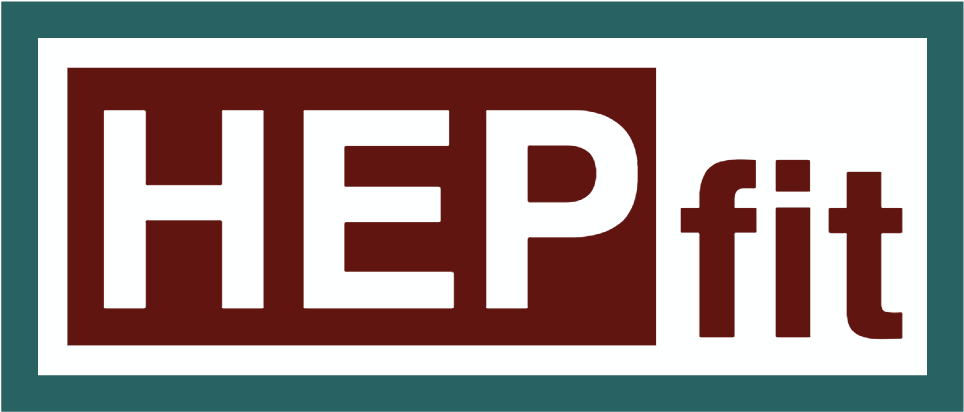}}]{\texttt{HEPfit:} Effective Field Theory Analyses with \texttt{HEPfit}}{Jorge de Blas, Angelica Goncalves, Víctor Miralles, Laura Reina, and Luca Silvestrini}{}
%%%%%%%%%%%%%%%%%%%%%%%%%%%%%%%%%%%%%%%%%%%%%%%%%%%%%%%%%%%%%%%%%%%%%%%%%%%%%%%%%%%%%%%%%%%%%%%%

\HEPfit is a tool developed to facilitate the combination of all different types of available constraints that can be used to learn from the parameter space of the SM or any new physics model. In the case of new physics, these constraints include experimental searches looking for the direct production of new particles, i.e. {\it direct searches}, or to find deviations from SM predictions in measurements of SM processes, i.e. {\it indirect searches}. The code has great flexibility in the form in which experimental likelihoods for these searches can be implemented, allowing e.g. correlations, binned measurements or non-Gaussian likelihoods. Theory constraints such as, e.g. unitarity, and theory uncertainties (including correlations) can also be taken into account in the analysis of a desired model. 

The above-mentioned types of information can be combined and used to sample the model parameter space via the built-in Bayesian Markov Chain Monte Carlo (MCMC) engine, which uses the Bayesian Analysis Toolkit ({\tt BAT}) library \cite{Caldwell_2009}. This enables the possibility of doing Bayesian statistical inference of the model parameters. This Bayesian analysis framework is parallelized with MPI so it can be run in clusters and CPUs capable of multi-thread computing. 
Alternatively, \HEPfit can also be used in library mode to compute predictions for observables. These can then be used to perform inference in any other statistical framework. 
To use {\tt HEPfit}'s Bayesian framework, the user only needs to provide the priors for the different model input parameters, those observables to be included in the likelihood calculation and the settings of the MCMC. Examples can be found in Section 7 of Ref.~\cite{DeBlas:2019ehy}.

Another important feature of \HEPfit is that, aside from the observables and models currently implemented in the code, the latter including the SM and several new physics scenarios, the user can implement their own custom observables and/or models as external modules.

On the technical side, \HEPfit is developed in {\tt C++} and it requires a series of mandatory dependencies such as the GNU Scientific Library, the {\tt BOOST} libraries, and {\tt ROOT}. To use the \HEPfit MCMC engine {\tt BAT} is also required. Finally, to enable the parallel use of \HEPfit  one needs {\tt OpenMPI}. See the Installation section in \cite{DeBlas:2019ehy} for more details. 

Aside from the SM, the current version of \HEPfit already includes several BSM models, such as Two-Higgs doublet models~\cite{Chowdhury:2017aav}, as well as several model-independent frameworks for the phenomenological description of new physics effects using EFTs. The implementation of these EFT is briefly described in the next section, following the status of the most up-to-date (developer's) version of the code, which can be found in the {\it Downloads} area of \href{https://hepfit.roma1.infn.it}{https://hepfit.roma1.infn.it}. These features are expected to appear in the next public release of {\tt HEPfit}.

%-------------------------------- DOCUMENT: SECTIONS -----------------------------------------%

\subsubsection{Effective Field Theory implementation in HEPfit}
\label{section_EFT}

Assuming that BSM physics is characterized by a mass scale $\Lambda$ and that for energies $E\ll \Lambda$ the particle spectrum and symmetries of nature are those of the SM, two types of EFT can be used to describe the physics at such energies: the SMEFT (see e.g. Ref.~\cite{Brivio:2017vri}), where the Higgs-like boson is embedded in a $SU(2)_L$ doublet as in the SM; and the HEFT, where the Higgs boson is described by a singlet scalar state, i.e. not belonging to an $SU(2)_L$ doublet. Most of the current development in \HEPfit is focused on the SMEFT, whose power counting follows an expansion in operators of increasing canonical mass dimension, and thus BSM effects are suppressed by correspondingly larger powers of the EFT cut-off scale $\Lambda$. Hence, the effective Lagrangian expansion takes the following form,
\begin{equation}
{\cal L}_{\mathrm{eff}}={\cal L}_{\mathrm{SM}} + \frac{1}{\Lambda}{\cal L}_5 + \frac{1}{\Lambda^{2}}{\cal L}_6 + \ldots,~~~~{\cal L}_d=\sum_i C_i^{(d)} {\cal O}_i^{(d)},
\label{eq:EffLag}
\end{equation}
where ${\cal O}_i^{(d)}$ are operators of mass dimension $d$ and $C_i^{(d)}$ the corresponding WCs. 
The first term, ${\cal L}_5$, only contains the lepton-number-violating Weinberg operator. In a lepton-number preserving theory the leading order (LO) new physics effects are therefore given by the dimension-six operators in ${\cal L}_6$ 
and these are the effects implemented in {\tt HEPfit}.

In the current implementation of SMEFT effects in {\tt HEPfit}, which can be found in the so-called {\tt NPSMEFTd6} model class, new physics contributions from dimension-six operators are considered for several types of observables:
\begin{itemize}
{\item Electroweak precision measurement ($Z$-pole observables at LEP/SLD and measurements of the $W$ mass and decay widths). These are implemented to the state-of-the-art precision in the SM and to LO in the SMEFT~\cite{deBlas:2021wap,deBlas:2022hdk}.}
{\item Diboson production at LEP2~\cite{Berthier:2016tkq} and the LHC~\cite{Baglio:2020oqu}.}
{\item LHC Higgs measurements, including the signal strengths for the different production and decay modes, as well as the Simplified Template Cross Section bin parameterization from~\cite{ATLAS-CONF-2020-053}. A comprehensive set of Higgs observables at future $e^+ e^-$ or $\mu^+ \mu^-$ colliders at different energies, with or without polarization, is also available in the code, for future collider studies~\cite{deBlas:2019rxi,DeBlas:2019qco,deBlas:2022ofj}.}
\end{itemize}
The current version of the code allows the use of either the $\left\{M_Z, \alpha, G_F\right\}$ or the $\left\{M_Z, M_W, G_F\right\}$ schemes for the SM electroweak input parameters for most of these observables.

\begin{figure}[t!]
  \centering
  \includegraphics[width=0.45\textwidth]{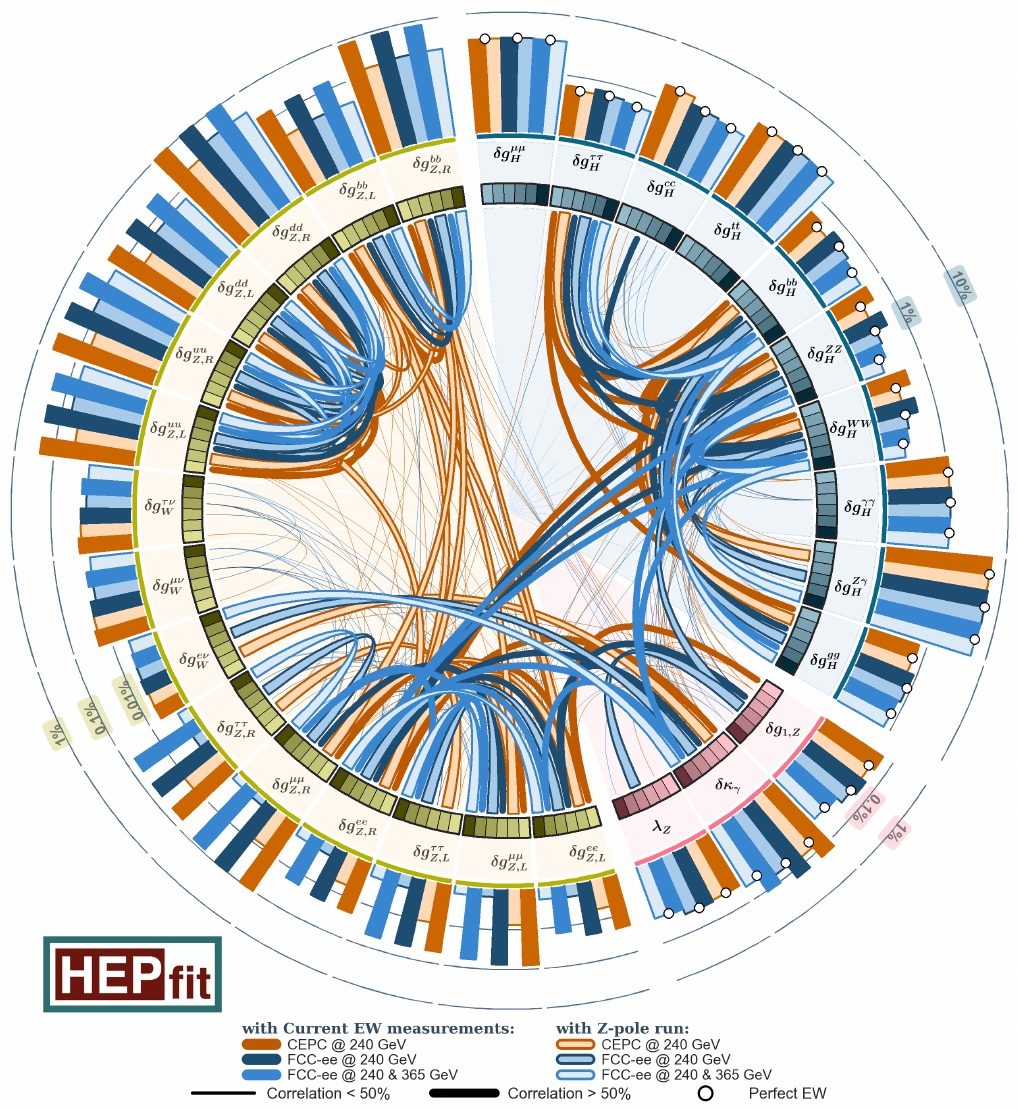}~~~~~~
  \includegraphics[width=0.45\textwidth]{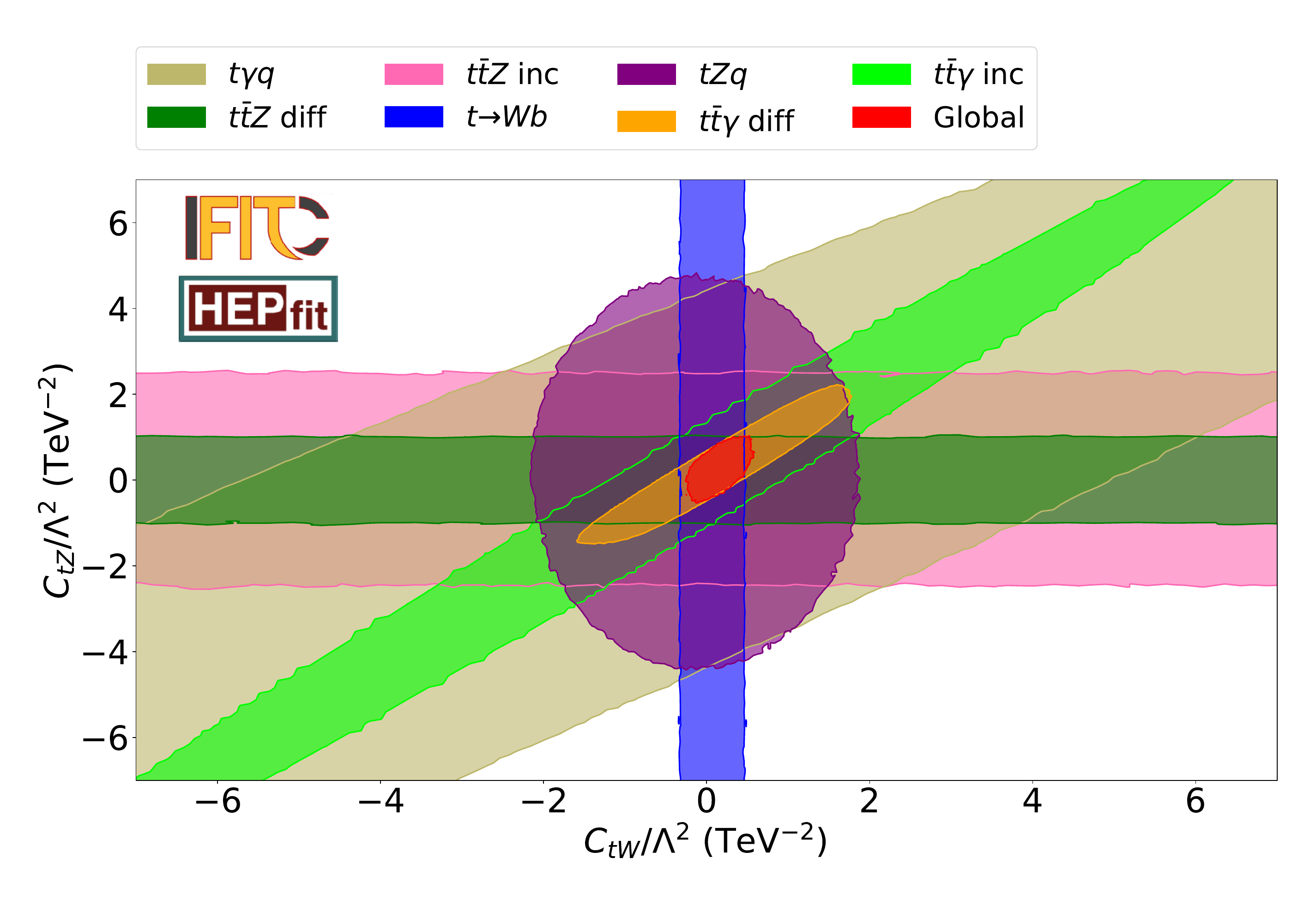}
  \caption{(Left) Constraints on SMEFT modifications of electroweak, Higgs and anomalous triple gauge couplings at future circular $e^+ e^-$ colliders. From Ref.~\cite{DeBlas:2019qco}.
(Right) Constraints from different top processes on Top dipole operators. From Ref.~\cite{Miralles:2021dyw}.}
  \label{fig:SMEFTresults_1}
\end{figure}

A comprehensive set of top-quark observables at the LHC is also available in {\tt HEPfit}, via the {\tt NPSMEFT6dtopquark} model class used in \cite{Miralles:2021dyw}. These include differential cross section measurements of $t\bar{t}Z$ and $t\bar{t}\gamma$ processes and inclusive cross sections for $t\bar{t}W$, $t\bar{t}H$ and single top processes. (See Fig.~\ref{fig:SMEFTresults_1} right.) 
These top-quark observables are also being implemented as part of the main {\tt NPSMEFTd6} class for global analyses. 

For the above-mentioned set of observables, new physics corrections are currently implemented at the linear level in $1/\Lambda^2$,
\begin{equation}
O=O_{\mathrm{SM}}+ \sum_i F_i \frac{C_i}{\Lambda^2}.
\label{eq:Odim6}
\end{equation}
The coefficients $F_i$ parametrizing the dependence on the WCs $C_i$ are computed at leading order, either analytically, as in the case of the electroweak precision measurements or, for LHC Higgs and top-quark observables, numerically, by fitting Eq.~(\ref{eq:Odim6}) to the results of {\tt MadGraph5\_aMC@NLO}~\cite{Alwall:2014hca} simulations using our own UFO implementation of the SMEFT or any of the models available in the literature, e.g. {\tt SMEFTsim}~\cite{Brivio:2017btx} or {\tt SMEFT@NLO}~\cite{Degrande:2020evl}. Our expressions are given in the so-called {\it Warsaw} basis~\cite{Grzadkowski:2010es}, but we give the possibility of choosing as model parameters some operators in other bases, in which case the corresponding expressions are obtained via the SM equations of motion. 
Different flavor assumptions can be chosen for fermionic operators, not restricted to flavor universality. 

\begin{figure}[t!]
  \centering
  \includegraphics[width=0.85\textwidth]{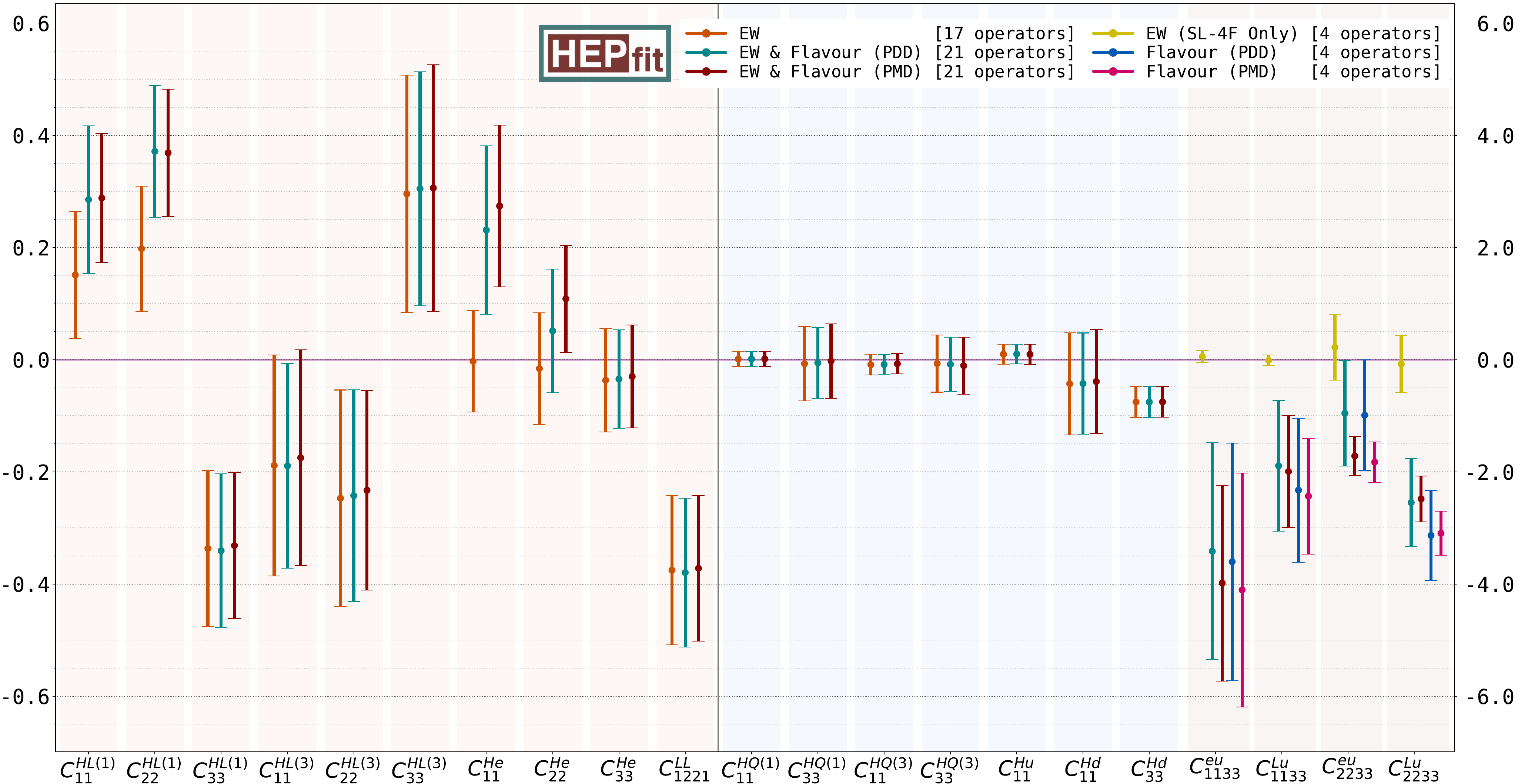}
  \caption{Results from a combined fit to electroweak precision data and flavor observables. From Ref.~\cite{Alasfar:2020mne}.  }
  \label{fig:SMEFTresults_2}
\end{figure}

Flavor physics is another sector that has been the focus of attention during the development of {\tt HEPfit}~\cite{Ciuchini:2015qxb,Silvestrini:2018dos,Ciuchini:2019usw,Ciuchini:2021smi,Ciuchini:2022wbq}, with multiple $\Delta F=2$ and $\Delta F=1$ observables included in the code. As in the case of the electroweak precision measurements the SM prediction has been implemented including all available corrections. New physics corrections are implemented as a function of the WCs of the LEFT, and the full matching with the SMEFT is currently work in progress (so far it is only implemented for interactions relevant for the analysis of $B$ anomalies~\cite{Ciuchini:2021smi,Ciuchini:2022wbq}). Combined analyses of flavor physics with electroweak precision observables can be found in, e.g. \cite{Silvestrini:2018dos,Alasfar:2020mne}, see Fig.~\ref{fig:SMEFTresults_2}.

As mentioned above, most of the SMEFT effects implemented in {\tt HEPfit} are currently available at leading order and at ${\cal O}(1/\Lambda^2)$. Part of the work to extend such calculations include the implementation of ${\cal O}(1/\Lambda^4)$ effects, see e.g. \cite{Corbett:2021eux}, and the full renormalization group running~\cite{Jenkins:2013zja,Jenkins:2013wua,Alonso:2013hga} via the integration of {\tt RGESolver}~\cite{DiNoi:2022ejg} in {\tt HEPfit}. The remaining contributions needed to obtain the full next-to-leading order (NLO) calculation of given observables are becoming increasingly available in the literature (e.g.~\cite{Dawson:2019clf,Dawson:2018pyl}) and will be gradually implemented. 

Finally, aside from the SMEFT implementation, a model describing the HEFT corrections to single Higgs processes is also available in {\tt HEPfit}.
These corrections include the effects from the leading order HEFT Lagrangian, using a power counting in terms of chiral dimensions~\cite{Buchalla:2013eza}. These include all operators of chiral dimension two, but we also include several operators of chiral dimension four, to parameterize local contributions from new particle loops in $H\to gg, \gamma\gamma$ and $Z\gamma$. The results from a global analysis using LHC run 1 and 2 Higgs data in this HEFT formalism can be found in Ref.~\cite{deBlas:2018tjm}.

%%%%%%%%%%%%%%%%%%%%%%%%%%%%%%%%%%%%%%%%%%%%%%%%%%%%%%%%%%%%%%%%%%%%%%%%%%%%%%%%%%%%%%%%%%%%%%%%
\contribution{\sfr v3: A tool for creating and handling vertices in SMEFT}{Athanasios Dedes, Janusz Rosiek and Michal Ryczkowski} 
%%%%%%%%%%%%%%%%%%%%%%%%%%%%%%%%%%%%%%%%%%%%%%%%%%%%%%%%%%%%%%%%%%%%%%%%%%%%%%%%%%%%%%%%%%%%%%%%

The abundance of parameters and interaction vertices in SMEFT requires automation.  
The scope of the \sfr v3
code~\cite{Dedes:2023zws} is to derive the Feynman rules for
interaction vertices from dimension-5, and -6, and, so far, all
bosonic dimension-8 operators which can easily be imported into other
codes, such as {\tt FeynArts}~\cite{Hahn:2000kx} and {\tt
  MadGraph5}~\cite{Alwall:2014hca}, for further symbolic or numeric
calculations of matrix elements and cross-sections.

\sfr starts from the most commonly used dimension-5,6 ``Warsaw"
basis~\cite{Grzadkowski:2010es} and dimension-8 basis of
ref.~\cite{Murphy:2020rsh} of operators in the unbroken phase, and,
following the steps of ref.~\cite{Dedes:2017zog} generates all
relevant Feynman rules in the physical mass basis quantized in Unitary
or $R_\xi$-gauges.  It is written in {\it Mathematica} language and
uses the package \frules\cite{Alloul:2013bka}.  The code \sfr is an
open access, publicly available code and can be downloaded
from \webpage.

There are several advances in \sfr v3~\cite{Dedes:2023zws} compared to
its predecessor \sfr v2~\cite{Dedes:2019uzs}.  Apart from
general optimizing and speeding up the code,
\sfr v3 can calculate vertices consistently up to the order
  $1/\Lambda^4$ in the EFT expansion, including terms quadratic in
  dim-6 WCs and linear in bosonic dim-8 WCs.
  What is particularly important is that \sfr v3 is able to express the SMEFT
  interaction vertices directly in terms of the chosen set of {\em
    observable input parameters}, avoiding the need of
  reparametrizations of transition amplitudes calculated in terms of
  SM gauge and Higgs couplings.  For convenience, \sfr v3 is
augmented with two predefined\footnote{In fact, also user-defined
  input schemes can straightforwardly be implemented in \sfr v3.}
input parameter schemes in the electroweak sector including
corrections of order $O(1/\Lambda^4)$:
%%%%%%%%%%%%%%%%%%%%%%%%%%%%%%%%%
\begin{itemize}
    \item The {\tt GF}-scheme with input parameters
      $(G_F,M_Z,M_W,M_H)$,
    \item The {\tt AEM}-scheme with input parameters
      $(\alpha_{em},M_Z,M_W,M_H)$.\footnote{A more customary choice for the {\tt AEM}-scheme would be the $({\alpha_{em}, G_F, M_Z, M_H})$-scheme, which can also be added in {\tt SmeftFR}. We must note, however, that the choice of this scheme leads to technical complications, such as having WC-dependent mass in the denominator of the $W$-boson propagator, which makes consistent expansion to a given EFT order much more involved.}
\end{itemize}
%%%%%%%%%%%%%%%%%%%%%%%%%%%%%%%%%%
Moreover, \sfr v3 employs the flavour input scheme of
ref.~\cite{Descotes-Genon:2018foz} which inserts the SMEFT corrected
CKM matrix elements starting directly from flavour observable
processes.\footnote{The CKM corrections are limited to
  $O(1/\Lambda^2)$, but this should suffice for most applications in
  flavour physics.}

\subsubsection*{\sfr v3 by an example}

All details about physics and usage of \sfr v3 are presented
in~\cite{Dedes:2023zws}.  To get the essence of what \sfr can do in
practice, it is better to study a step-by-step example for a given set
of dim-6 and dim-8, CP-even, operators.  The processes we have in mind
are vector-boson scattering at the LHC.  The subsequent steps follow the
{\it Mathematica} notebook file given in the \sfr distribution, {\tt
  SmeftFR-init.nb}.  After loading \frules and \sfr codes, we need
first to set the operator's set (in gauge basis).  For the processes
we have in mind, we set:
\begin{mmaCell}{Input}
OpList6=\{"phi","phiBox","phiD","phiW","phiWB","phiB","W"\};
OpList8=\{"phi8","phi6Box","phi6D2","phi4n1","phi4n2","phi4n3"\};

\end{mmaCell}
%
%%%%%%%%%%%%%%%%%%%%%%%%%%%
%\begin{center}
%    \includegraphics[scale=0.7]{Oplist-SmeftTools.png}
%\end{center}
%%%%%%%%%%%%%%%%%%%%%%%%%%%%%%
%
The naming of operators is given in App.~B of
ref.~\cite{Dedes:2023zws}, e.g.   $Q_{\varphi\Box} \rightarrow \texttt{ ``phiBox"}$, $Q_{\varphi^4 D^4}^{(1)} \rightarrow \texttt{ ``phi4n1"}$,
etc.   The next step is to initialize the SMEFT Lagrangian with a
chosen set of available options:
\begin{mmaCell}{Input} 
  SMEFTInitializeModel[Operators\(\pmb{\to}\)OpList, Gauge\(\pmb{\to}\)Rxi, 
      ExpansionOrder\(\pmb{\to}\)2, WCXFInitFile\(\pmb{\to}\)WCXFInput, 
      InputScheme\(\pmb{\to}\)"GF", CKMInput\(\pmb{\to}\)"no", 
      RealParameters\(\pmb{\to}\)True, MaxParticles\(\pmb{\to}\)4];
  
\end{mmaCell}
%%%%%%%%%%%%%%%%%%%%%%%%%%%
%\begin{center}
%    \includegraphics[scale=0.6]{Initialize-SmeftTools.png}
%\end{center}
%%%%%%%%%%%%%%%%%%%%%%%%%%%%%%
Here we choose to generate vertices in the $R_\xi$-gauges up to the EFT
expansion order of $1/\Lambda^4$ and with maximal 4 external legs
  (this option does not affect the UFO and FeynArts file generation
  where there is no such restriction).  We have also chosen to use
the $G_F$-input parameter scheme, and no SMEFT corrections to the CKM
matrix.  Moreover, we use real numerical parameter values for WCs
({as required by} {\tt MadGraph5}) taken from the {file named,}
{\tt WCxfInput}.  The next step is to load the parameters' model-file
and calculate the Lagrangian in the gauge basis, find field-bilinears and
diagonalize mass matrices to maximal order $1/\Lambda^4$, and finally,
find the SMEFT Lagrangian in the mass basis and generate the Feynman
rules, at this stage keeping the field redefinitions necessary to
  canonicalize the Lagrangian as symbols, without expanding them in
  $1/\Lambda$ powers.  Up to now, the program takes $\sim 7$ mins on
a typical laptop.\footnote{Running times throughout are referring to a
  {\tt i7,2.8GHz,16GB-RAM} computer with {\tt Linux Ubuntu 22.04}.}
%JR: I think next sentence is not precise (canonical ALL TERMS, not
%just kinetic) and hard to understand without much longer
%explanations.  So I would comment it out.
%
%The key to this fast calculation time success is the abbreviation so
%far of the field redefinitions needed for canonical kinetic terms for
%every vertex.
%
The obtained vertices in this form are stored in {\tt
  "/output/smeft\_feynman\_rules.m"} file.

Now we are ready to expand the field-redefinition parameters and read
the full vertices in user's $G_F$-scheme, previously adopted.  We use
the \frules command to select the $h\gamma\gamma$-vertex and obtain:
\begin{mmaCell}{Input}
  SMEFTExpandVertices[Input\(\pmb{\to}\)"user", ExpOrder\(\pmb{\to}\)2]; 
  SelectVertices[GaugeHiggsVerticesExp,SelectParticles\(\pmb{\to}\)\{H,A,A\}]
  
\end{mmaCell}
%\begin{center}
%\includegraphics[scale=0.33]{AAH_Michal.png}
%\end{center}
\begin{mmaCell}{Output}
 \{\{\{\{A,1\},\{A,2\},\{H,3\}\},\
 \mmaSup{\Big(\mmaFrac{1}{\mmaSup{\(\Lambda\)}{2}}\Big)}{2}\
 \mmaFrac{i}{\mmaSup{2}{3/4} \mmaSubSup{G}{F}{3/2} \mmaSubSup{M}{Z}{2} \
(\mmaSubSup{M}{W}{2}-\mmaSubSup{M}{Z}{2})}\newline\
\Bigg(-\mmaSup{C}{\(\phi\)D} \mmaSup{C}{\(\phi\)W} \mmaSubSup{M}{W}{4} \
-8 \mmaSup{(\mmaSup{C}{\(\phi\)W})}{2} \mmaSubSup{M}{W}{4} \
+4 \mmaSup{(\mmaSup{C}{\(\phi\)WB})}{2} \mmaSubSup{M}{W}{4} \
+16 \mmaSup{(\mmaSup{C}{\(\phi\)W})}{2} \mmaSubSup{M}{W}{2} \mmaSubSup{M}{Z}{2}\newline\
 -4 \mmaSup{(\mmaSup{C}{\(\phi\)WB})}{2} \mmaSubSup{M}{W}{2} \mmaSubSup{M}{Z}{2} \
 +\mmaSup{C}{\(\phi\)D} \mmaSup{C}{\(\phi\)W} \mmaSubSup{M}{Z}{4} \
 -8 \mmaSup{(\mmaSup{C}{\(\phi\)W})}{2} \mmaSubSup{M}{Z}{4} \newline\
 -\mmaSup{C}{\(\phi\)D} \mmaSup{C}{\(\phi\)WB} \mmaSubSup{M}{W}{3} \mmaSqrt{-\mmaSubSup{M}{W}{2}+\mmaSubSup{M}{Z}{2}} \
 -12 \mmaSup{C}{\(\phi\)W} \mmaSup{C}{\(\phi\)WB} \mmaSubSup{M}{W}{3} \mmaSqrt{-\mmaSubSup{M}{W}{2}+\mmaSubSup{M}{Z}{2}} \newline\
 +12 \mmaSup{C}{\(\phi\)W} \mmaSup{C}{\(\phi\)WB} \mmaSub{M}{W} \mmaSubSup{M}{Z}{2} \mmaSqrt{-\mmaSubSup{M}{W}{2}+\mmaSubSup{M}{Z}{2}}\
 +8 \mmaSup{(\mmaSup{C}{\(\phi\)B})}{2} (\mmaSubSup{M}{W}{4}-\mmaSubSup{M}{W}{2} \mmaSubSup{M}{Z}{2})\newline\
 +\mmaSup{C}{\(\phi\)B} \mmaSub{M}{W} (\mmaSubSup{M}{W}{2}-\mmaSubSup{M}{Z}{2}) (4 \mmaSup{C}{\(\phi\)Box} \mmaSub{M}{W}+\mmaSup{C}{\(\phi\)D} \mmaSub{M}{W}-4 \mmaSup{C}{\(\phi\)WB} \mmaSqrt{-\mmaSubSup{M}{W}{2}+\mmaSubSup{M}{Z}{2}})\newline\
 -4 \mmaSup{C}{\(\phi\)Box} (\mmaSubSup{M}{W}{2}-\mmaSubSup{M}{Z}{2}) (\mmaSup{C}{\(\phi\)W} \mmaSubSup{M}{W}{2}-\mmaSup{C}{\(\phi\)W} \mmaSubSup{M}{Z}{2}+\mmaSup{C}{\(\phi\)WB} \mmaSub{M}{W} \mmaSqrt{-\mmaSubSup{M}{W}{2}+\mmaSubSup{M}{Z}{2}})\Bigg)\newline\
  *(\mmaSubSup{p}{1}{\mmaSub{\(\mu\)}{2}} \mmaSubSup{p}{2}{\mmaSub{\(\mu\)}{1}}-\mmaSub{\(\eta\)}{\mmaSub{\(\mu\)}{1},\mmaSub{\(\mu\)}{2}} \mmaSub{p}{1}.\mmaSub{p}{2})\newline\
 +\mmaFrac{1}{\mmaSup{\(\Lambda\)}{2}}\mmaFrac{2 i \mmaSup{2}{3/4}}{ \mmaSqrt{\mmaSub{G}{F}} \mmaSubSup{M}{Z}{2}} \Big(\mmaSup{C}{\(\phi\)B} \mmaSubSup{M}{W}{2}-\mmaSup{C}{\(\phi\)WB} \mmaSub{M}{W} \mmaSqrt{-\mmaSubSup{M}{W}{2}+\mmaSubSup{M}{Z}{2}}+\mmaSup{C}{\(\phi\)W} (-\mmaSubSup{M}{W}{2}+\mmaSubSup{M}{Z}{2})\Big)\newline\
  *(\mmaSubSup{p}{1}{\mmaSub{\(\mu\)}{2}} \mmaSubSup{p}{2}{\mmaSub{\(\mu\)}{1}}-\mmaSub{\(\eta\)}{\mmaSub{\(\mu\)}{1},\mmaSub{\(\mu\)}{2}} \mmaSub{p}{1}.\mmaSub{p}{2})\}
 
\end{mmaCell}
%%%%%%%%%%%%%%%%%%%%%%%%%%%
%\begin{center}
%   \includegraphics[scale=0.5]{Hgg1-SmeftTools.png}
%\end{center}
%%%%%%%%%%%%%%%%%%%%%%%%%%%%%%

As expected from gauge invariance, the resulting vertex is
proportional entirely to the Lorentz factor $(p_1^{\mu_2}p_2^{\mu_1} -
g^{{\mu_1}{\mu_2}} p_1\cdot p_2)$.  In this vertex, there are terms
linear in dim-6 WCs plus quadratic terms of the form (dim-6)$^2$.
Although, the chosen set of WCs associated to dim-8 operators does not
appear in $H\gamma \gamma$-vertex, in the following example they do:
\begin{mmaCell}{Input}
  SMEFTExpandVertices[Input\(\pmb{\to}\)"user",ExpOrder\(\pmb{\to}\)2];
  SelectVertices[GaugeSelfVerticesExp,SelectParticles\(\pmb{\to}\)\{Z,Z,Z,Z\}]
  
\end{mmaCell}
\begin{mmaCell}{Output}
 \{\{\{\{Z,1\},\{Z,2\},\{Z,3\},\{Z,4\}\},\
 2 i (\mmaSup{C}{\(\phi\)4n1}+\mmaSup{C}{\(\phi\)4n2}+\mmaSup{C}{\(\phi\)4n3})\mmaSup{\Big(\mmaFrac{1}{\mmaSup{\(\Lambda\)}{2}}\Big)}{2}\mmaSubSup{M}{Z}{4} \newline\
 *(\mmaSub{\(\eta\)}{\mmaSub{\(\mu\)}{1},\mmaSub{\(\mu\)}{4}} \mmaSub{\(\eta\)}{\mmaSub{\(\mu\)}{2},\mmaSub{\(\mu\)}{3}}+\mmaSub{\(\eta\)}{\mmaSub{\(\mu\)}{1},\mmaSub{\(\mu\)}{3}} \mmaSub{\(\eta\)}{\mmaSub{\(\mu\)}{2},\mmaSub{\(\mu\)}{4}}+\mmaSub{\(\eta\)}{\mmaSub{\(\mu\)}{1},\mmaSub{\(\mu\)}{2}} \mmaSub{\(\eta\)}{\mmaSub{\(\mu\)}{3},\mmaSub{\(\mu\)}{4}})\}
 
\end{mmaCell}
%\begin{center}
%  \includegraphics[scale=0.6]{ZZZZ_Michal.png}
%\end{center}
%%%%%%%%%%%%%%%%%%%%%%%%%%%
%\begin{center}
%    \includegraphics[scale=0.6]{ZZZZ-SmeftTools.png}
%\end{center}
%%%%%%%%%%%%%%%%%%%%%%%%%%%%%%

The quartic $Z$-vertex is generated for the first time at dim-8 level
from the chosen operators $Q_{\phi^4 D^4}^{(1),(2),(3)}$ ! The user
could enjoy investigating further new vertices that did not appear up
to dim-6 level.  Finally, we note here that analogous vertices can be
extracted at this stage in {standard SM parametrization
  ($(\bar{g},\bar{g}',v)$-scheme)} or even in the
unexpanded-field-redefinition version of this scheme.

If we now want to continue with interfaces to LaTeX, {\tt UFO}, {\tt
  FeynArts} and {\tt WCxf} formats we have to {\tt Quit[]} {\it
  Mathematica} kernel and open the notebook {\tt
  SmeftFR\_interfaces.nb} {located} in the home-directory of the \sfr
distribution.  We again have to load \frules and \sfr engines and
reload the mass basis Lagrangian by typing:
%\begin{mmaCell}{Input}
%SMEFTInitializeMB[Expansion\(\pmb{\to}\)"user",\\Include4Fermion\(\pmb{\to}\)Tr%ue,\\IncludeBL4Fermion\(\pmb{\to}\)False];
%\end{mmaCell}
\begin{mmaCell}{Input}
  SMEFTInitializeMB[Expansion\(\pmb{\to}\)"user", Include4Fermion\(\pmb{\to}\)True];
\end{mmaCell}
%%%%%%%%%%%%%%%%%%%%%%%%%%%
%\begin{center}
%    \includegraphics[scale=0.6]{InitializeMB_SmeftTools.png}
%\end{center}
%%%%%%%%%%%%%%%%%%%%%%%%%%%%%%
where the ``{\tt user}"-{input scheme} from the previous session
is used (i.e.  the $G_F$-scheme), expansion is up to $1/\Lambda^4$,
{\it etc.}  (we do not include
%baryon or lepton number violating operators \red{(incompatible with
%MadGraph 5)}, nor input of
4-fermion operators, so
%both these options are
this option is irrelevant for the chosen set of operators in this
example).  The whole SMEFT Lagrangian in the mass basis is finally stored
in variable {\tt SMEFT\$MBLagrangian} for further use by interface
routines.

At this point, we can continue by exporting numerical values of WCs from the \frules model-file to a {\tt WCxf}-file format.
The created file can be used to transfer numerical values of WCs to
other codes that also support {\tt WCxf}-format.  In addition, as
already {possible} in \sfr v2, we can generate a LaTeX~file with
vertices and {corresponding} Feynman graphs.  Since the resulting
{expressions} for (dim-6)$^2$ and dim-8 {contributions} are
(usually) too long, we have kept in the LaTeX~output only the linear
dim-6 terms.
%JR: certainly not so easily!
%(although this can easily be changed).

%\textcolor{magenta}{MR: add here latex output example from the manual? E.g.  only hWW because it's much shorter}
%\textcolor{blue}{AD: it will be the same as in the FR paper, no?}
%\textcolor{magenta}{MR: yes, but we can also do e.g.  AAH, or just don't show it at all}

\subsubsection*{\sfr {\tt UFO} and {\tt MadGraph5}}

{\sfr v3 provides} a new routine for producing {\tt UFO} model
files  that may be useful in running realistic Monte Carlo simulations and replaces the standard \frules one. It can assign the correct
  ``interaction orders'' for both the SM couplings and the higher
  order operators, as required by MC generators to properly truncate
  transition amplitude calculations and reads:
%%%%%%%%%%%%%%%%%%%%%%%%%%%%%%%%%%%%%%
\begin{mmaCell}{Input}
  SMEFTToUFO[SMEFT\$MBLagrangian, CorrectIO\(\pmb{\to}\)True];
\end{mmaCell}
%%%%%%%%%%%%%%%%%%%%%%%%%%%
%\begin{center}
%    \includegraphics[scale=0.55]{SMEFTToUFO.png}
%\end{center}
%%%%%%%%%%%%%%%%%%%%%%%%%%%%%%
For details, including several comparisons to other existing codes
such as, {\tt SMEFT@NLO}~\cite{PhysRevD.103.096024}, {\tt
  Dim6Top}~\cite{Aguilar-Saavedra:2018ksv} and {\tt
  SMEFTsim}~\cite{Brivio:2020onw}, the user must consult
ref.~\cite{Dedes:2023zws}.  The generation of the {\tt UFO} model
files (especially in the ``user” scheme) is a time-consuming process.
For this particular example, it took about 2 hrs to generate the {\tt
  "/output/UFO”} directory.
%\textcolor{magenta}{MR: \textit{Add comment on 2 ways to produce
%it—one model or separate models, which faster.}
Moreover, the resulting {\tt UFO} model-file may lead to lengthy
calculations in {\tt MadGraph5} itself.  If the goal of the user is to
examine the influence of a single SMEFT operator on the chosen set of
processes at a time, one may either start with the model containing
several SMEFT operators and manually set {only one of them to be
  non-zero} by using {\tt MadGraph5}'s, {\tt set} command (e.g.  {\tt
  set CW 1e-06}) or produce separate models, each containing one of
the SMEFT operators and load different models before each run.  Both
options lead to the same results, but the latter one may be especially
attractive to users with limited CPU facilities.  Whichever we choose,
one must copy the produced {\tt UFO} model-directory to the {\tt
  models}-directory of {\tt MadGraph5} and then import it with the
command {\tt import model UFO}.  We are now ready to generate matrix
elements and cross-sections with {\tt MadGraph5}.

For example, the cross-section for vector-boson scattering at LHC is
calculated with: {\tt generate p p > w+ w+ j j QCD=0} (\& {\tt NP=0} -
SM, {\tt NP<=1} - $\mathcal{O}(\Lambda^{-2})$ and {\tt NP<=2} -
$\mathcal{O}(\Lambda^{-4})$ order).  In order to highlight the
significance of the quadratic (dim-6)$^2$ corrections, we adopt for
the input WCs {large values which could arise} from a
hypothetical strongly coupled sector.  The resulting cross-sections
are given in Table~\ref{table:cross_section_UFO}, with further
definitions in its caption.  As we can see, the quadratic effects for
$(C_W)^2$ are by a factor of $4400$ bigger than the linear
contributions.  For the pure scalar operators, the effects of
(dim-6)$^2$ terms depend on the sign of $C_{\varphi \Box}$, while the
effect of dim-8 coefficient $C_{\varphi^4 D^4}^{(i)}$, has an impact
of about 100 in the cross-section.  The tendency in the results for
the pure scalar operators, presented in
Table~\ref{table:cross_section_UFO}, follow the analytic amplitude
expression for the $W_L^+W_L^+$-scattering in eq.~(\ref{eq1}) below.
To our knowledge, the effects of these (dim-6)$^2$ and dim-8 modifications
to the cross-section
appear for the first time in the literature.

%\textcolor{blue}{Michal: a table here as I suggested will be nice.}
%%%%%%%%%%%%%%%%%%%%%%%%%%%%%%%%%%%%%%%%
\begin{table}[t!]
\centering
{\small
\begin{tabular}{|c|c|c|}
\hline
&  {\tt SmeftFR} $\mathcal{O}(\Lambda^{-2})$& {\tt SmeftFR}
$\mathcal{O}(\Lambda^{-4})$ \\
\hline
\multicolumn{3}{|c|}{{\tt p p > w+ w+ j j QCD=0}} \\
\hline
SM &\multicolumn{2}{|c|}{ $0.12456 \pm 0.00029$ } \\
\hline
$C_{W}$ &$8.564 \pm 0.020$  &  $37161 \pm 83$ \\
\hline
$+C_{\varphi \Box}$ & $0.13387 \pm 0.00032$ & $0.20981 \pm 0.00059$  \\
\hline
$-C_{\varphi \Box}$ & $ 0.14670 \pm 0.00043$ & $0.12511 \pm 0.00035$\\
\hline
$C_{\varphi 6\Box}$ & - & $0.12868 \pm 0.00031$ \\
\hline
$C^{(i)}_{\varphi^4 D^4}$ & - &  $10.891 \pm 0.024$ \\
\hline
\end{tabular}
}%end of \small
\caption{\sl Cross-sections (in pb) obtained using {\tt MadGraph5
    v3.4.1} with UFO models provided by \sfr v3 at the orders
  $\mathcal{O}(\Lambda^{-2})$ and $\mathcal{O}(\Lambda^{-4})$ in the
  EFT expansion for the {\tt p p > w+ w+ j j QCD=0} process at the Large Hadron Collider (LHC)
  with $\sqrt{s}=13$ TeV and cuts: $\Delta \eta_{jj}> 2.5$,
  $m_{jj}>500$ GeV.  Simulations are performed in the default ``$G_F$"
  electroweak input scheme with default numerical values of input
  parameters.  For each run, only one of the WCs has non-zero value
  assigned, equal to $\frac{C_i}{\Lambda^2}=\frac{4\pi}{\text{TeV}^2}$
  for dim-6 and $\frac{C_i}{\Lambda^4}=\frac{(4\pi)^2}{\text{TeV}^4}$
  for dim-8 operators.}
\label{table:cross_section_UFO}
\end{table}
\subsubsection*{\sfr to {\tt FeynArts} }

\sfr can generate a {\tt FeynArts} output by just using the native
\frules command,
\begin{mmaCell}{Input}
  WriteFeynArtsOutput[SMEFT\$MBLagrangian,Output\(\pmb{\to}\)\newline\
  FileNameJoin[\{SMEFT\$Path,"output","FeynArts","FeynArts"\}]];
\end{mmaCell}
%%%%%%%%%%%%%%%%%%%%%%%%
%\begin{center}
%  {\tt WriteFeynArtsOutput[SMEFT\$MBLagrangian]} .
%\end{center}
%%%%%%%%%%%%%%%%%%%%5
This is also a time-consuming stage (about double the {\tt UFO} file
generation).  The generated file is stored in the {\tt
  "output/FeynArts"}-directory.  We can use the files suffixed {\tt
  *.gen},{\tt *.mod} and {\tt *.pars} in the patched {\tt FeynArts}
program, {\tt FormCalc}~\cite{Hahn:2016ebn} or {\tt
  FeynCalc}~\cite{Shtabovenko:2020gxv}.  As an example, we create tree-level diagrams for the vector boson scattering, $W^+ W^+ \to W^+ W^+$
and isolate the longitudinal $W$-bosons, $W_L^+$.  We obtain the tree
amplitude at high energies expanded for $s\gg M_W^2$, with $\theta$
being the scattering angle,
%%%%%%%%%%%%%%%%%%%%
\begin{eqnarray}
&& \mathcal{M}_{W_L^+W_L^+ \to W_L^+ W_L^+}(s,\theta) = -2 \sqrt{2}
  G_F M_H^2 \left [ 1 - \frac{M_Z^2}{M_H^2}\left (
    1-\frac{4}{\sin^2\theta} \right ) \right ]\qquad \mathbf{(SM)}
  \nonumber \\[2mm]
&+& \left (2 C_{\varphi\Box} + C_{\varphi D} \right ) \frac{s}{\Lambda^2} \qquad \mathbf{(dim-6)}
  \nonumber \\[2mm]
&+& \left [ 8 C_{\varphi^6 \Box} + 2 C_{\varphi^6 D^2} + 16
    (C_{\varphi \Box})^2 + (C_{\varphi D})^2 - 8 C_{\varphi \Box}\,
    C_{\varphi D} \right.  \nonumber \\
&-& \left.  16 (C_{\varphi^4 D^4}^{(1)}+2 C_{\varphi^4
      D^4}^{(2)}+C_{\varphi^4 D^4}^{(3)}) G_F M_W^2 \right]
  \frac{\sqrt{2}}{8\, G_F \Lambda^2} \frac{s}{\Lambda^2} \qquad
  \mathbf{(dim-6)^2} \nonumber \\[2mm]
&+& \left [ (3 + \cos 2\theta) ( C_{\varphi^4 D^4}^{(1)} +
    C_{\varphi^4 D^4}^{(3)} ) + 8 C_{\varphi^4 D^4}^{(2)} \right
  ]\frac{s^2}{8 \Lambda^4} \qquad \mathbf{dim-8} \;.
\label{eq1}
\end{eqnarray}
%%%%%%%%%%%%%%%%%%%%%
This result, up to linear dim-6 operators, agrees with
ref.~\cite{Dedes:2020xmo} whereas all other contributions, the
quadratic (dim-6)$^2$ and the linear dim-8 effects, are new.  The
advantage of using a $R_\xi$-gauge (here Feynman gauge), is that we
can confirm this result by using the Goldstone-Boson Equivalence
Theorem comparing eq.~(\ref{eq1}) with the amplitude for charged
Goldstone boson scattering, $G^+ G^+ \to G^+ G^+ $.  Indeed, we find
agreement.  This is a serious non-trivial check since the Feynman
diagrams involved in $W_L^+ W_L^+$ elastic scattering contain in
addition, the coefficients $C_W, C_{\varphi WB}, C_{\varphi B},
C_{\varphi W}$ in a complicated way, but in the end their
contributions cancel out.  We have also verified, that in the {\tt
  "AEM"} input scheme the combination of WCs appearing in (\ref{eq1})
are exactly the same and therefore, numerically, the result is
identical.  This is another check towards correctness of SMEFT
vertices generated by \sfr v3.

\subsubsection*{Conclusions}
 We briefly presented a step-by-step example illustrating  the practical use and capabilities of the recently released \sfr v3 code~\cite{Dedes:2023zws}. \sfr  brings forward Feynman rules for a desired set of WCs by consistently including corrections of up-to order  $O(1/\Lambda^4)$ in the EFT expansion.  
 \sfr  generates interaction vertices in terms of chosen physical input parameters. Furthermore, 
 \sfr offers LaTeX~output, as well as  {\tt UFO} and {\tt FeynArts} model-files useful for numerical and analytical calculations. 
 In Table~\ref{table:cross_section_UFO} and in eq.~(\ref{eq1}), we show an example in which,
  (dim-6)$^2$ and dim-8 operator effects should not be ignored when mapping experimental data
 onto their associated WCs. For such research, \sfr v3 is a requisite.

 %%%%%%%%%%%%%%%%%%%%%%%%%%%%%%%%%%%%%%%%%%%%%%%%%%%%%%%%%%%%%%%%%%%%%%%%%%%%%%%%%%%%%%%%%%%%%%%%
\contribution{Application of EFT tools to the study of positivity bounds\label{section:chala}}{Mikael Chala}
%%%%%%%%%%%%%%%%%%%%%%%%%%%%%%%%%%%%%%%%%%%%%%%%%%%%%%%%%%%%%%%%%%%%%%%%%%%%%%%%%%%%%%%%%%%%%%%%

Positivity bounds are restrictions on the S-matrix of well-defined relativistic-quantum theories that follow from locality, unitarity and crossing-symmetry. 
In order to discuss the findings from Refs.~\cite{Chala:2021wpj,Chala:2023jyx}, let us first consider any such theory with a low-energy spectrum coinciding with that of the SM and with heavy fields of mass $\sim M$. Let us focus on two-to-two Higgs scattering for simplicity, $\phi\phi\to\phi\phi$. In the forward limit, the corresponding scattering amplitude satisfies that $\mathcal{A}(s)=\mathcal{A}(-s)$ due to crossing-symmetry, and it is analytic everywhere in the complex plane of the Mandelstam invariant $s$ up to certain ``mild'' singularities (definitely not as severe as delta functions~\cite{Weinberg:1995mt}). 

In first approximation, the only singularities of $\mathcal{A}(s)$ are single poles at $s=\pm M^2$, from where it can be easily proven that $\mathcal{A}^{\prime\prime}(s=0) \geq 0$~\cite{Adams:2006sv}. But this positivity restriction is actually much more widely satisfied. For example, let us assume that the singularities of $\mathcal{A}(s)$ are branch-cuts sitting along the $\text{Re}(s)$ axis, with branch points at $s\sim M^2$.
% ; .
Then, following Fig.~\ref{fig:singularities}, we can compute the quantity
\begin{align}
 \Sigma &\equiv \frac{1}{2\pi\ii}\int_\Gamma \text{d}s\, \frac{\mathcal{A}(s)}{s^3} \,,
\end{align}
which fulfils
\begin{align}
 \Sigma &= \frac{1}{\pi\ii} \int_{M^2}^\infty \text{d}s\, \frac{1}{s^3}\lim_{\epsilon\to 0}\left[\mathcal{A}(s+\ii\epsilon)-\mathcal{A}(s-\ii\epsilon)\right]\nonumber\\
 &= \frac{1}{\pi\ii} \int_{M^2}^\infty \text{d}s\, \frac{1}{s^3}\lim_{\epsilon\to 0}\left[\mathcal{A}(s+\ii\epsilon)-\mathcal{A}(s+\ii\epsilon)^*\right]\nonumber\\
 &= \frac{2}{\pi}\int_{M^2}^\infty \text{d}s\, \frac{\sigma(s)}{s^2}\geq 0\,,
\end{align}
where in the first equality we have used that, by virtue of the Froissart's bound~\cite{Froissart:1961ux}, the integral over the circular paths of $\Gamma$ vanishes; in the second equality we have relied on the Schwarz reflection principle $\mathcal{A}(s^*)=\mathcal{A}(s)^*$; and in the last step we have invoked the optical theorem, which relates the imaginary part of the forward amplitude to the total cross section $\sigma(s)$.

\begin{figure}[t]
     \begin{center}
     \includegraphics[width=0.4\columnwidth]{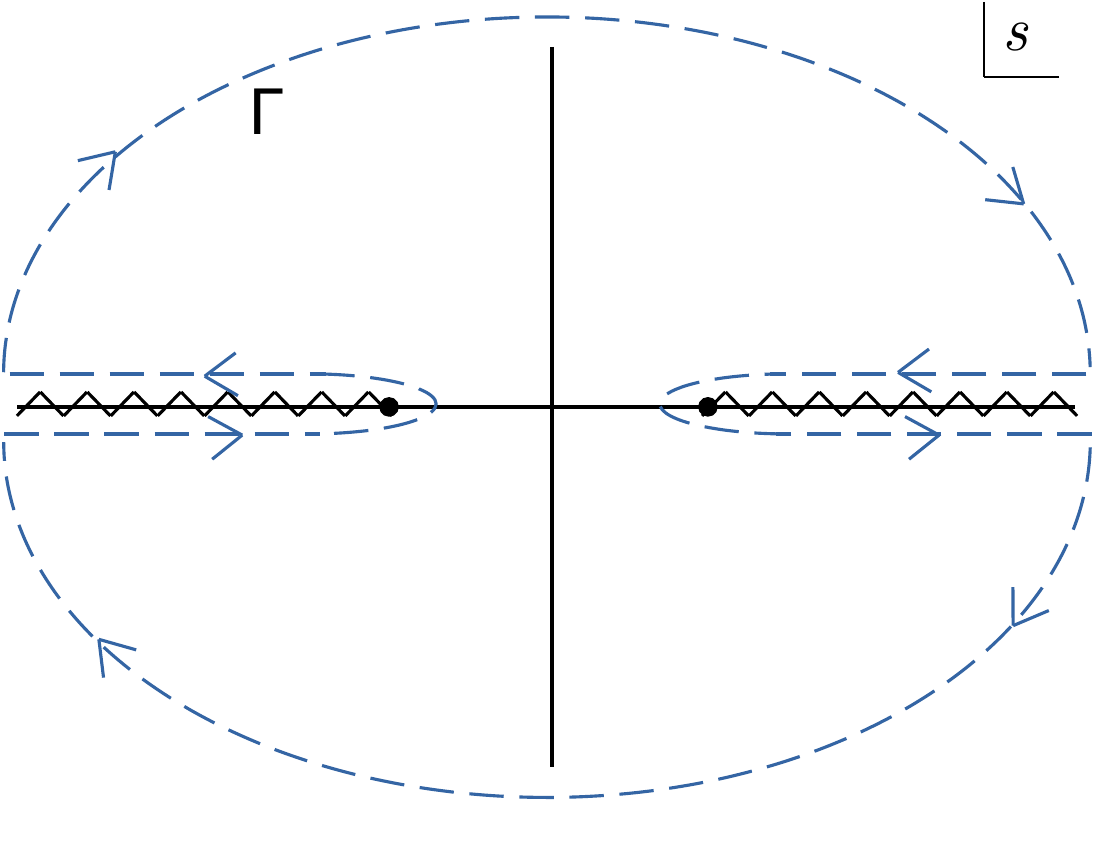}
     \end{center}
     \caption{\it Singularities of the amplitude for scalar two-to-two scattering in the forward limit, and contour of integration used in the derivation of positivity bounds.}\label{fig:singularities}
\end{figure}

Now, by analyticity, and using Cauchy's theorem, $\Sigma$ can be also computed from the residue of $\mathcal{A}(s)/s^3$ in the origin, which is nothing but the second derivative of the amplitude itself at $s=0$, from where we conclude again that $\mathcal{A}^{\prime\prime}(s=0)\geq 0$. Similar results can be drawn even in the case in which the branch cut extends all the way to $s=0$~\cite{Herrero-Valea:2020wxz}.

Because the amplitude in the vicinity of the origin can be computed within the EFT, the aforementioned restriction translates to bounds on the parameters of the EFT. Thus, if this process occurs already at tree level, then
\begin{equation}
 \mathcal{A}(s) = a_0 + a_2 \frac{s^2}{\Lambda^4} + a_4 \frac{s^4}{\Lambda^8}+\cdots
\end{equation}
where $\Lambda\sim M$ represents the cutoff of the EFT.
(Note that odd terms in $s$, and in particular the linear one and hence contributions from dimension-six EFT interactions, are absent due to the invariance of the amplitude under $s\to -s$.) From this equation and $\mathcal{A}^{\prime\prime}(s=0)\geq 0$, it can be concluded that $a_2\geq 0$.

If the EFT amplitude vanishes at tree level, then the WCs $a_i$ must be understood as evaluated at a scale $\mu\ll\Lambda$. For $a_2$ in particular:
\begin{equation}
 a_2(\mu) \sim a_2(\Lambda) + \frac{1}{16\pi^2} (\beta_{2} + \beta_{2}^\prime)\log{\frac{\mu}{\Lambda}}\,,
\end{equation}
where $\beta$ and $\beta^\prime$ are the one-loop beta functions induced by dimension-eight and pairs of dimension-six operators, respectively. From this schematic point of view several interesting conclusions can be drawn~\cite{Chala:2021wpj,Chala:2023jyx}:

\begin{enumerate}
 \item The matching contribution $a_2(\Lambda)$ must be always non-negative at tree level.\label{cond:1}
 \item It must be also non-negative if it does not run at one loop.\label{cond:2}
 \item On the contrary, if it does run, then it can be negative.\label{cond:3}
 \item $\beta_2$ must be non-positive, because $\beta_2'$ can be neglected in the limit of vanishing gauge and Yukawa couplings of the UV.\label{cond:4}
 \item $\beta_2^\prime$ must be non-positive whenever $\beta_2$ is zero.\label{cond:5}
\end{enumerate}

\subsubsection{Explicit computations using EFT tools}\label{explicit_computations_using_EFT_tools}
The conclusions \ref{cond:1}--\ref{cond:5} can be explicitly checked within different UV completions of the SM.
For concreteness, we focus mostly on the restrictions ensuing from the processes $\varphi_i\varphi_j\to\varphi_i\varphi_j$, with $\varphi_i$ representing any of the real degrees of freedom of the Higgs doublet $\phi$. We assume that the Higgs is massless.

The condition $a_2\geq 0$ translates into the bounds $c_{\phi^4 D^4}^{(2)}\geq 0$, $c_{\phi^4 D^4}^{(1)}+c_{\phi^4 D^4}^{(2)}\geq 0$ and $c_{\phi^4 D^4}^{(1)}+c_{\phi^4 D^4}^{(2)}+c_{\phi^4 D^4}^{(3)}\geq 0$, where $c_{\phi^4 D^4}^{(1,2,3)}$ are the WCs of the only three $\phi^4 D^4$ SMEFT dimension-eight operators in the basis of Ref.~\cite{Murphy:2020rsh}:
\begin{align}
 \mathcal{O}_{\phi^4 D^4}^{(1)} &= (D_\mu \phi^\dagger D_\nu \phi) (D^\nu\phi^\dagger D^\mu\phi)\,,&
 \mathcal{O}_{\phi^4 D^4}^{(2)} &= (D_\mu \phi^\dagger D_\nu \phi) (D^\mu\phi^\dagger D^\nu\phi)\,,\nonumber\\
 \mathcal{O}_{\phi^4 D^4}^{(3)} &= (D_\mu \phi^\dagger D^\mu \phi) (D^\nu\phi^\dagger D_\nu\phi)\,.
\end{align}
The benefit of testing conclusions \ref{cond:1}--\ref{cond:5} with explicit computations within concrete UV models is that it strengthens the confidence on results that might be hard to follow at the pure abstract level.
In turn, a careful validation of these conclusions implies a thorough cross-check of the EFT tools (which entails performing highly non-trivial computations of matching and running up to dimension eight) against robust results supported by very fundamental physics principles.

In what follows, we describe the different ways in which we have tested the conclusions \ref{cond:1}--\ref{cond:5} using \texttt{matchmakereft}~\cite{Carmona:2021xtq}, \texttt{SuperTracer}~\cite{Fuentes-Martin:2020udw}, and \texttt{MatchingTools}~\cite{Criado:2017khh}.

\paragraph{Tree-level matching} 
Let us consider five different single-field extensions of the SM that induce $\phi^4 D^4$ operators at tree level: 
\begin{align}
 \mathcal{S}&\sim (1,1)_0 \rightarrow c_{\phi^4 D^4}^{(1,2,3)} \sim (0,0,1)\,,&
 \Xi&\sim (1,3)_0 \rightarrow c_{\phi^4 D^4}^{(1,2,3)}\sim (2,0,-1)\,,\nonumber\\
 \mathcal{B}&\sim (1,1)_0 \rightarrow c_{\phi^4 D^4}^{(1,2,3)} \sim (-1,1,0)\,,&
 \mathcal{B}_1&\sim (1,1)_1 \rightarrow c_{\phi^4 D^4}^{(1,2,3)} \sim (1,0,-1)\,,\nonumber\\
 \mathcal{W}&\sim (1,3)_0 \rightarrow c_{\phi^4 D^4}^{(1,2,3)}\sim (1,1,-2)\,.
\end{align}
This should be read as follows: $\mathcal{S}$ is a full singlet of $SU(3)_c\times SU(2)_L$ with hypercharge $Y=0$ (in sub-index), which when integrated out produces the WCs (in arbitrary units) specified in the last parenthesis; likewise for the scalar triplet $\Xi$ and for the three vectors. In these and all cases hereafter, the omitted UV couplings appear squared, so they are always positive.

The WCs above satisfy the positivity relations in a non-trivial way. For example, in the $\mathcal{W}$ case, $c_{\phi^4 D^4}^{(3)}$ and $c_{\phi^4 D^4}^{(2)}+c_{\phi^4 D^4}^{(3)}$ are negative, but precisely $c_{\phi^4 D^4}^{(2)}$, $c_{\phi^4 D^4}^{(1)}+c_{\phi^4 D^4}^{(2)}$ and $c_{\phi^4 D^4}^{(1)}+c_{\phi^4 D^4}^{(2)}+c_{\phi^4 D^4}^{(3)}$ are non-negative.

We have obtained these results with \texttt{matchmakereft}. Despite being ``simply'' a tree-level computation, the task is not as easy as it might seem. Within \texttt{matchmakereft}, where the matching is performed by computing one-light-particle irreducible (1PI) Green's functions off-shell in both the UV and in the IR, one needs to specify the full set of EFT operators independent up to field redefinitions, as well as their reduction to physical ones in on-shell observables. Fortunately, these results, for the bosonic sector of the dimension-eight SMEFT, can be found in Ref.~\cite{Chala:2021cgt}; see also Ref.~\cite{Ren:2022tvi}. But even implementing this into \texttt{matchmakereft} can be very cumbersome.

As an alternative cross-check, we have verified the values of the WCs by using \texttt{MatchingTools}, which performs the matching by solving for the classical equations of motion. The advantage is that no EFT basis needs to be provided \textit{a priori}, but the problem is that the final result involves operators related by all kind of redundancies (field redefinitions, integration by parts, different names of same indices, ...). As a matter of example, integrating out $\mathcal{W}$ within \texttt{MatchingTools} gives (suppressing couplings)~\cite{Chala:2021cgt}:
\begin{align}
 \mathcal{L}_\text{EFT} &= (D_\mu\phi^\dagger D_\nu\phi) (D^\mu \phi^\dagger D^\nu\phi)
 +\underbrace{\cdots}_{\text{17 terms}}-\,\frac{1}{4} (D_\nu D_\mu\phi^\dagger\phi)(D^\mu D^\nu\phi^\dagger\phi)\,. 
\end{align}
Our approach to reduce this Lagrangian consists of using dedicated routines to export the output of \texttt{MatchingTools} to \texttt{Feynrules}~\cite{Alloul:2013bka}, where it is in turn exported to \texttt{FeynArts}~\cite{Hahn:2000kx} and \texttt{FormCalc}~\cite{Hahn:1998yk}, in which 1PI amplitudes are computed and matched onto the basis of Green's functions of Ref.~\cite{Chala:2021cgt}. The final result is finally reduced onto a physical basis using the relations obtained from equations of motion therein. It reads:
\begin{equation}
 \mathcal{L}_\text{EFT} = 2\mathcal{O}_{\phi^4 D^4}^{(1)} + 2\mathcal{O}_{\phi^4 D^4}^{(2)} - 4\mathcal{O}_{\phi^4 D^4}^{(3)} + \cdots
\end{equation}
in agreement with \texttt{matchmakereft} (the ellipses stand for higher-point interactions).
 
\paragraph{One-loop matching}
Let us now take the scalar singlet and triplet cases up to one loop, in the limit in which the only relevant couplings in the UV are the trilinear terms (which we set to unit). Working with \texttt{matchmakereft}, we get:
\begin{equation}
 c_{\phi^4 D^4}^{(1)} = c_{\phi^4 D^4}^{(2)} = -\frac{13}{48\pi^2}\,,
\end{equation}
for the scalar case, and
\begin{equation}\label{eq:posbreaking}
 c_{\phi^4 D^4}^{(2)} = -\frac{61}{144\pi^2}\,,
\end{equation}
in the triplet case. Here, we ignore the WCs that arise already at tree level. In both models, at least the condition $c_{\phi^4 D^4}^{(2)}\geq 0$ is broken, as expected from conclusion~\ref{cond:3}.

We have cross-checked this result with the help of \texttt{SuperTracer}~\cite{Fuentes-Martin:2020udw}. To this aim, the output of \texttt{SuperTracer} is simplified within the code itself, and the final result is processed following the same strategy as with \texttt{MatchingTools}. This provides a very strong and robust test of the validity of both \texttt{matchmakereft} and \texttt{SuperTracer}.

Let us now take scalar quadruplet extensions of the SM, with $Y=1/2$ and $Y= 3/2$. These scalars couple linearly to three Higgses. The only operators that they induce at tree level are of the form $\phi^6 D^{2n}$, which do not renormalize $\phi^4 D^4$. Consequently, following conclusion~\ref{cond:2}, we expect the positivity bounds to hold. Indeed, from \texttt{matchmakereft} we obtain (we ignore couplings again):
\begin{equation}
 c_{\phi^4 D^4}^{(1)} = \frac{1}{9\pi^2}\,, \,\, c_{\phi^4 D^4}^{(2)} = \frac{1}{36\pi^2}\,, \,\, c_{\phi^4 D^4}^{(3)} = -\frac{1}{18\pi^2}\,,
\end{equation}
for $Y=1/2$ and
\begin{equation}
 c_{\phi^4 D^4}^{(1)} = c_{\phi^4 D^4}^{(3)} = 0\,, \quad c_{\phi^4 D^4}^{(2)} = \frac{1}{4\pi^2}\,,
\end{equation}
for $Y=3/2$.

\paragraph{One-loop running}
Despite the breaking of positivity in one-loop matching, as for example highlighted in Eq.~\eqref{eq:posbreaking}, the amplitude for $\varphi_i\varphi_j\to\varphi_i\varphi_j$ is non-negative in the deep IR because there it is dominated by the running of $c_{\phi^4 D^4}^{(2)}$ induced by tree-level operators. It can be indeed checked that $\beta_{\phi^4 D^4}^{(2)}$ is always non-positive~\cite{Chala:2021pll,Chala:2021wpj}.

On the other hand, this implies that $\beta_{\phi^4 D^4}^{(2)\prime}$ does not need to be negative. Computed again with \texttt{matchmakereft} as well as with \texttt{FeynArts}+\texttt{FormCalc}, we obtain for example:
\begin{equation}
 \beta_{\phi^4 D^4}^{(2)\prime}= \frac{1}{6}  (28c_{\phi^4 D^4}^{(1)}+ 43 c_{\phi^4 D^4}^{(2)}+ 15 c_{\phi^4 D^4}^{(3)}) g_2^2 +\cdots
\end{equation}
which is positive, for example, in the scalar singlet case ($c_{\phi^4 D^4}^{(1,2)}=0$, $c_{\phi^4 D^4}^{(3)}=1$). In the equation above, $g_2$ stands for the $SU(2)_L$ gauge coupling and the ellipses represent terms proportional to other SM couplings.

In cases where $\beta_2$ vanishes, we do expect $\beta_2'$ to be non-positive; see conclusion~\ref{cond:5}. One such case is given by the renormalisation of $W^2\phi^2 D^2$ operators (where $W$ is the $SU(2)_L$ gauge boson) by $\phi^4 D^4$ operators. We know that $\beta_2$ vanishes in this case because loops with two insertions of $\phi^4 D^{2n}$ operators must have at least four Higgses.

Among the $W^2\phi^2 D^2$ operators, there is one that is restricted by the positivity of the amplitude for $W\phi\to W\phi$. The corresponding $\beta_2$ function reads:
\begin{equation}
 \beta_{W^2\phi^2 D^2}^{(1)} = -\frac{g_2^2}{6} (2 c_{\phi^4 D^4}^{(1)} + 3 c_{\phi^4 D^4}^{(2)} + c_{\phi^4 D^4}^{(3)})+\cdots\,.
\end{equation}
The ellipses encode non-$\phi^4 D^4$ operators. This quantity is necessarily non-positive, because the parenthesis is non-negative (at tree level). Indeed, we can recast it in the form
\begin{equation}
 (c_{\phi^4 D^4}^{(1)}+c_{\phi^4 D^4}^{(2)}+c_{\phi^4 D^4}^{(3)}) + (c_{\phi^4 D^4}^{(1)}+c_{\phi^4 D^4}^{(2)}) + c_{\phi^4 D^4}^{(2)}\,,
\end{equation}
which is non-negative because the three terms in the sum are non-negative, as we saw before.

For all these calculations, we have relied on \texttt{matchmakereft} with full cross-check using \texttt{FeynArts}+\texttt{FormCalc}.

\subsubsection{Towards fully-automated one-loop matching}\label{sec:towards_fully_automated_one-loop_matching}
Even with the help of current EFT tools, the explicit computations  described before can become extremely tedious. This is because the simplification of the Lagrangian resulting from integrating out the heavy degrees of freedom is highly redundant. To the best of our knowledge, there is no generic and publicly available method to reduce the effective Lagrangian to a physical basis in an automated way.~\footnote{After this work was presented in the SMEFT-Tools workshop, and prior to the publication of this manuscript, the tool \texttt{Matchete}~\cite{Fuentes-Martin:2022jrf} was released, which, among many other features, makes progress towards reducing redundant Lagrangians in an automated fashion.}
In this final section, we comment briefly on the approach we have adopted to face this problem, and on the progress we have made so far.

Our idea for automating the process of reducing a redundant Lagrangian to a physical basis of operators consists in requiring explicitly that both Lagrangians provide exactly the same S-matrix for all different processes that can be computed within the EFT (up to the corresponding order in the expansion in inverse powers of the cutoff).

In practice, this amounts to equating all needed tree-level on-shell connected and amputated Feynman graphs. As a matter of example, let us focus here on the SMEFT Higgs sector up to dimension eight. For the redundant Lagrangian, we consider that comprised by all Higgs operators in the Green's basis of Ref.~\cite{Gherardi:2020det} (dimension six), together with those in Ref.~\cite{Chala:2021cgt} (dimension eight). For the physical one, we stick to the basis of Ref.~\cite{Murphy:2020rsh}. The notation below follows the conventions in these references.

Upon equating the resulting calculations in both theories, one obtains a set of equations from where the physical WCs in the physical theory can be solved in terms of the physical and redundant WCs in the redundant EFT.

The main complication of on-shell matching, besides the huge number of diagrams, is the presence of light propagators, which manifest as non-local combinations of momenta in the S-matrix, which implies that solving the aforementioned system of equations analytically and in an automated way becomes very complicated. It can be instead solved numerically, upon giving concrete values to the different momenta involved in the process. In order to do so, without conflicting with (i) momentum conservation, (ii) on-shellness of the external legs and (iii) the fact that, as soon as the number of momenta in the amplitude is larger than four, not all them can be linearly independent; namely to ensure that we are in the physical region, we use Monte Carlo algorithms (currently \texttt{RAMBO}~\cite{Kleiss:1985gy}) to sample the phase space.

Our current results for the Higgs sector of the SMEFT (up to the operator $\phi^8$), expressed as shifts on physical WCs ($c_i$) induced by the presence of redundant ones ($r_j$), read:
\begin{align}
 c_{\phi\Box}&\to c_{\phi\Box}+\frac{1}{2}r_{\phi D}'\,,\\
 c_{\phi^6}&\to c_{\phi^6} + 2\lambda r_{\phi D}'\,,\\
 c_{\phi^6 D^2}^{(1)}&\to c_{\phi^6 D^2}^{(1)}+2\lambda (2 r_{\phi^4 D^4}^{(12)}-2 r_{\phi^4 D^4}^{(4)}-r_{\phi^4 D^4}^{(6)})-4 c_{\phi\Box} r_{\phi D}'-\frac{1}{2} c_{\phi D}r_{\phi D}'\textcolor{blue}{-\frac{7}{4} r_{\phi D}'^2 + r_{\phi D}''^2}\,,\\
 c_{\phi^6}^{(2)}&\to c_{\phi^6}^{(2)} + 2\lambda (r_{\phi^4 D^4}^{(12)} -r_{\phi^4 D^4}^{(6)})-c_{\phi D} r_{\phi D}'\,,
\end{align}
where we omit those that remain invariant. These shifts coincide fully with previous results derived from using equations of motion~\cite{Gherardi:2020det,Chala:2021pll,Chala:2021cgt}, with the exception of the terms in blue. 
These terms are of higher order in the power counting used in the mentioned references, but in other scenarios they must be considered and they can be only captured by field redefinitions~\cite{Criado:2018sdb} or via our explicit matching of scattering amplitudes. Both of these approaches guarantee exactly the invariance of the S-matrix, while equations of motion do not.

As of now, we have successfully applied this method to the SMEFT and other EFTs up to dimension eight including (massless or massive) scalars and gauge bosons, and we are working on extending it to fermions as well~\cite{Chala:2023zzz}.

%%%%%%%%%%%%%%%%%%%%%%%%%%%%%%%%%%%%%%%%%%%%%%%%%%%%%%
%%%%%%%%%%%%%%%%%%%%%%%%%%%%%%%%%%%%%%%%%%%%%%%%%%%%%%
\section{Summary and outlook}
%%%%%%%%%%%%%%%%%%%%%%%%%%%%%%%%%%%%%%%%%%%%%%%%%%%%%%
%%%%%%%%%%%%%%%%%%%%%%%%%%%%%%%%%%%%%%%%%%%%%%%%%%%%%%

Effective field theories are basic tools in our description of nature. On the one hand, they let us calculate physical quantities without knowing the underling theory; e.g., the SMEFT provides an efficient way to characterize new physics at the EW scale in terms of coefficients of higher-dimension operators
without knowing the underlying UV completion. On the other hand, even if we want to consider specific NP models, EFTs offer a comprehensive approach to compare with data while, at the same time, providing a way to sum large logarithms via the use of the RG-improved perturbation theory. 

However, implementing the EFT approach in an automated way suited for systematic phenomenological analyses is a formidable task. This approach includes, for instance, the identification of higher-dimensional operators appearing in an EFT, the extraction of Feynman rules, the calculation of the matching coefficients between EFTs valid at different energy scales, and the calculation of anomalous dimensions to evolve the WCs between different energy scales. Many of these tasks are not feasible in a reasonable amount of time without computer codes as they can involve hundreds or even thousands of WCs, like in the SMEFT and the LEFT.

The interpretation of experimental data in terms of constraints on effective couplings also requires automation, in all but a few restricted cases.
Extensions of the SM are typically parameterized in terms of the SMEFT. Each WC, however, can appears in many observables. 
This is also a challenging task, as it involves computing predictions for a large number of observables and performing global fits with hundreds of experimental constraints. 

With this report, we have documented the large efforts in the theory community in automating calculations within EFTs framework.
They have been presented at the SMEFT-Tools 2022 workshop held at the University of Zurich from 14th-16th September 2022.
The milestones reached so far can be summarized as follows:
\begin{itemize}
    \item For the two major extensions of the SM, SMEFT and LEFT below the electroweak scale, the anomalous dimensions of dimension-six operators have been calculated to leading order and implemented in user-friendly programs such as \texttt{DsixTools} and \texttt{Wilson}. 
    They also provide the complete matching between SMEFT and LEFT at tree level and one-loop order.
    Higher order operators (dimension sever or higher) can be studied with \texttt{Sym2Int}, which automatically build explicit bases of operators for EFTs, given their fields and symmetries.
    Derivation of Feynman rules is enabled by \texttt{SmeftFR}, which also generate
    \texttt{UFO} model files suitable for further symbolic or numerical calculation of matrix elements and cross-sections.
    \item The viability study of specific BSM scenarios can be largely simplified 
    by first matching the NP models to the SMEFT and then determining the experimental constraints on the SMEFT WCs.
    Such matching between any realization of NP can be performed in an automated way at tree level with tools like \texttt{Matchmakereft}, \texttt{Machete} and \texttt{CoDEx}.
    These programs allow to get all effective operators at the EW scale.
    Matching at one-loop is possible in many cases. A completely generic implementation at one-loop level in currently under development.
    \item Several programs (e.g., \texttt{smelli}, \texttt{HighPT}, \texttt{HEPfit}, \texttt{EOS}) have been developed for phenomenological studies, implementing predictions in the presence of EFT WCs which are then used to constraints NP effects in global fits.
    They include observables from a wide range of high-energy physics, 
    such as flavor physics, physics at hadron colliders,
    electroweak precision tests, Higgs physics, and other precision tests of the SM.
\end{itemize}

Many issues and development directions were also addressed during the workshop.
Most probably, the next-to-leading logarithms in SMEFT and LEFT will have to be tackled. 
Given the large number of operators, it is inconvenient to consider separately the two-loop running of
the various EFTs separately. In fact, it is more advantageous to calculate the anomalous dimensions for a generic
EFTs with an arbitrary number of real scalars and left-handed fermions, invariant under a generic gauge structure. 
Result for NLO running of the SMEFT or LEFT WCs can be derived in a second step by specifying the
field content and the gauge group. An ongoing project aims at evaluating the one-loop RG equations for all the dimension-six operators in such generic EFT.

Another relevant development is the complete classification of contributions to dimension six operators in SMEFT which arise from BSM models only at the one-loop level. Such one-loop matching program is using on-shell matching to avoid the use of (redundant) Green's basis.

Also a consistent treatment of $\gamma_5$ in running of four-fermion operators is necessary since it gives rise to evanescent structures in dimensional regularization. Renormalization of chiral Abelian theory up to two loops has been developed recently in the so called BMHV scheme for $\gamma_5$ and it will be extended to the non-Abelian case in preparation for the application to the SM.

%%%%%%%%%%%%%%%%%%%%%%%%%%%%%%%%%%%%%%%%%%%%%%%%%%%%%%
%%%%%%%%%%%%%%%%%%%%%%%%%%%%%%%%%%%%%%%%%%%%%%%%%%%%%%
\section*{Acknowledgements}
%%%%%%%%%%%%%%%%%%%%%%%%%%%%%%%%%%%%%%%%%%%%%%%%%%%%%%
%%%%%%%%%%%%%%%%%%%%%%%%%%%%%%%%%%%%%%%%%%%%%%%%%%%%%%

The organizers are grateful for the generous support from the Pauli Center for Theoretical Studies, the UZH Alumni foundation and the ICCUB-Universitat de Barcelona through the Maria de Maeztu excellence program. Furthermore, we thank Denise Caneve and Monika Röllin for their help with the workshop organization.

\bigskip
Jason Aebischer acknowledges financial support from the European Research Council (ERC) under the European Union’s Horizon 2020 research and innovation programme under grant agreement 833280 (FLAY), and from the Swiss National Science Foundation (SNF) under contract 200020-204428.

\bigskip
The work of Matteo Fael was supported by the European Union’s Horizon 2020 research and innovation program under the Marie Sklodowska-Curie grant agreement No.\ 101065445 -- PHOBIDE.

\bigskip
The work of Javier Fuentes-Mart\'in is supported by the Spanish Ministry of Science and Innovation (MCIN) and the European Union NextGenerationEU/PRTR under grant IJC2020-043549-I, by the MCIN grant PID2019-106087GB-C22, and by the Junta de Andaluc\'ia grants P21\_00199, P18-FR-4314 (FEDER) and FQM 101.

\bigskip
The work of Anders Eller Thomsen has received funding from the Swiss National Science Foundation (SNF) through the Eccellenza Professorial Fellowship “Flavor Physics at the High Energy Frontier” project number 186866.

\bigskip
Javier Virto acknowledges funding from the European Union's Horizon 2020 research and innovation programme under the Marie Sk\l odowska-Curie grant agreement No 700525 `NIOBE', from the Spanish MINECO through the ``Ram\'on y Cajal'' program RYC-2017-21870, the “Unit of Excellence Mar\'ia de Maeztu 2020-2023” award to the Institute of Cosmos Sciences (CEX2019-000918-M) and from the grants PID2019-105614GB-C21 and 2017-SGR-92, 2021-SGR-249 (Generalitat de Catalunya).

% Lukas Allwicher

%
\bigskip
Supratim Das Bakshi and Sunando Kumar Patra are thankful to Joydeep Chakrabortty for numerous suggestions and discussions. Their work is supported by the Department of Science and Technology, Government of India, under the Grant IFA12/PH/34 (INSPIRE Faculty Award);  the Science and Engineering Research Board, Government of India, under the agreement SERB/PHY/2016348 (Early Career Research Award), and Initiation Research Grant, agreement no. IITK/PHY/2015077, by IIT Kanpur. SDB is supported by SRA (Spain) under Grant No.\ PID2019-106087GB-C21 / 10.13039/501100011033 and PID2021-128396NB-100; by the Junta de Andalucía (Spain) under Grants No.\ FQM 101, A-FQM-467-UGR18, and P18-FR-4314 (FEDER).

\bigskip
The work of  Herm\`{e}s B\'{e}lusca-Ma\"{i}to and Paul K\"{u}hler is supported by the Croatian Science Foundation (HRZZ) under the project ``PRECIOUS'' \verb|HRZZ-IP-2016-06-7460|, the project ``Basic interactions and related systems in statistical physics'', as well as the German Science Foundation DFG, grant STO 876/8-1.

\bigskip
The work of Jorge de Blas has been supported by the FEDER/Junta de Andaluc\'ia project grant P18-FRJ-3735. The work of Angelica Goncalves and Laura Reina has been supported by the U.S. Department of Energy under grant DE-SC0010102. Additional support from Italian Ministry of Research (MIUR) under grant PRIN 20172LNEEZ is also acknowledged.

\bigskip
The work of Mikael Chala is supported by SRA under grants PID2019-106087GB-C21 and PID2021-128396NB-I00, by the Junta de Andaluc\'ia grants FQM 101, A-FQM-211-UGR18, P21-00199 and P18-FR-4314 (FEDER), as well as by the Spanish MINECO under the Ram\'on y Cajal programme.

% Juan Carlos Criado

% Athanasios Dedes

%
\bigskip
Renato Fonseca acknowledges the financial support from the Ministerio de Economía y Competitividad through grant number PID2019-106087GB-C22, and from the Junta de Andalucía through grants number P18-FR-4314 (FEDER) and P21\_00199.

% Matthias König

%
\bigskip
The contribution of Miko{\l}aj Misiak and Ignacy Na{\l\c{e}}cz is part of a project that is carried out together with Jason Aebischer, Patryk Mieszkalski and Nudzeim Selimovi\'c. Miko{\l}aj Misiak has been partially supported by the National Science Center, Poland, under the research project 2020/37/B/ST2/02746.

\bigskip
The work of Michal Ryczkowski was supported in part by Polish National Science Center under research grant DEC-2019/35/B/ST2/02008. The research of  Janusz Rosiek has received funding from the Norwegian Financial Mechanism for years 2014-2021, under the grant no 2019/34/H/ST2/00707.

\bigskip
The work of José Santiago is supported by grants FQM 101, P18-FR-4314 (Junta de Andalucía, Fondos FEDER), and PID2019-106087GB-C22 (AEI/10.13039/501100011033).

% Peter Stangl

\bigskip
Peter Stoffer gratefully acknowledges financial support by the Swiss National Science Foundation (Project No.~PCEFP2\_194272).

\bigskip
The work of Avelino Vicente is supported by the Spanish grants PID2020-113775GB-I00 (AEI/10.13039/501100011033) and CIPROM/2021/054 (Generalitat Valenciana), and from MINECO through the Ramón y Cajal contract RYC2018-025795-I.

%%%%%% Bibliography %%%%%%
\bibliographystyle{JHEP}
\bibliography{bibliography}
%%%%%%%%%%%%%%%%%%%%%%%%%%

\end{document}